\documentclass[fleqn,usenatbib]{mnras}

\usepackage[T1]{fontenc}

\DeclareRobustCommand{\VAN}[3]{#2}
\let\VANthebibliography\thebibliography
\def\thebibliography{\DeclareRobustCommand{\VAN}[3]{##3}\VANthebibliography}

\usepackage{graphicx}	%
\usepackage{amsmath}	%
\usepackage{amssymb}	%
\usepackage{float}      %
\usepackage{xcolor}

\usepackage[normalem]{ulem} %

\newcommand{\apc}[1]{\textcolor{black}{#1}}

\newcommand{\refrep}[1]{#1}

\newcommand{\insitu}{\textit{in situ}}

\newcommand{\kpc}{\,\mathrm{kpc}}
\newcommand{\Mpc}{\,\mathrm{Mpc}}

\newcommand{\sdunits}{\,\mathrm{M_{\sun} \, kpc^{-2}}}

\newcommand{\galform}{{\sc galform}}

\newcommand{\stings}{{\sc stings}}

\newcommand{\coco}{\textsc{coco}}

\newcommand{\dhalo}{\textsc{dhalo}}

\newcommand{\tnghundred}{\textsc{tng}-100}

\DeclareRobustCommand{\Msol}{\,\mathrm{M_{\sun}}}
\DeclareRobustCommand{\Mvir}{\,M_\mathrm{\mathrm{vir}}}

\DeclareRobustCommand{\fmb}{f_{\mathrm{mb}}}

\DeclareRobustCommand{\lcdm}{{$\Lambda$}CDM}
\DeclareRobustCommand{\kms}{\mathrm{km\,s^{-1}}}

\defcitealias{Cooper:2010aa}{C10}
\defcitealias{Cooper:2013aa}{C13}
\defcitealias{Cooper:2015ab}{C15}
\defcitealias{Cooper:2017wv}{C17}
\defcitealias{Lacey:2016aa}{L16}

\usepackage{newtxtext,newtxmath}

\title[Stellar haloes of low-mass field galaxies]{Simulations of the accreted stellar haloes 
 of low-mass field galaxies}

\author[Cooper et al.]{Andrew P. Cooper$^{1,2}$\thanks{E-mail:
apcooper@gapp.nthu.edu.tw}, 
Carlos S. Frenk$^{3}$,
Wojciech A. Hellwing$^{4}$,
Sownak Bose$^{3}$
\\
$^{1}$Institute of Astronomy and Physics Department, National Tsing Hua University, 101 Section 2, Kuang Fu Road, Hsinchu 30013, Taiwan\\
$^{2}$Center for Informatics and Computation in Astronomy, National Tsing Hua University, Taiwan\\
$^{3}$Institute for Computational Cosmology, Department of Physics, Durham University, South Road, Durham DH13LE, UK\\
$^{4}$Center for Theoretical Physics, Polish Academy of Sciences, Al. Lotnikow 32/46, PL-02-668 Warsaw, Poland}
\begin{document}

\date{Accepted 2025 May 14. Received 2025 May 10; in original form 2025 January 21}

\pagerange{\pageref{firstpage}--\pageref{lastpage}} \pubyear{2011}

\maketitle

\label{firstpage}

\begin{abstract} 

  \noindent We predict the properties of stellar haloes in galaxies of present-day virial mass $10^8 < M_{200} < 10^{12}\,\Msol{}$ by combining the \galform{} semi-analytic model of galaxy formation, the \coco{} cosmological $N$-body simulation, and the \stings{} particle tagging technique. Galaxies in low-mass haloes have a wide range of stellar halo properties. Their diversity is much greater at fixed stellar mass than fixed virial mass. In the least massive DM haloes capable of supporting galaxy formation, accreted mass fractions are $\lesssim 10$ per cent, and the typical density profile of accreted stars is similar to that of stars formed \insitu{}. \apc{In low-mass galaxies, the radial scale lengths of accreted stellar haloes are smaller than those of stellar discs formed \insitu{}, but the accreted component is still diffuse, not bulge-like. At the scale of galaxies like M33, the surface density of accreted stars exceeds that of the \insitu{} component at  $\approx15\,\mathrm{kpc}$; the accreted surface density at this radius is $3.5\lesssim\ \log_{10}\,M_\star/\Msol\,\mathrm{kpc^{-2}}\lesssim6$}, varying systematically with virial mass. We compare our predictions to observations and other cosmological simulations. A small but significant number of $\sim10^{12}\,\Msol$ haloes with exceptionally inefficient star formation -- `failed
   Milky Ways' -- are more prominent in our model than others; the true abundance of this population is a potential constraint on galaxy formation physics and could be measured by low surface brightness surveys targeting field galaxies with M33-like stellar mass. The stellar particle data for our simulation is publicly available.

\end{abstract}


\section{Introduction}
\label{sec:intro}
In the Lambda cold dark matter ($\Lambda$CDM) cosmogony, the fraction of the present-day stellar mass of a galaxy built up by the accretion of stars from tidally disrupted satellites depends on its total (virial) mass \citep[][hereafter \citetalias{Cooper:2013aa}]{Purcell:2007aa, Guo:2008aa, Cooper:2013aa}. That expectation follows almost directly from the tight relationship between total stellar mass and halo (virial) mass implied, for example, by comparisons of the present-day galaxy luminosity function to the theoretical CDM halo mass function \citep{Frenk88,White:1991aa,Guo:2010aa}. Observations of how accreted stars contribute to the masses and morphologies of galaxies can therefore constrain the masses and mass accretion histories of their dark matter haloes. Such information may, in turn, complement probes of the small-scale clustering of dark matter,  as traced, for example, by the satellite galaxy luminosity function or the abundance and morphology of coherent tidal streams, which are also sensitive to the mass and assembly history of the host \apc{\citep[e.g.][]{Bose:2017aa, Lovell:2021aa, Deason:2022aa}}. Theoretical models have only recently started to explore the observable consequences of galactic accretion on the level of the galaxy population as a whole. Much remains to be understood, particularly 
\apc{how sensitive those predictions are} to uncertainties in models of the galactic baryon cycle. The majority of work to date with cosmological simulations has focused on systems of similar mass to the Milky Way (virial mass $M_\mathrm{vir} \gtrsim 5\times10^{11}\,\Msol$) and above (\citealt{Font:2011uz}; \citetalias{Cooper:2013aa}; \citealt{Pillepich:2014aa, Elias:2018aa, Proctor:2024ab,Proctor:2024aa}). 

Low-mass field galaxies (stellar mass, $M_{\star}\lesssim 10^{10}\,\Msol$), below the knee of the galaxy stellar mass function, which corresponds to a typical \apc{virial mass}\footnote{\apc{We use $M_\mathrm{vir}$ to refer in general terms to the virial mass of bound gravitating systems, comprising both baryonic and dark matter. When reporting the masses of systems in \coco{} and other simulations, we use the approximation $M_\mathrm{vir} \simeq M_{200}$, the mass enclosing a region with 200 times the critical density for closure.}} \refrep{$M_\mathrm{vir} \lesssim 5\times10^{11}\,\Msol$}, are thought to have formed most of their present-day stars `\insitu{}', from the inflow of radiatively cooling gas virialized within their host dark matter halo. The efficiency of \insitu{} star formation is thought to decline rapidly towards lower virial masses, such that no stars form \insitu{} in dark haloes of present-day mass $M_{\mathrm{vir}}\lesssim 3\times 10^{8} \Msol$ \citep{Benitez-Llambay:2020uy}.  Consequently, in this regime, the few most massive satellites that merge with a given galaxy \apc{are expected to} account for most of its accreted stellar mass. Even those most-massive progenitors have substantially fewer stars per unit total mass than the primary galaxies they merge into, such that less massive galaxies are expected to have systematically smaller fractions of their total stellar mass associated with accreted stars (\citealt{Bullock:2001aa,Purcell:2007aa}; \citetalias{Cooper:2013aa}; \citealt{DSouza:2018aa}).

\apc{Two straightforward consequences of this picture have been emphasized in previous work. First, the tight correlation between stellar mass and metallicity means that the accreted component will always be significantly more metal poor than the {in situ} component. Secondly, the total mass and spatial distribution of the accreted component will show substantial variation from galaxy to galaxy, at fixed virial (or stellar) mass. At the scale of the Milky Way and below, these properties are sensitive to the stochastic nature of the mass assembly histories and  orbits of $\lesssim10$ individual progenitors \citep{Amorisco:2017aa}}. There may also be a large intrinsic scatter in the star formation histories of individual progenitor galaxies (again, even at fixed mass).

These expectations are broadly supported by observations of the Milky Way (MW) and M31, where distinct chemodynamical components can be most confidently identified with \insitu{} formation and galactic accretion \citep[although see e.g.][]{Gomez:2017aa, Proctor:2024aa}. The extended, dynamically hot and metal-poor stellar haloes of these galaxies, which are believed to be comprised of mostly accreted stars, account for only a few percent of their total stellar mass. This is a lower limit; the inner regions of these galaxies host dynamically hot, metal-rich components, which may correspond to a particular mode of \insitu{} formation, such as starbursts \citep{Cole:2000aa},  or early and/or massive mergers\footnote{In our usage, at the level of this paper, there is no significant difference between a `classical bulge' and a `stellar halo', or between `accretion' and `merging'.} \citep{Belokurov:2018aa, Belokurov:2023aa}. The MW and M31, ostensibly similar galaxies, appear to have very different stellar haloes. 

Much observational effort is now focused on exploring the range of variation among the accreted stellar haloes of Milky Way analogues \citep[loosely and variously defined, but usually understood to mean isolated, disc-dominated $L^{*}$ galaxies; ][]{Radburn-Smith:2011aa,  Monachesi:2016aa, Merritt:2016aa, Harmsen:2017aa}. This effort is motivated by the need to provide a broader context for the uniquely detailed studies of accreted structure that are possible in the Milky Way and M31. It also reflects the technical challenge of surveying large numbers of galaxies at sufficient depth to detect their stellar haloes \citep{Abraham:2014aa}. Work to date highlights the diversity among galaxies of similar stellar mass; there is some evidence that sufficiently low mass galaxies, such as M33 and M101, may have almost no detectable stellar haloes extending beyond the bulk of stars formed \insitu{}, even if they have a non-negligible mass of accreted stars overall \citep[e.g.][]{van-Dokkum:2014aa, Merritt:2016aa,Jang:2020aa,Jang:2020ab, Gilhuly:2022aa,Gilbert:2022aa}. Recent and upcoming ultra-low surface brightness imaging surveys, including the ARRAKIHS satellite mission\footnote{The \href{https://www.cosmos.esa.int/documents/7423467/7423486/ESA-F2-ARRAKIHS-Phase-2-PUBLIC-v0.9.2.pdf/61b363d7-2a06-1196-5c40-c85aa90c2113?t=1667557422996}{ARRAKIHS proposal (PDF)} is available at \href{https://www.cosmos.esa.int}{www.cosmos.esa.int}.} \citep{guzman:2022aa} are \apc{focused on the extension of this work to larger numbers of galaxies \citep{Martinez-Delgado:2023aa} and lower stellar masses \citep[e.g.][]{Carlin:2016aa,Pucha:2019ab, Annibali:2020aa,Higgs:2021aa,Trujillo:2021aa,Zaritsky:2024aa,Hunt:2024aa}. In parallel, a growing body of evidence points to the existence of extended stellar components around some of the dwarf satellites of the Milky Way, suggestive of accreted stellar haloes \citep[e.g.][]{Kang:2019aa, Chiti:2021aa, Zhang:2021ab, Jensen:2024aa, Tau:2024ab}. The interpretation of these observations is complicated by the possible effects of tidal interaction with the Milky Way, and would benefit from more comparable data on the outskirts of field dwarfs \citep[e.g.][]{Higgs:2021aa, Goater:2024aa}.}

Here, we describe cosmological simulations of accretion onto galaxies across a wide range of virial mass, from $\Mvir \sim10^{8}$ to $10^{13}\,\Msol$. The overall purpose of these models is to provide predictions for \apc{low-mass} galaxies that are anchored in comparisons between observations and predictions of the same model at the Milky Way mass scale. This requires a large dynamic range, and a model of baryonic astrophysics that is well-calibrated and numerically robust across that range. We use the \stings{}  (\textit{Stellar Tags in N-body Galaxy Simulations}) method, which comprises an $N$-body particle tagging technique in combination with semi-analytic modelling of galaxy formation \citep[][respectively, \citetalias{Cooper:2010aa}, 
\citetalias{Cooper:2013aa}, \citetalias{Cooper:2015ab}, and \citetalias{Cooper:2017wv} hereafter]{Cooper:2010aa,Cooper:2013aa,Cooper:2015aa,Cooper:2017wv}. Like other semi-analytic methods, this approach lacks dynamical self-consistency in comparison to hydrodynamical simulations but enables alternative models of baryonic physics and assumptions about the nature of the dark matter to be explored in cosmological volumes much more rapidly at spatial resolutions generally greater than in the hydrodynamical state of the art. We review the approximations involved and the limitations of the method below. 

In this paper, we concentrate on describing \apc{central (i.e.~field) galaxies} in a fiducial model. This will serve as the basis for subsequent \apc{investigation of} alternative baryonic and dark matter realizations of our underlying cosmological simulation, the Copernicus Complexio \citep[][\coco{}]{Hellwing:2016aa}. \apc{In future work, we plan to use this model as a reference for further comparisons to alternative models of galaxy formation and dark matter}. We will make data from these simulations publicly available through incremental releases, the first of which we describe in Appendix~\ref{appendix:datarelease}. We hope that this will assist with future comparisons to other simulations, and with the interpretation of observations. The data we describe here have previously been used for observational comparisons in \citet{Tanaka:2018aa}, \citet{Martinez-Delgado:2023aa}, and \citet{Miro-Carretero:2024aa}.

We proceed in two stages: first describing the construction of our models and their predictions on scales that are well-studied in the literature, then exploring predictions \apc{down to} dwarf galaxy scales. After describing the simulations themselves in section~\ref{sec:simulations}, we begin in section~\ref{sec:galaxy_statistics} by comparing our model to observations and earlier theoretical work, mainly in the context of Milky Way analogues. \coco{} allows us to explore variations with virial mass and stellar mass more comprehensively than previous zoom simulations of individual galaxies. As we discuss in section~\ref{sec:accreted_fraction_trends}, this highlights several potentially significant points concerning the interpretation of accreted mass fractions for Milky Way analogues, which extend and update conclusions from \citetalias{Cooper:2010aa} and \citetalias{Cooper:2013aa}.

In the second part of the paper, starting in section~\ref{sec:low_mass_haloes}, we examine the stellar haloes of galaxies less massive than the Milky Way. Although we consider all isolated galaxies down to the mass resolution limit of our model (effectively the lower mass limit of the field galaxy population), we focus on M33 analogues in order to make contact with recent observations at that scale (section~\ref{sec:low_mass_observations}). In section~\ref{sec:progenitors_and_populations} we present predictions for the mass density profiles of stellar haloes in low-mass galaxies, their progenitor diversity, and their stellar population gradients. Section~\ref{sec:conclusions} summarizes our conclusions. Appendices describe, respectively, the public data set associated with this paper (\ref{appendix:datarelease}), details of our semi-analytic model (\ref{appendix:technical}), a comment on the galaxy size distribution in our models (\ref{appendix:m33_sizes}) and comparison of our models with hydrodynamical Milky Way zoom simulations (\ref{appendix:mhalo_mw_compare}).

\section{Simulations}
\label{sec:simulations}

We use \coco{}, a collisionless zoom simulation of an approximately spherical volume of radius $\sim25\Mpc$ \citep{Hellwing:2016aa}. \coco{} is based on the WMAP7 cosmological parameters (matter, dark energy, and baryon density
parameters $\Omega_{0} = 0.272$, $\Omega_{\Lambda} = 0.728$, $\Omega_{b} =
0.04455$ respectively; Hubble parameter $h = 0.704$, matter overdensity on a
scale of $8 \Mpc$ $\sigma_{8} = 0.81$; and power-law spectral index $n_{s}= 0.967$;
\citealt{Komatsu:2011aa}).  The high-resolution `zoom' region is embedded in a
cubic cosmological box of side-length $100\Mpc$. \apc{The high-resolution region was chosen to 
match the cosmic mean abundance of haloes with viral mass $0.5 < M_{200} < 2\times10^{12}\Msol$, to contain no haloes with $M\gtrsim7\times10^{13}\Msol$, and to have no haloes with
$M\gtrsim7\times10^{14}\Msol$ within $7\Mpc$ of the boundary. This resulted in a region with an overdensity density slightly below the cosmic mean, somewhat similar to the Local Volume.}
The particles in this region have mass $m_{p} = 1.612\times10^{5}\Msol$ and Plummer-equivalent
softening length $\epsilon=0.327\kpc$. Haloes were identified with the Frieds-of-Friends (\textsc{FoF}) 
\citep{Davis:1985tb} and decomposed iteratively into self-bound subhaloes using
\textsc{subfind} \citep{Springel:2001aa}. \apc{As noted above, we report the total masses of haloes using} $M_{200}$, the mass enclosed by a sphere centred on the halo with mean density 200 times the critical density for closure. We use the \dhalo{} algorithm \citep{Jiang:2014aa} to distinguish between `central' and `satellite' subhaloes within FoF overdensities and  to construct merger trees. The \dhalo{} algorithm `promotes' a fraction of \textsc{subfind} subhaloes back to the status of central haloes, based on their spatial isolation and accretion history\footnote{The total masses reported for satellite subhaloes are simply the sum of their constituent particle masses, as determined by \textsc{subfind}, rather than $M_{200}$.} . We provide further details in Appendix~\ref{sec:appendix:definition_dhalo}.

Our dark matter particle mass resolution is \apc{slightly higher than that of} the Illustris TNG-50 simulation ($m_{p,\mathrm{TNG}} = 4.5 \times 10^{5}\Msol$) which has approximately twice the volume of \coco{}. \apc{It is comparable to} the recent ARTEMIS suite of hydrodynamical zoom simulations of individual Milky Way-like haloes \citep[$m_{p} = 1.7 \times 10^{5}\Msol$][]{Font:2020aa}. Both of those simulations used a separate set of less massive collisionless particles to represent stellar mass. 
Our previous work with the Millennium 2 simulation \citepalias{Cooper:2013aa}
had a particle mass of $9.4\times10^{6}\Msol$ (comparable to the
Illustris and \textsc{Eagle} cosmological volume simulations). This earlier model could resolve essentially all the star-forming progenitor haloes of MW analogues. The higher resolution of \coco{} allows us to resolve in more detail the distribution of streams and other structures, and also the internal structure of dwarf galaxies down to the mass of the classical MW satellites. \coco{} can resolve a halo of $\sim2\times10^{8}\Msol$ with 1000 particles, whereas only haloes of $\sim1\times10^{10}\Msol$ were resolved with the same number of particles in Millennium 2. 

\begin{figure*}
  \includegraphics[width=168mm, trim=0.0cm 0cm 0.0cm 0cm,clip=True]{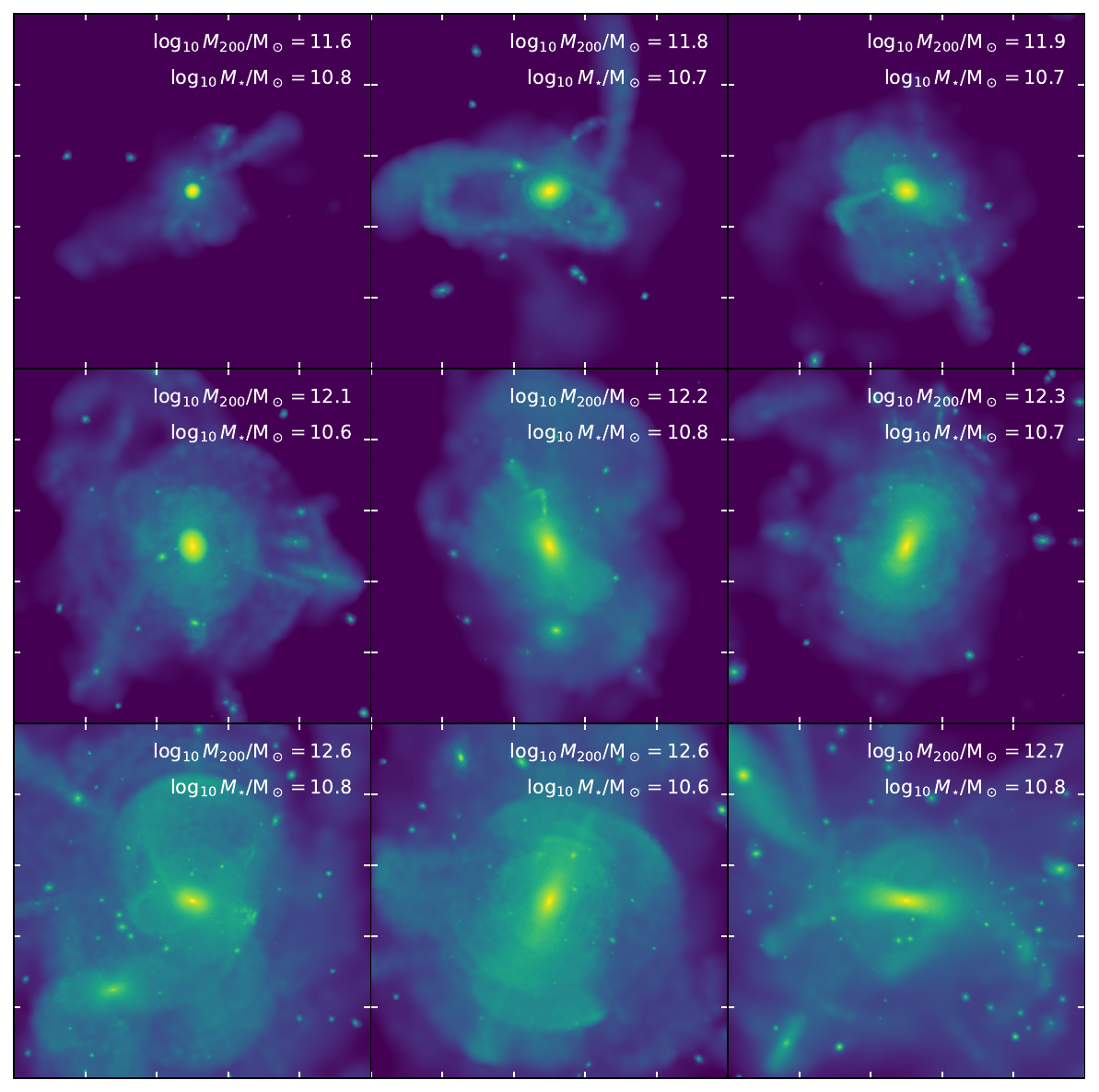}
  \caption{Images of logarithmic stellar mass surface density for 9 of the 14 galaxies in our fiducial \coco{} simulation having stellar mass similar to the MW. Each panel shows a $400\times400$~kpc region.  The logarithmic range of surface density is $0<\log_{10}\Sigma/\mathrm{\Msol\,kpc^{-2}}<9$.}
\label{fig:mw_images}
\end{figure*}

\subsection{Galform models}
\label{sec:galform_models}

Galform is a semi-analytic galaxy formation code \citep{Cole:2000aa,Bower:2006aa,Lacey:2016aa}. It processes dark matter halo merger trees (i.e.\ mass assembly histories) obtained either from collisionless $N$-body simulations or Monte Carlo methods, evolving forwards in time a system of coupled differential equations that describe the evolution of baryonic mass within dark matter haloes. A particular Galform model comprises a specific combination of physical prescriptions and parameter choices within the general Galform framework. Galform provides a comprehensive, internally consistent forward-modelling approach to predicting the star formation histories of present-day galaxies and their high redshift progenitors, constrained by observations predominantly at low and high redshift. 

In this paper we \apc{use only the Galform model} %
described in \citet[][hereafter \citetalias{Lacey:2016aa}]{Lacey:2016aa}\footnote{The \galform{} model we use here is not strictly identical to the published \citetalias{Lacey:2016aa} model, because we alter the prescription for merging satellite galaxies with their hosts (see Section 3.6.1 of \citetalias{Lacey:2016aa}). For consistency with the assumptions of our particle tagging approach, in our variant of the \citetalias{Lacey:2016aa} model, satellite galaxies merge with their hosts at soon as their subhalo is lost in the $N$-body simulation (lost means that no descendant can be identified by searching up to five snapshots ahead). We do not investigate the differences associated with this change in detail.}. The \citetalias{Lacey:2016aa} model was constrained primarily by comparison to optical and near-IR luminosity functions at low redshift; the evolution of the near-IR luminosity function; the present-day galaxy H\textsc{i} mass function and morphological distribution; sub-mm galaxy number counts and redshift distributions; far-IR number counts; and the UV luminosity function of Lyman-break galaxies. The Tully-Fisher, mass-sizes, and mass--metallicity relations at $z\approx0$ were used as secondary constraints (for details see \citetalias{Lacey:2016aa}). In future work, we plan to explore how different choices for astrophysical parameters in  \galform{} affect the results we show here. 
We note that \citetalias{Cooper:2010aa} used an earlier  \galform{} model \citep[close to but not exactly that of][]{Bower:2006aa}, while \citetalias{Cooper:2013aa} and \citetalias{Cooper:2015ab} used a different semi-analytic galaxy formation code, L-Galaxies \citep{Guo:2011aa}. %

\subsection{Particle Tagging} 
\label{sec:tagging}

The idea of the particle tagging approach is to extend the predictions of semi-analytic models (run on merger trees from an $N$-body simulation)
to a six-dimensional phase space, analogous to that of the collisionless `star' particles in a hydrodynamical simulation. This is done by selecting subsets of particles from the $N$-body simulation that follow phase-space trajectories similar to those that would be populated by star particles. Each of these subsets is `tagged' with the mass and other properties of the associated single-age stellar population. The tagging of $N$-body particles is done in post-processing, and therefore does not account self-consistently for the response of the total gravitational potential to the implied motion of baryons relative to dark matter. However, implemented with sufficient care, this approach serves as a good approximation to hydrodynamical methods \citepalias{Cooper:2017wv}. The most significant advantage of particle tagging is that higher resolution can be achieved for much lower computational cost, at the price of less accurate dynamics. 

Particle tagging methods in the literature differ in two important respects: the criteria used to select appropriate subsets of dark matter particles, and the time at which those criteria are applied. Both are important for achieving a reasonable approximation to hydrodynamical methods. In the following subsections, we briefly discuss these issues in the context of \stings{} and the potential limitations of particle tagging in general.

\subsection{Fiducial particle tagging model} 
\label{sec:fiducialmodel}

\apc{In this paper we examine a fiducial \coco{}-\citetalias{Lacey:2016aa} \stings{} model. Our tagging method is controlled by a single free parameter, $\fmb$, the `most-bound fraction' (see Section~\ref{sec:fmb}). In our fiducial model, we adopt $\fmb=3$~per~cent (for reasons given in Section \ref{sec:mass_size}). Fig.~\ref{fig:mw_images} shows $400\times400~\mathrm{kpc}$ images\footnote{To create these images, we smooth the stellar mass of each particle using a cubic spline kernel with compact support (the projected form of the smoothing kernel used by \citealt{Springel05}), as in \citetalias{Cooper:2013aa} and \citet{Martinez-Delgado:2023aa}. The kernel scale for each particle is determined adaptively by the root-mean-square distance to its 16th nearest neighbour; this choice is a compromise between under- and oversmoothing. Note that \insitu{} stars are included in these images; the bright central region of each images corresponds to the \insitu{}-dominated region of the galaxy.} of stellar mass surface density for 9 of the 14 galaxies in this fiducial model with central stellar mass $10.6 \le \log_{10}M_\star/\mathrm{M_{\sun}} \le 10.9$, i.e.\ the set of approximate Milky Way analogues by stellar mass. In Fig.~\ref{fig:mw_images}, the examples are ordered by $M_{200}$ from top to bottom and left to right. As we describe in detail in Section~\ref{sec:accreted_fraction_trends}, the wide range of virial mass corresponding to this stellar mass selection ($11.5 \lesssim \log_{10}M_{200}/\mathrm{M_{\sun}} \lesssim 13.2$) gives rise to the considerable diversity of overall brightness, substructure and satellite counts in the outskirts of the galaxies shown in the figure.}

\subsubsection{Applicability and potential limitations}
\label{sec:tagging_limitations}

In \stings{}, we select an independent\footnote{That is, regardless of whether those particles are being used to trace other populations; each dark matter particle can be tagged with multiple single age stellar populations.} set of dark matter particles to represent every single-age stellar population formed in the semi-analytic model (with ages discretized on the grid of simulation output times). Tagging in \stings{} is therefore carried out `continuously' along every branch of every halo merger tree in the simulation, with  tagged particles selected immediately after a given stellar population forms. This `continuous tagging' circumvents the most serious limitations of naive particle tagging implementations, which tag each branch only once, at the time at which the associated halo becomes a satellite of a more massive system (`tag at infall'). In a `tag at infall' method, the entire star formation history of the branch is collapsed to a single point, which makes it much harder to specify an appropriate energy distribution function. A tagging scheme based on a simple relative binding energy criterion (a `fixed fraction' scheme, see below) is not appropriate in that case. As shown by \citet{Le-Bret:2017aa}, continuous tagging more naturally represents the diffusion of populations in phase space over time, even with the straightforward tagging scheme described in Section~\ref{sec:fmb}. 

\citet{Bailin:2014aa} clearly illustrate the limitations of a `tag at infall' scheme combined with a simple binding energy criterion, by comparing hydrodynamical simulations to $N$-body particle tagging realizations with identical initial conditions. \citetalias{Cooper:2017wv} use the same approach to show that \stings{} achieves good agreement with hydrodynamical methods, even for the central galaxies of Milky Way-mass haloes, where adiabatic contraction and other baryonic effects on the gravitational potential are likely to \apc{have the greatest effect} (see below). 

\citetalias{Cooper:2017wv} further show that the limitations of particle tagging models can be exaggerated if they are compared to hydrodynamical star formation and feedback models that violate the assumptions of the tagging approach (for example, by \apc{significantly} modifying the dark matter distribution in response to star formation); see also \citet{Bailin:2014aa}. Such models are not currently favoured. Early simulations predicted rapid cooling and intense star formation at early times in the major progenitors of present-day Milky Way-like galaxies \citep[e.g.][]{Abadi:2003aa}; in contrast, results over recent decades have favoured weaker, quasi-equilibrium feedback, lower peak star formation rates, more extended star formation histories and limited net modification of the present-day dark matter distribution \citep[e.g.][]{Dutton07, Duffy:2010aa,Brook:2011aa, Brook:2012aa,Scannapieco:2012aa,Crain:2015aa, Grand:2017aa}.

In that regime -- of modest mass-to-light ratios and limited changes to the structure of dark matter haloes due to the motion of baryons -- \citetalias{Cooper:2017wv} argue that the most significant  uncertainty in predicting the properties of stellar haloes comes from the wide range of parameter choices allowed by the underlying galaxy formation model, rather than from the choice of simulation technique. It is therefore reasonable, and informative, to compare stellar halo observations to predictions from semi-analytic galaxy formation models, post-processed with particle tagging, alongside similar results from hydrodynamical simulations.

\subsubsection{Numerical and physical effects on satellite disruption}

The spherically averaged stellar mass density profiles presented in \citetalias{Cooper:2013aa} were shown to be numerically converged at all radii of interest to within an order of magnitude, at the scale of the MW (see appendix A in \citetalias{Cooper:2013aa}). This suggests the internal structure and tidal disruption history of the major accreted progenitors of such systems (in the collisionless case) are robustly predicted at the resolution of Millennium 2. However, the progenitors of less massive galaxies (not studied by \citetalias{Cooper:2013aa}) may experience artificially rapid disruption due to limited resolution \citep[e.g.][]{van-den-Bosch:2018aa,van-den-Bosch:2018ab,Green:2021aa}. This \apc{could} result in overpredicted stellar halo mass surface densities at larger radii. This problem is alleviated but not eliminated by the higher resolution of \coco{}.

\apc{Particle tagging does not account for the possibility that satellite potentials may be modified significantly by baryons. In a CDM model, baryons could enhance the dark matter density cusp (if their central density is sufficiently high) or erase it through dynamical effects \citep[e.g.][]{Navarro:1996aa, Pontzen:2012aa}. These effects are still under debate, with a wide variety of behaviours found in different simulations \citep[e.g.][]{Benitez-Llambay:2019aa, Orkney:2021aa}. Satellites with central cores disrupt rapidly, whereas satellites with central cusps are highly resilient to total disruption, although they may lose a substantial fraction of their mass \citep[][and references therein]{Errani:2023aa,Errani:2024aa}. Particle tagging models are therefore `conservative', in the sense of favouring the survival of bound satellite remnants. Particle tagging may even overestimate the rate of stripping, if baryons enhance the total binding energy of satellites.}

The particle tagging technique may 
\apc{also underestimate the disruption of} satellites that fall to the innermost regions of Milky Way-like  haloes, because they do not experience the stronger tidal effects that would arise from a concentrated, planar disc potential. The impact of \apc{discs on satellite survival} is still under debate \citep[e.g.][]{Zhu:2016aa, Errani:2017aa, Richings:2020aa, Green:2022aa}. \citetalias{Cooper:2017wv} found that differences in orbital evolution appeared to be the main source of differences in mass loss rates of otherwise similar SPH and particle tagging models, although these differences were not attributed specifically to the disc.

Comparisons of simulated stellar haloes to observations may help to distinguish among the wide range of satellite disruption models in the literature. 
\refrep{For example, \citet{Miro-Carretero:2024aa} report statistical differences in the morphology of bright tidal features between the \coco{} models we describe here and the TNG50 simulations; such differences may be related to the factors described above.}
Particle tagging provides a particularly straightforward `benchmark model' in this context, in which the stellar mass, star formation history and scale of progenitor haloes can be calibrated statistically, without any modification of the gravitational potential.

\subsubsection{The $\fmb{}$ parameter}
\label{sec:fmb}

Our particle tagging method introduces a free parameter, $\fmb{}$. This is the mass fraction of dark matter particles, taken in rank order of binding energy, to which each newly-formed stellar population is associated. The stars are assumed to follow the energy distribution function of the dark matter selected by this criterion (that is, equal stellar weight is given to each dark matter particle within a set defined by  $\fmb{}$). The characteristic size and concentration of each newly-formed stellar population therefore varies from halo to halo.
In this respect, the density profiles of newly formed stellar populations are determined by the properties of their dark matter haloes. This is similar to the model of \citet{Mo:1998tf}, but here we assume no net effect on the potential due to the collapse of baryons into galaxies (or the subsequent expulsion of baryons by feedback). Whereas \citet{Mo:1998tf} impose an exponential disc profile and consider angular momentum support explicitly, our model simply specifies the relative binding energies of stars and dark matter at the end of the collapse, and ignores the angular momentum distribution of the star-forming gas. 

As described by \citetalias{Cooper:2013aa} and \citetalias{Cooper:2017wv}, a (spherical) `King-model'-like profile is a natural consequence of imposing an energy threshold on the DM distribution function. The projected surface density profile of each population approaches an exponential profile over time, because tagged particles diffuse across the initial energy threshold (this is the effect captured by `continuous' tagging, as described above). The result is a population of galaxies that have approximately exponential surface density profiles, which scale with the size and concentration of their host haloes.  As described in \citetalias{Cooper:2013aa}, the stellar mass surface density profiles of low mass, isolated late-type galaxies are dominated by stars formed \insitu{} over a few gigayears, at late epochs ($z \lesssim 2$), in potentials that are not strongly affected by subsequent mergers. They are therefore simple to start with, and remain simple to $z=0$. They are almost directly determined by the `initial conditions' set by the time of star formation and the choice of $\fmb$. An appropriate value of $\fmb$ can therefore be determined  with reference to the size--mass relation for isolated low-mass galaxies (see Section \ref{sec:mass_size}).

We note that a fixed, universal value of $\fmb{}$ is a significant approximation for massive galaxies, in part because dynamical instabilities lead to multi-modal binding energy distribution functions for newly formed stars \citepalias[see][]{Cooper:2017wv}. Nevertheless, a constant value works well to approximate the observed size--mass relation. \citet{Rey:2019aa} demonstrate how variations across the plausible range of $\fmb{}$ may affect the distribution of accreted stars in Milky Way analogues and dwarf galaxies.

\section{Galaxy population statistics}
\label{sec:galaxy_statistics}

We first summarize the halo mass, stellar mass and half-light radius distributions predicted by our model at $z=0$. We compare these distributions to data and to hydrodynamical models. These predictions underpin the accreted mass fraction and density profile results shown later in the paper.

\begin{figure*}
  \includegraphics[width=177mm, trim=0.2cm 0.2cm 0.0cm 0cm,clip=True]{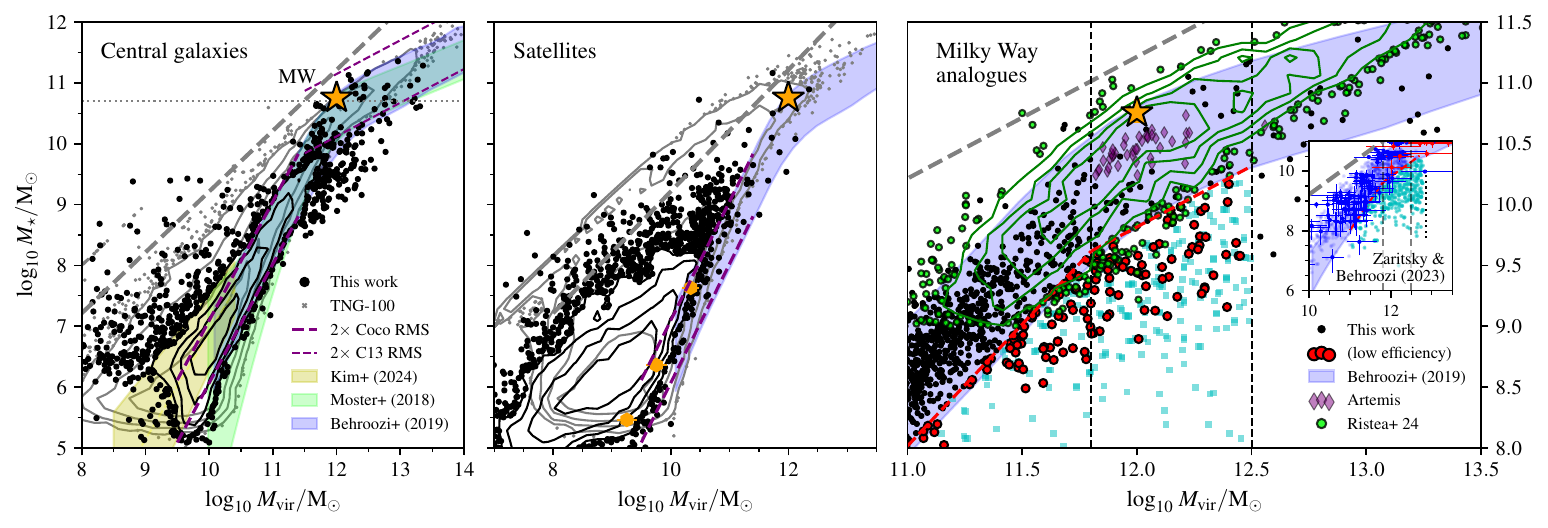}
  \caption{\textit{Left panel}: The relation between stellar mass and virial mass for central galaxies in our fiducial model (black contours/points\apc{; see footnote~\ref{footnote:contours} in the text}) and Illustris TNG-100 (grey contours/points). For masses below the break at $10^{12}\Msol{}$, thick purple dashed lines trace the median of our fiducial model $\pm2$ RMS (beyond the break, thinner lines trace the same locus in the model of C13). Green, purple, and yellow bands show the empirical relations of \citet{Moster:2018aa}, \citet{Behroozi:2019aa}, and \citet{Kim:2024aa} (with their $2\sigma$ ranges). The diagonal grey dashed line corresponds to the conversion of 100\% of available baryons to stars at a given virial mass. An orange star marks the approximate location of the Milky Way. \textit{Central panel}: The same relation for satellite galaxies in our model and TNG-100; satellite mass is defined as the bound mass at the present day. Additional orange points mark approximate values for Milky Way satellites Fornax, Draco, and Sculptor. The  \citet{Behroozi:2019aa} relation and limiting stellar mass for central galaxies are shown for reference. \textit{Right panel}: A zoom-in on the central galaxy relation. Vertical dashed lines indicate an approximate range for virial mass analogues of the Milky Way (see the text). Red points indicate simulated galaxies that fall significantly below the Behroozi et al.\ relation (the region occupied by these `failed Milky Ways' is bounded by the dashed red line). For comparison, we show corresponding points for MW analogues in the Artemis simulation suite \citep[purple][]{Font:2020aa}. The green contours and points show stellar and virial masses derived for MaNGA galaxies by \citet{Ristea:2024aa}. The cyan points, reproduced from fig.~8 of \citet{Zaritsky:2023aa}, are estimates of stellar and virial mass for a sample of field galaxies in the same low-efficiency region. The inset panel shows these data on a larger scale, including galaxies from the same data set that fall outside the low efficiency region (darker blue points).}
\label{fig:mstar_mhalo}
\end{figure*}

\subsection{The stellar mass -- halo mass relation}
\label{sec:mstar_mhalo}

Fig.~\ref{fig:mstar_mhalo} (left panel) shows the stellar mass -- halo (virial) mass (SMHM) relation for central galaxies in our model\footnote{\label{footnote:contours}\apc{In this figure, for our own data and several other data sets, we draw contours of number density in the highest density regions of the distribution and show individual points in sparsely populated regions. The contours correspond to different number densities for different data sets (rather than fractions of the total number of points).}}. We compare our results to the \tnghundred{} simulation, and to empirical fits to observations from \citet{Moster:2018aa} and \citet{Behroozi:2019aa}\footnote{When quoting results from the literature, we do not correct for the  different definitions of halo mass between the simulations ($M_{200}$) and the empirical fits ($M_\mathrm{vir}$) or (small) differences in cosmological parameters other than $H_{0}$, which we rescale to the value assumed by the \coco{} simulations.}. As described in the Introduction, the most prominent general feature of this distribution in a concordance \lcdm{} cosmogony is the well-known break in slope around $10^{12}\Msol$, the approximate virial mass of the Milky Way (we indicate the stellar mass and virial mass of the Milky Way with an orange star in Fig.~\ref{fig:mstar_mhalo}). This break is not very clear in 
our model, because the \coco{} volume is small and somewhat underdense, and hence \apc{contains few haloes more massive than $10^{12}\Msol$}.

The grey dashed line in Fig.~\ref{fig:mstar_mhalo} shows the total mass of baryons available in a halo at the universal fraction, $\Omega_\mathrm{b}/\Omega_{\mathrm{DM}}\approx0.14$. This corresponds to a theoretical maximum stellar mass at a given virial mass. Most haloes are extremely inefficient at converting baryons into stars. Only a handful of central galaxies approach this limit in our model and in TNG. This is consistent with the high-efficiency tail of the distribution suggested by the empirical SMHM relations (the low efficiency tail is discussed below; see also Appendix~\ref{appendix:mstar_mhalo}). Satellites (shown in the central panel of Fig.~\ref{fig:mstar_mhalo}) scatter towards this limit, and can exceed it because a large fraction of their dark matter ( $\gtrsim90$~per cent of their total mass) \apc{can be} lost to tidal stripping before the more tightly bound stars are affected.

A handful of points from our simulation and TNG scatter above the theoretical maximum \apc{stellar mass at lower virial} masses. In the case of TNG-100, there is a clear sequence corresponding to a $1:1$ relation below $M_{\mathrm{vir}}\sim10^{9}\Msol$, most prominent for satellite galaxies. These are likely self-bound clumps of stellar particles with no associated dark matter halo \citepalias[e.g.][]{Cooper:2017wv}.

The slope of the SMHM relation above the break mass is steeper for the two simulations (\coco{} and \tnghundred{}) than for the empirical fits. Haloes in this mass range are not the focus of this paper, so we provide only a brief explanation. In the simulations, the total stellar mass includes all stars bound to the halo (but not to its subhaloes). However, empirical models are fit to observed luminosities, which \apc{may neglect diffuse light below the surface brightness limit}. This can be a substantial fraction of the total luminosity in the most massive systems \citep[e.g.][]{Lauer07,DSouza:2015aa,Bernardi:2017ab,Moster:2018aa}. A range of approaches have been proposed to correct for (or directly measure) the `missing' light. The situation is made more complicated by the open question of whether or not such light should be defined as part of the central galaxy for this purpose, regardless of how accurately it can be measured (see for example our discussion in \citepalias{Cooper:2013aa} and \citepalias{Cooper:2015ab} in addition to the references above). The fiducial models of \citet{Moster:2018aa} and \citet{Behroozi:2019aa} do not include diffuse, low surface brightness regions in their definition of stellar mass; hence, they predict relatively lower central galaxy stellar masses at the highest virial masses. 

This paper is concerned with virial masses less than $\sim10^{12}\Msol$. In this regime, the mean efficiency and its scatter in our model are comparable to the results of \citet{Behroozi:2019aa}. \citet{Moster:2018aa} find an increasing tail of low-efficiency haloes at lower virial masses. 
The scatter towards low efficiency at these masses is considerably smaller in our model than found by \citet{Moster:2018aa}, but similar to that of \citet{Behroozi:2019aa}. It is also broadly consistent with the SMHM relation obtained by \citet{Kim:2024aa}, based on a physically motivated empirical model calibrated against observed and simulated dwarf galaxies. The high-efficiency envelope is similar in both models and our simulations. The distribution from TNG-100 is similar to our model overall,\footnote{This is expected, because both models are calibrated in a similar way to the $z=0$ stellar mass or optical/IR luminosity functions.} although with a somewhat greater fraction of highly efficient galaxies at all masses. In the low virial mass regime, TNG has a broader scatter in efficiency towards lower stellar mass, in better agreement with \citet{Moster:2018aa}.\footnote{\citet[][see their section 4.2]{Moster:2018aa} note that they find lower efficiency at these masses compared to subhalo abundance matching (SHAM) approaches. They explain this as a consequence of allowing the scatter in efficiency to vary with virial mass. Relative to their method, the conventional SHAM assumption of fixed (logarithmic) scatter underestimates the contribution of a tail of highly efficient low-mass haloes to higher stellar mass bins and compensates for this with a higher mean efficiency.} The temporary or permanent suppression or reduction of cooling in slowly-growing haloes after reionisation makes a significant contribution to the greater range of efficiency at lower virial masses  \citep[see e.g.,][]{Rey:2019ab,Benitez-Llambay:2020uy, Wang:2023aa}, as does the wider range of halo formation times at this scale \citep{Kim:2024aa}.

In the right-hand panel of Fig.~\ref{fig:mstar_mhalo}, we zoom in on a region of the diagram around the likely virial mass of the Milky Way (indicated by vertical dashed lines). 
In our model, most galaxies in Milky Way-mass haloes fall in the low-efficiency tail of the distribution (marked by the dashed red line; individual points below this line are shown in red). These are `failed Milky Ways'\footnote{This use of `failed' differs from that in \citet{Bose:2023aa}, where the same term refers to objects that have LMC-like mass at $z\approx2$ but fail to grow because they become a satellite of another, more massive system.}. They are a generic feature of \apc{the most recent generation of} \galform{} models,\footnote{We discuss the origin of this population in \galform{} further in Appendix~\ref{appendix:mstar_mhalo}.} including \citet{Lacey:2016aa}.
 This population has a significant \apc{effect on the distributions of} stellar halo properties \apc{we predict for} stellar-mass selected samples of field dwarf galaxies. We \apc{say more about} this population in the discussion of subsequent figures. 

\apc{These exceptionally low stellar mass galaxies ($8.5\lesssim \log_{10} M_{\star}/\Msol \lesssim9.5\,\Msol$) in Milky Way-mass haloes are not as common in} other cosmological simulations. \apc{These are also not} apparent \apc{in current galaxy surveys}. \apc{Our models are therefore} inconsistent with the \apc{narrower distribution of galaxy formation efficiency in TNG-100}, and \apc{with the empirical} models such as \citet{Behroozi:2019aa}. Such systems are also not seen in suites of Milky Way-analogue zoom simulations, such as Artemis \citep{Font:2020aa} and Auriga \citep{Grand:2024aa}, although these may not \apc{include} a sufficiently wide range of accretion histories to \apc{sample} the low-efficiency tail.

The extent of the failed Milky Way population in our model also appears inconsistent with the range of virial mass estimates for rotation-supported galaxies in the MaNGA IFU survey, recently derived from their sizes and kinematics by \citet[][green contours and points; see also \citealt{Posti:2019aa}]{Ristea:2024aa}. Nevertheless, those data do support a broad scatter in $M_\star$, comparable to that inferred by \citet{Behroozi:2019aa}, such that a fraction of galaxies with Milky Way-like virial mass have only one tenth of the Milky Way stellar mass. Since being very inefficient may be unusual in other respects, it is possible that they fall outside the selection criteria of surveys such as MaNGA. The existence of extremely inefficient galaxies at this mass scale is also consistent with the distribution of virial masses at a given stellar mass inferred by \citet{Zaritsky:2023aa}; a representative sample of \apc{those} data (\apc{see their fig.~8}) are shown in the inset panel.

In contrast to these inefficient galaxies, most models (including ours) and empirical fits suggest the Milky Way itself has been conspicuously successful at forming stars \insitu{} -- it probably lies in the upper envelope of star formation efficiency for a typical $10^{12}\Msol$ halo. For example, the \citeauthor{Moster:2018aa} SMHM relation predicts a typical stellar mass $M_\star \approx 3.2 \times 10^{10}\Msol$ at $M_{200} = 10^{12}\Msol$, roughly half the stellar mass of the Milky Way. As we discuss further below, a corollary of the Milky Way's apparently high efficiency is that its virial mass of the Milky Way appears low compared to most galaxies of the same stellar mass. 

\begin{figure*}
\centering
    \includegraphics[height=79mm,trim=0.0cm 0cm 0.0cm 0cm,clip=True]{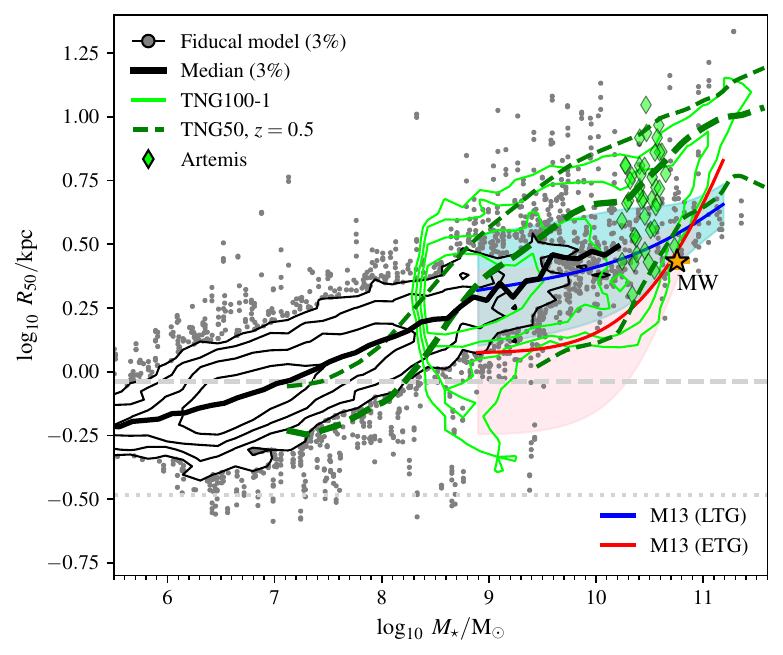}
  \includegraphics[height=79mm,trim=1.8cm 0cm 0.0cm 0cm,clip=True]{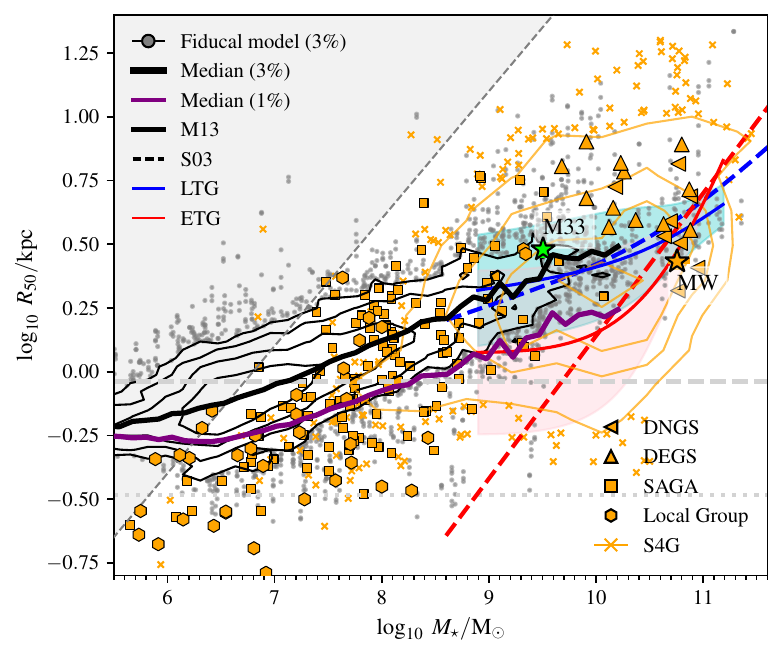}
  \caption{Relation between stellar mass, $M_{\star}$, and projected stellar half-mass
  radius, $R_{50}$, for central galaxies in our fiducial model (black contours, grey points in both panels; \apc{contours are drawn as in Fig.~\ref{fig:mstar_mhalo}}). A thicker black line shows the median, and a purple line the median of a model with $f_\mathrm{mb}=1\%$. The medians are taken over equally spaced bins of stellar mass, requiring more than 10 points per bin. Left: comparison to the TNG-100, TNG-50 and Artemis simulations, showing also the observed relation and $1\sigma$ scatter from \citet{Mosleh:2013aa} for early types (red) and late types (blue). Right: Comparison to a variety observational data, repeating the relations of \citet{Mosleh:2013aa} and also showing the relations of \citet[][without scatter]{Shen:2003aa}. The shaded grey region corresponds to a stellar mass surface density $<10^{6}\,\sdunits$. The gravitational force softening scale, $\epsilon$, and the Plummer-equivalent scale, $2.8\epsilon$, are indicated by dotted and dashed horizontal lines respectively. }
\label{fig:sizemass_central}
\end{figure*}

Estimates of the Milky Way's stellar mass and virial mass are thought to be accurate to $\sim20$~per cent \citep[e.g.][]{Callingham:2019aa,Wang:2020aa, Deason:2021aa}. A virial mass greater than $10^{12}\Msol{}$ would \apc{be more typical of its} stellar mass, but there is relatively little support for higher \apc{virial} masses in the literature \citep[examples include the presence of the Magellanic clouds and the MW-M31 timing argument without correction for low tangential velocity or environmental bias;][]{Li:2008ab,Boylan-Kolchin:2010aa,Wang:2020aa}. In contrast there are many claims in support of a \textit{lower} MW virial mass ($\sim8\times10^{11}\Msol$), which would help to reconcile the observed MW satellite luminosity function with the predictions of galaxy formation models (excepting the Magellanic clouds). Recent studies of stellar halo kinematics also favour masses no larger than $1 \times 10^{12}\Msol$. A lower \apc{virial} mass would \apc{make the high stellar mass of the Milky Way even more atypical in comparison to SMHM relations}. There is circumstantial evidence that the Milky Way is indeed in the high efficiency tail for its virial mass, including the relatively old age of its stellar population, the low mass of its central supermassive black hole, and its compact size, which may indicate a highly concentrated (hence early-forming) DM halo \citep[see e.g.][]{Boylan-Kolchin:2010aa,Licquia:2016aa,Boardman:2020aa}. \apc{ As we discuss further below, the apparently high \insitu{} star formation efficiency of the Milky Way is relevant for interpreting observations of its stellar halo.}

\begin{figure*}
  \includegraphics[width=175mm, trim=0.0cm 0cm 0.0cm 0cm,clip=True]{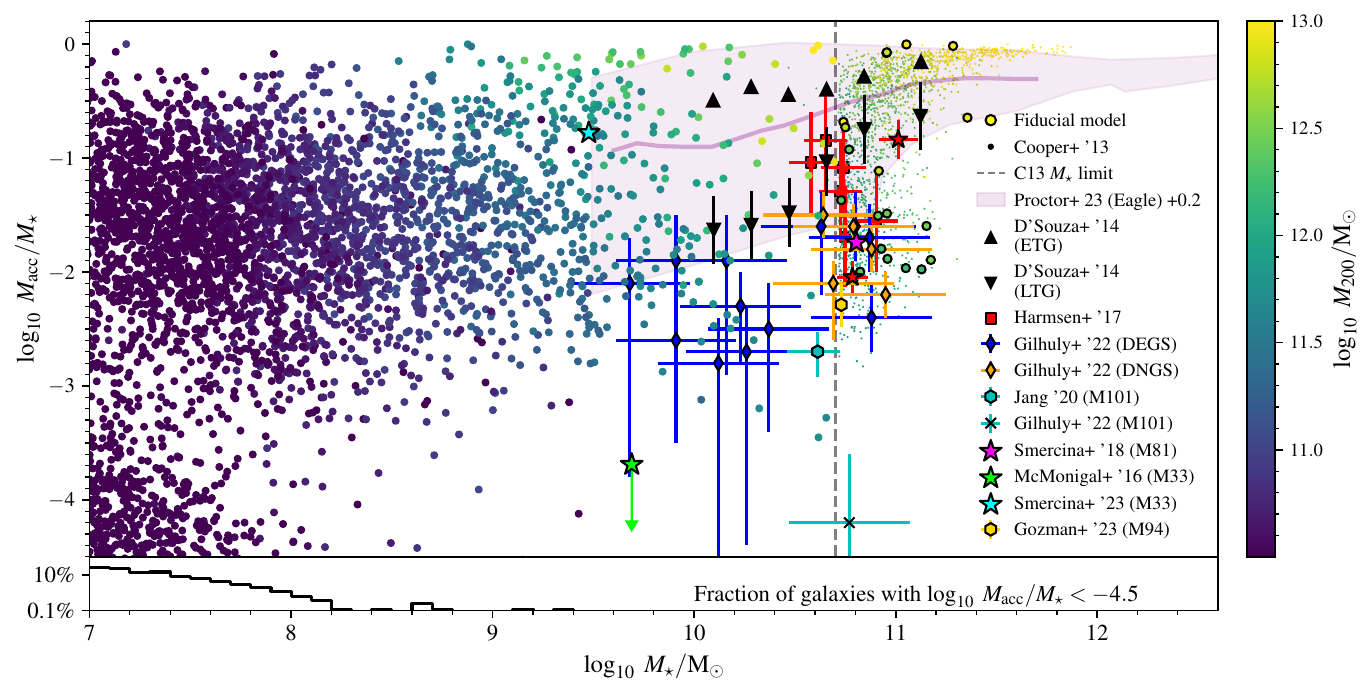}
  \caption{Fraction of stellar mass accreted by galaxies in our fiducial model as a function of stellar mass. The dashed grey line indicates the lower stellar mass limit of the analysis in \citetalias{Cooper:2013aa}. To the right of this line, points from our fiducial model are larger and outlined, to distinguish them from galaxies in the C13 model, which we show as smaller points in the same region. Other symbols show estimates of `stellar halo mass' from \citet{DSouza:2014aa}, the GHOSTS survey \citep{Radburn-Smith:2011aa,Harmsen:2017aa} and the Dragonfly Edge-on Galaxy Survey \citep[DEGS;][]{Gilhuly:2022aa}; the latter includes updated results from the Dragonfly Nearby Galaxy Survey, previous reported by  \citet{Merritt:2016aa}. \citet{DSouza:2015aa} separate their results by morphology, with late types (downward-pointing triangles) having substantially larger error bars than early types (upward-pointing triangles). We do not account for the differences in the definition of accreted mass between these methods (see for example the discussion in \citealt{Gilhuly:2022aa}). In the case of \citet{Gilhuly:2022aa}, the estimates correspond only to mass in the outer region of the galaxy, hence are effectively lower limits. Their result for M101 is consistent with a non-detection (see \citealt{Gilhuly:2022aa} for details). MW and M31 estimates from \citeauthor{Harmsen:2017aa} are shown with red star symbols, the M81 estimate from \citet{Smercina:2018aa} with a pink star symbol and the M94 estimate from \citet{Gozman:2023aa} with a yellow hexagon. For M33 we show two estimates, from \citet{McMonigal:2016ab} (green star; upper limit) and \citet{Smercina:2020aa} (light blue star). The difference between these two values emphasizes the uncertainty of  stellar halo mass estimates in this regime. The shaded pink region and median line shows the distribution of individual galaxies from the Eagle and C-Eagle hydrodynamical simulations \citep[][]{Proctor:2024aa}, with a constant offset of $+0.2$~dex to account for the difference between the $f_\mathrm{IHL}$ quantity in that work and our definition of the total accreted mass (see the text). The lower axis shows the percentage of galaxies in our model that have accreted mass fractions below the lower limit of the plot.}
\label{fig:macc_mstar_g13_3}
\end{figure*}

\subsection{Stellar mass versus half-mass radius}
\label{sec:mass_size}

Fig.~\ref{fig:sizemass_central} shows the projected half-mass radius of central galaxies as a function of stellar mass in our model, compared to data and the Illustris TNG simulations.

As described in Section~\ref{sec:tagging}, the sizes of low-mass central galaxies in our model are determined almost directly by our choice of $\fmb{}$, the free parameter of \apc{\stings}. The stellar mass surface density profiles of these galaxies are dominated by stars formed \insitu{} at $z \lesssim 1$, which are associated with tightly-bound regions of the host potential that are not strongly perturbed by subsequent mergers \citepalias{Cooper:2013aa,Cooper:2017wv}. The slope and scatter of the relation at low stellar mass therefore follows from the range of host virial masses occupied by galaxies of a given stellar mass and the (evolving) relation between concentration and virial mass. Different choices of $\fmb$ \apc{rescale} the typical galaxy size at a given mass, but %
\apc{do not change} the slope of the relation (above the resolution limit of the simulation). For this reason, we use the observed size--mass relation in the low-mass regime \apc{to determine $\fmb$}. 

We show contours and medians for the size--mass distribution in our fiducial $\fmb=3$~per~cent model, and the median alone for an  $\fmb=1$~per~cent variant model (with \apc{exactly the} same star formation histories). 
These distributions count each galaxy in the simulation three times, projected along each of the three coordinate directions of the simulation box. There is a clear systematic shift in size between the two values of $\fmb$. The predicted relation flattens at stellar masses below $10^{6}\Msol$, corresponding to virial masses of approximately $10^{8}$--$10^{9}\Msol$. Although haloes of this mass are well resolved in \coco{}, the regions picked out by \apc{these values of $\fmb$} correspond to $\sim10$ particles and thus are close to the resolution limit. The median relation flattens between the gravitational force softening scale, $\epsilon$, and the Plummer-equivalent scale, $2.8\epsilon$. The flattening occurs at higher stellar mass for the $\fmb{}=1$~per~cent model, indicating the approximate resolution limit. In hydrodynamic simulations, where the stellar mass distribution is represented by a separate set of particles, the sizes of galaxies in low-mass haloes may be affected by the 
transfer of energy from dark matter particles to less massive stellar particles in two-body interactions \citep[][]{Ludlow:2023aa}. This \apc{spurious} effect does not occur in a particle tagging model with a single particle species.

We compare our model with observed size--mass relations from \citet[][S03]{Shen:2003aa} and \citet[][M13]{Mosleh:2013aa}. Both provide separate relations for early and late types. These relations were fit to data at stellar masses greater than $10^{9}\Msol$ and therefore have only limited overlap with our sample of simulated galaxies. The slope and scatter of both late-type size--mass relations are in broad agreement with our results for stellar masses greater than $10^{9}\Msol$ (the M13 relation implies a shallower slope if extrapolated to lower stellar mass). Massive, compact galaxies are rare in our model. Almost no model galaxies fall below the median size--mass relation for early types from M15, \apc{or close to} the much steeper locus of S03. This is the case even for satellite galaxies in our model. One reason for this may be the assumption of a universal value of $\fmb$. Star formation in compact galaxies may be dominated by merger-driven central starbursts, which may have a higher characteristic binding energy than star formation in discs \citepalias[see][]{Cooper:2017wv}.

We also compare our results with the size--mass relation from Illustris TNG-100, which has a similar mass limit to the observed relations. TNG-100 predicts a tail to relatively large sizes compared to the observed relations at the MW mass. In this regime we have few galaxies with which to compare. We note that the Artemis galaxies, which are somewhat lower mass than the MW, are mostly larger than the observed median size. At masses below $\sim2\times10^{8}$, TNG-100 tends towards constant size (the data show a similar trend). Our model predicts a steady decline in size towards lower mass. The TNG-50 simulation shows an even steeper decline, suggesting that the flattening of this relation in TNG-100 
may be due in part to limited resolution.

The shaded region on Fig.~\ref{fig:sizemass_central} corresponds to a mean surface mass density $M_{\star}/\pi R_{50}^2 < 10^{6}\Msol\,\mathrm{kpc}^{-3}$. This is the approximate surface mass density limit of dwarf galaxies observed by the SAGA survey\footnote{Most SAGA dwarfs are likely to be satellites, so perhaps should not be directly compared to predictions for central galaxies; although in our model, most surviving dwarf satellites have similar average sizes to centrals of comparable stellar mass.} and in the Local Group outside the MW and M31 \citep{McConnachie:2012aa}. \apc{Below $M_{\star}\sim10^{7}\Msol$, this limit means we cannot compare upper envelope of sizes in our model with observations. Galaxies less massive than $M_{\star}\sim10^{7}\Msol$ in our model may have artificially large sizes as the result of limited numerical resolution}. %

In Appendix \ref{appendix:mstar_mhalo} we provide a more detailed comparison of the size distribution in a narrow range \apc{around the stellar mass of}
M33, where our data, Illustris TNG and the observed relations \apc{are well sampled}.
\apc{We conclude that}
our fiducial value of $\fmb=3$~per~cent produces a distribution \apc{of} 
sizes in 
reasonable agreement with observations. A more thorough comparison would need to account for effects including projection and the broadening of the observed distribution by 
\apc{measurement}
uncertainties.
Since particle tagging with a universal value of $\fmb$ is a significant approximation to begin with, we consider this level of agreement to be sufficient.

\begin{figure*}
  \includegraphics[width=175mm, trim=0.0cm 0cm 0.0cm 0cm,clip=True]{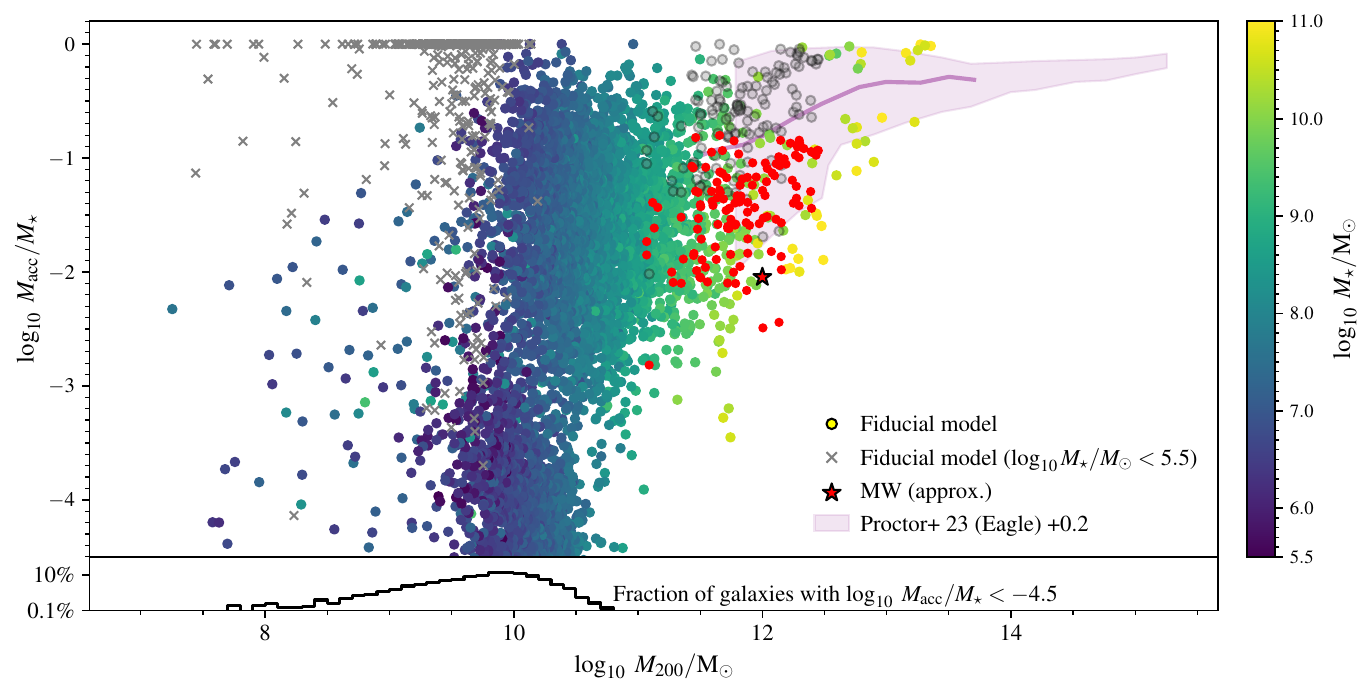}
  \caption{Fraction of stellar mass accreted by galaxies in our fiducial model as a function of virial mass. We mark the approximate location of the Milky Way and the envelope of Eagle/C-Eagle galaxies from \citet{Proctor:2024aa} with an offset in $M_\mathrm{acc}/M_{\star}$ of $+0.2$~dex as in Fig.~\ref{fig:macc_mstar_g13_3}. Grey crosses correspond to haloes with stellar masses $M_{\star}<10^{5.5}\Msol$. \refrep{Predictions for galaxies on this scale may not be reliable in our simulation (see footnote~16) and they are extremely difficult to observe in the field beyond the Local Group. The grey circles highlight "failed Milky Ways" (see Section~\ref{sec:failedmws}); the red points show where  these failed Milky Ways  appear when their \textit{in situ} stellar masses are set to the mean value for their halo mass (see e.g.~Fig.~\ref{fig:loweff_smhm_breakdown}).}}
\label{fig:macc_mhalo}
\end{figure*}

\section{Trends in accreted stellar mass fraction across the galaxy population}
\label{sec:accreted_fraction_trends}

Fig.~\ref{fig:macc_mstar_g13_3} shows the fraction of accreted (`\textit{ex situ}') stellar mass associated with central galaxies in our fiducial model. The colour of each point indicates the virial mass associated with each central galaxy at $z=0$. Accreted stars are defined as those that are bound to the main halo at $z=0$ but which were \textit{not} bound to the main progenitor branch of its merger tree when they formed. Stars that form in subhaloes of the main branch and are later stripped are therefore counted as accreted in our model\footnote{This is in contrast to other studies that treat all stars formed within the virial radius of the host as \insitu{}, even those that formed in satellite subhaloes.}. Since \coco{} has few high-mass galaxies, we also plot results from \citetalias{Cooper:2013aa} for reference (small dots, $M_{\star}>5\times 10^{10}\Msol$). These were generated from the \citet{Guo:2011aa} semi-analytic model applied to the Millennium 2 simulation, also \apc{using \stings{}}.
The two models are consistent with each other in the range of mass for which they overlap.

It is extremely difficult to measure the total accreted mass fraction for real galaxies \citep[e.g.][]{Merritt:2016aa, Kluge:2021aa, Gilbert:2022aa, Gilhuly:2022aa}. We emphasize that a high accreted mass fraction does not equate to a high surface brightness stellar halo component: in projection, much of the accreted mass may overlap with the innermost part of the galaxy, dominated by a much larger mass of stars formed \insitu{}. For that reason, observational estimates of accreted mass are often limited to regions outside a large physical radius (e.g. 20~kpc). Such estimates are lower limits. 
Criteria to isolate an accretion-dominated region have, in general, been motivated by scales relevant only to Milky Way-like systems. \apc{They may not apply} across a wider range of halo and stellar mass. A detailed account of these issues from an observational perspective is given by  \citet{Gilhuly:2022aa}, and in the context of hydrodynamical simulations by \citet{Proctor:2024aa} . In the next section, we present comparisons with aperture-based definitions and a direct comparison of surface brightness profiles, which are much more informative. These issues notwithstanding, Fig.~\ref{fig:macc_mstar_g13_3} illustrates several important features of our simulation. 

Below the \apc{break in the SMHM relation}, the scatter in accreted mass fraction increases sharply, such that galaxies
may accrete from $0.1$~per~cent to more than $90$ per cent of the their present-day stellar mass. The range of our predictions is consistent with observational estimates for nearby late type galaxies by \citet{Harmsen:2017aa} and for stacked \apc{photometry of} early- and late-type galaxies by \citet{DSouza:2014aa}. Massive galaxies with accreted fractions $\lesssim 1$~percent are common in our model. Our findings are in broad agreement with \apc{recent analysis} of the Eagle simulation by \citet[][]{Proctor:2024aa} based on a kinematic definition of `intra-halo light' (IHL). The envelope of \apc{the results} in that work, and their median relation, are shown by the pink shaded region in Fig.~\ref{fig:macc_mstar_g13_3}. We have added a constant offset of $+0.2$~dex to those data to account for the difference between their \apc{definition} of IHL mass and the total accreted mass fraction (see their fig.~6). This consistency suggests that (at high masses) the predictions of 
\citetalias{Lacey:2016aa}-\coco{}, processed with \stings{},
are similar to those of cosmological hydrodynamical models at comparable resolution. \citet{Proctor:2024aa} also find that galaxies with IHL fractions above the median of their relation have lower disc fractions, and vice versa.

Fig.~\ref{fig:macc_mstar_g13_3} demonstrates that, at fixed stellar mass, higher accreted fractions correspond to higher virial mass, on average. Our model %
\apc{does not show any}
significant decline in the maximum accreted fraction at a given stellar mass \apc{for galaxies} below the stellar mass limit of
\citetalias{Cooper:2013aa}. We therefore predict that many field dwarf galaxies with masses in the range $10^{7} <M _{\star} <10^{9} \Msol$ have accreted fractions greater than 10 per cent, and in some cases $\sim100$~per~cent. There is only a weak trend in the lower limit of $f_{\mathrm{acc}}$ down to $M_{\star}\sim10^{8}\Msol$. At even lower stellar masses, a  population appears with extremely low accreted fractions. We discuss this population below. 

Fig.~\ref{fig:macc_mhalo} is a simple permutation of Fig.~\ref{fig:macc_mstar_g13_3}, plotting accreted mass fraction as a function of virial mass, with points now coloured according to their stellar mass\footnote{\label{footnote:tinygals}\refrep{In Fig.~\ref{fig:macc_mhalo}, we use grey crosses to show galaxies with $M_{\star}<10^{5.5}\Msol$. This is our order-of-magnitude estimate of the limit to which our model is reliable at the resolution of \coco{}. In particular, many model `galaxies' (according to our particle-based definition) below this mass scale correspond to individual tagged DM particles accreted by otherwise star-free dark matter haloes. This creates an obvious sequence at 100~per~cent accreted fraction in Fig.~\ref{fig:macc_mhalo}. It is not totally implausible that something similar could happen in reality \citep[e.g.][]{Wang:2023aa}, but the examples we see in the model, at the resolution of \coco{}, are almost certainly not robust. We also exclude galaxies below this mass limit from the averages in virial mass bins shown in Figs.~\ref{fig:density_profiles_by_halo_mass},
\ref{fig:density_parameters_by_halo_mass}, \ref{fig:ages_mhalo}, and \ref{fig:zmet_mhalo}.}}. Two results already shown in Fig.~\ref{fig:mstar_mhalo} 
\apc{are also apparent} in this figure: stellar mass correlates with virial mass below $M_{200}\sim10^{11}\Msol$ and galaxies in the stellar mass range $10^{7} < M_{\star} < 10^{9} \Msol$ are hosted by haloes with a much narrower range of virial mass ($10^{9.5} < M_{200} < 10^{11} \Msol$). The latter result follows directly from the steep slope of the SMHM relation.

Fig.~\ref{fig:macc_mhalo} %
highlights another important point more clearly than Fig.~\ref{fig:macc_mstar_g13_3}. In haloes of mass $M_{200} \gtrsim 10^{12}\Msol$, galaxies in the upper envelope of the conditional stellar mass distribution for their virial mass have the \textit{lowest} accreted mass fractions. In our model, this follows from the fact that galaxies in the upper envelope of the SMHM relation are not outliers because they accrete relatively more stellar mass, but rather because they form relatively more stars \insitu{}. %
\apc{The inefficiency of}
stellar mass growth through accretion is another direct consequence of the steep SMHM relation\footnote{This is apparent from the shallow slope of the SMHM relation above the transition mass, where accretion is the only mode of stellar mass growth.}. The distribution of $f_\mathrm{acc}$ narrows at the highest virial masses ($M_{200}>10^{13}\Msol$) because \insitu{} star formation is shut down completely by AGN feedback, hence $M_\mathrm{acc}$ inevitably saturates to $\sim100$~per cent for all haloes.   

We conclude that, at \textit{fixed virial mass}, the galaxies with the \textit{highest} 
stellar masses should have the \textit{least} contribution from accretion. This result is significant because many studies of stellar haloes focus on samples of Milky Way analogues defined only by stellar mass \apc{(perhaps with indirect limits on virial mass, for example through selections on morphology or star formation rate)}. \apc{This bias is} even more important because there are indications that the Milky Way itself experienced relatively efficient \insitu{} star formation relative to other galaxies of similar present-day virial mass (Fig.~\ref{fig:mstar_mhalo}). This underlines the distinction between Milky Way analogues selected in a narrow range of \textit{stellar mass} (most of which have somewhat more massive dark matter haloes than the Milky Way, hence higher accreted fractions) and those selected by \textit{virial mass} (most of which have considerably \refrep{lower stellar mass} than the Milky Way, hence \refrep{higher} accreted fractions). 
We emphasize that, in our model, a stellar halo mass  of $\sim10^{9}\Msol$ is not at all unusual for a galaxy with the viral mass of the Milky Way (see Fig.~\ref{fig:loweff_smhm_breakdown}). Rather, the unusual feature of the Milky Way is its apparently high \insitu{} stellar mass.

M101 is another notable case. The peak of M101's rotation curve is similar to that of the Milky Way \citep{Jang:2020aa}, indicating (very approximately) a similar virial mass. %
\apc{Observations suggest} that M101 has a total stellar mass comparable to the Milky Way, but an almost negligible stellar halo (or bulge) mass \citep{van-Dokkum:2014aa, Merritt:2016aa}. \citet{Jang:2020aa} estimate the stellar halo mass fraction to be $\sim0.2\%$, whereas \citet{Gilhuly:2022aa} prefer a much lower value, $\sim0.01\%$. The latter  would make M101 an outlier among our models, but the \citet{Jang:2020aa} value is compatible with the lower envelope of the simulated distribution.

In summary (keeping in mind the caveat at the start of this section about the difficulties of interpreting $f_\mathrm{acc}$), our model predicts that galaxies \apc{typically considered to be MW analogues -- those dominated by massive discs that formed \insitu{} --} will often have comparatively `feeble' stellar haloes\footnote{\citet{Proctor:2024ab} find that, in the \textsc{Eagle} simulations, the stellar mass of galaxies with high IHL fractions is typically dominated by the contribution from recent mergers, but the reverse is not always true - galaxies that have had recent massive mergers (`active' merger histories) may have low IHL fractions. They attribute this primarily to enhanced \insitu{} star formation associated with the additional gas supplied by massive mergers; our point here is less specific but essentially similar, namely that $f_\mathrm{acc}$, being defined as a ratio to the \insitu{}-dominated total mass, does not correspond in a direct way to `active' or `quiet' merger histories.}. It also predicts that \refrep{accretion-dominated (potentially early-type)} galaxies in haloes of mass $M_{200} \sim 10^{12}\Msol$ will, on average, have \textit{lower} total stellar masses than the Milky Way. We further predict that the clear correlation between $f_{\mathrm{acc}}$ and either stellar mass or virial mass vanishes for haloes with \apc{around $M_\mathrm{vir}\lesssim10^{12}\,\Msol$.} These conclusions depend on whether or not the ratio between \insitu{} star formation and accretion at a given virial mass in our model is realistic. In this regard, it is notable that a similar distribution of accreted mass fractions and qualitative trend with morphology at fixed virial mass \apc{are} also seen in the \textsc{Eagle} simulations \citep{Proctor:2024aa}. 

\subsection{Failed Milky Ways and other accretion-dominated field galaxies}
\label{sec:failedmws}

\begin{figure}
  \includegraphics[width=\linewidth, trim=0.0cm 0cm 0.0cm 0cm,clip=True]{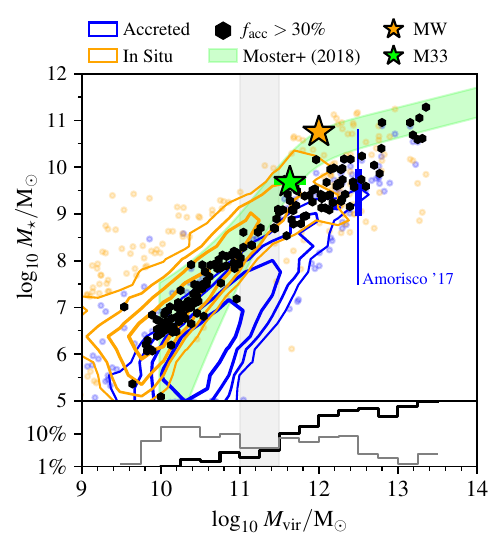}
  \caption{Relations between virial mass and stellar mass formed \insitu{} (orange contours and points) or accreted (blue contours and points). Individual points are used where the density is too low to draw meaningful contours. The black hexagons show the \textit{total} stellar mass for all galaxies that have accreted stellar mass fractions $f_\mathrm{acc} > 30\%$. The lower panel shows the fraction of all haloes of a given virial mass that have $f_\mathrm{acc}>30\%$ (black line) and the fraction of all $f_\mathrm{acc}>30\%$ galaxies with a given virial mass (grey line). %
  The grey band highlights the range $11<\log_{10}\,M_\mathrm{200}/\Msol<11.5$, over which the number of $f_\mathrm{acc}>30\%$ is somewhat lower (see the text). The total stellar masses of the Milky Way and M33 are shown for reference in the upper panel, along with the SMHM relation from \citet{Moster:2018aa}. 
  \apc{The blue contours also show that the stellar halo mass of the MW, $\sim10^9\Msol$, is typical for galaxies with virial mass $10^{12}\Msol$ in our model. The blue vertical line shows the 1\%-99\% (thicker segment, 20\% to 80\%) range of accreted mass in the model of \citet{Amorisco:2017ac}}.}
\label{fig:loweff_smhm_breakdown}
\end{figure}

Another interesting result highlighted by Fig.~\ref{fig:macc_mhalo} is an apparent minimum in the number of galaxies with accreted mass fractions $>10$~per cent around 
$M_\mathrm{200}\sim1$--$5\times10^{11}\Msol$, relative to higher and lower virial masses. This `dip' occurs below the virial mass associated with the small population of failed Milky Ways predicted by our \galform{} model (see Section \ref{sec:mstar_mhalo}). To emphasize this,  Fig.~\ref{fig:macc_mhalo} highlights the failed Milky Ways (as in Fig.~\ref{fig:mstar_mhalo}) and shows explicitly that the large-scale features of the distribution 
\apc{we show in that figure}
would not be changed by restoring these galaxies to the mean of \refrep{the} SMHM relation.

Instead, the `dip' is caused by a \apc{more} subtle effect, which we illustrate in Fig.~\ref{fig:loweff_smhm_breakdown}. This figure separates the contribution of \insitu{} and accreted stellar mass to the SMHM relation, plotting the trends for both components with virial mass. We pick out (with black hexagons) all galaxies with large accreted fractions, $f_\mathrm{acc} > 30\%$, and highlight the range of virial mass around which the `dip' occurs. It is clear that there are fewer  $f_\mathrm{acc} > 30\%$ systems in \apc{that} range (quantified by the grey histogram in the lower panel). Above the dip, the $f_\mathrm{acc} > 30\%$ population is dominated by 
`classical' accretion-dominated systems -- galaxies with virial masses around the \apc{SMHM break}, which have stopped forming stars \insitu{} but continued to grow through accretion. At $M_{200}\sim10^{13}\,\Msol$ most galaxies are accretion-dominated (black histogram in the lower panel). In our model, the failed Milky Ways also produce a sizable number of systems with high $f_\mathrm{acc}$ at $M_{200}\sim10^{12}\,\Msol$ ($10$--$50$ per cent of the total population at that virial mass). This is interesting, but likely model-dependent. As we have shown, the failed Milky Ways are not responsible for the `dip' (although they perhaps make it somewhat more prominent in Fig.~\ref{fig:macc_mhalo}). 

\begin{figure*}
  \includegraphics[width=168mm, trim=0.0cm 0cm 0.0cm 0cm,clip=True]{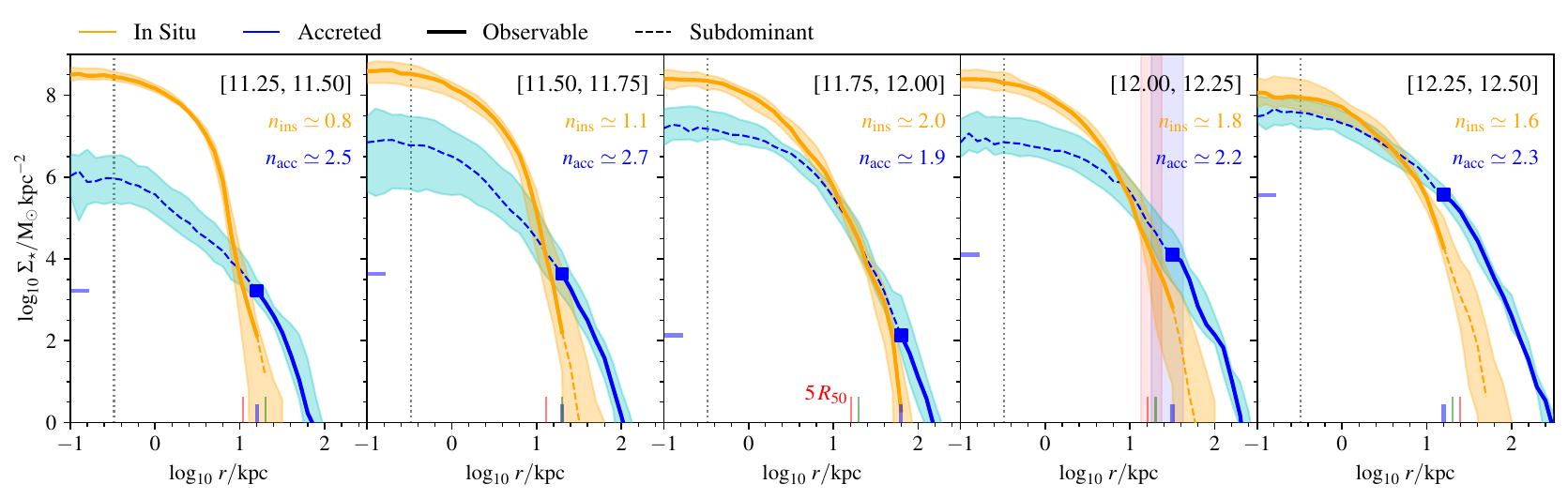}
  \caption{Stellar mass surface density profiles of M33 analogues defined by stellar mass (see the text and Fig.~\ref{fig:mstar_mhalo_accreted_properties}), separated into logarithmic bins of virial mass as indicated in the top right of each panel. The \insitu{} and accreted contributions are shown separately (for clarity, we do not show the sum of the two components). \refrep{Orange and blue labels give the approximate Sersic index of the median \insitu{} and accreted profiles, respectively.} Thick blue lines indicate the region of each profile in which the density of accreted stars is at least one order of magnitude greater than that of \insitu{} stars. This defines a `transition' point discussed in the text, beyond which accreted stars dominate observed profiles (on average). This point is marked with a square symbol. The corresponding transition radius and density are indicated with a short blue bar on each axis. Our definition of the `observable' accreted mass corresponds to the integral under the thick blue line. Likewise, the thick orange lines show the regions dominated by \insitu{} stars, or where the two components contribute similar mass. On the horizontal axis, a red bar marks $5R_{50}$ and a green bar a fixed radius of $20\,\mathrm{kpc}$. The shaded regions in the $[12,12.25]$ panel show the range of transition radii measured for galaxies in the EAGLE simulations by \citet{Proctor:2024aa} for discs (blue) and spheroids (red).}  
\label{fig:m33_profiles_by_mhalo}
\end{figure*}

The relevant question is therefore why the number of galaxies with high accreted mass fractions rises again at lower virial mass.  Fig.~\ref{fig:loweff_smhm_breakdown} shows this is simply due to the increasing scatter of \insitu{} formation efficiency towards the lowest masses. Below the transition mass, \insitu{} stars almost always dominate the total stellar mass. Approximately, the most significant accreted progenitors of a central galaxy with a given virial mass have stellar masses drawn from the conditional stellar mass function at a virial mass one decade lower than that of the central. \apc{For example,} the accreted components of systems with virial mass $\log_{10}\,M_{200}/\Msol\sim10.5$ are dominated by the distribution of progenitor stellar masses associated with virial mass $\log_{10}\,M_{200}/\Msol\sim9.5$. The range of \insitu{} star formation efficiency at a given stellar mass (orange contours) narrows with increasing virial mass \citep[see e.g.][]{Kim:2024aa}, hence the likelihood of a high accreted fraction diminishes, until \insitu{} formation efficiency is strongly suppressed by AGN feedback around the \apc{SMHM break}. At virial mass scale of the `dip', it is therefore relatively less common to find progenitors of comparable mass to their host galaxies (in other words, 
major mergers by stellar mass are less likely). This is a non-trivial and likely model-dependent conclusion, not least because the slope and scatter of the SMHM relation may evolve with redshift. We revisit this idea briefly in Sections~\ref{sec:observable_density_profiles} and \ref{sec:conclusions}.

\section{Stellar Haloes of low mass field galaxies}
\label{sec:low_mass_haloes}

We now turn to the main focus of the paper, the accreted stellar haloes of field dwarf galaxies. For massive galaxies, our fiducial \coco{} model predicts stellar mass surface density profiles at large radii that are similar to observational results and existing simulations in the same mass range. In Appendix~\ref{appendix:mhalo_mw_compare}, we validate this by comparing to observed surface density profiles for Milky Way-like galaxies and to hydrodynamical models. This motivates us to explore the a mass-limited sample of less massive haloes in \coco{}. In this section, we first provide illustrative results for central galaxies in a fiducial stellar mass range $3 < M_{\star} < 7 \times10^{9} \mathrm{\Msol}$. For convenience we refer to these as `M33 analogues'. We then examine variations with mass.

\subsection{The stellar haloes of M33 analogues}
\label{sec:m33_stellar_haloes}

It is well established that galaxies with stellar mass similar to the Milky Way have a wide range of stellar halo masses \citep[or accreted mass fractions;][]{Harmsen:2017aa,Merritt:2016aa}. This is primarily because the accreted mass fraction correlates with virial mass (as shown in Fig.~\ref{fig:loweff_smhm_breakdown}), and the shallow slope of the SMHM relation above the SMHM break (associated with AGN `radio mode' feedback) implies a broad range of virial mass for galaxies with Milky Way-like stellar mass. The opposite is true for the least massive field galaxies; the steepness of the SMHM relation in this regime means that a narrow range in stellar mass corresponds to a relatively narrow range in virial mass as well. For our fiducial M33 analogues, \apc{which have stellar masses slightly below the SMHM break}, the picture is more complicated and potentially model-dependent. \apc{This is particularly true in our model} due to the population of `failed Milky Ways' described above and in Appendix~\ref{appendix:mstar_mhalo}. \apc{Due to that population,} galaxies selected from our model in a narrow range around the M33 stellar mass span a range in virial mass that is wider than expected from empirical SMHM relations, such as \citet{Moster:2018aa} and \citet{Behroozi:2019aa}. It may be helpful to keep this point in mind when reading the following sections.

\subsubsection{Density profiles and observable stellar haloes}
\label{sec:observable_density_profiles}

\begin{figure*}
  \includegraphics[width=168mm, trim=0.0cm 0cm 0.0cm 0cm,clip=True]{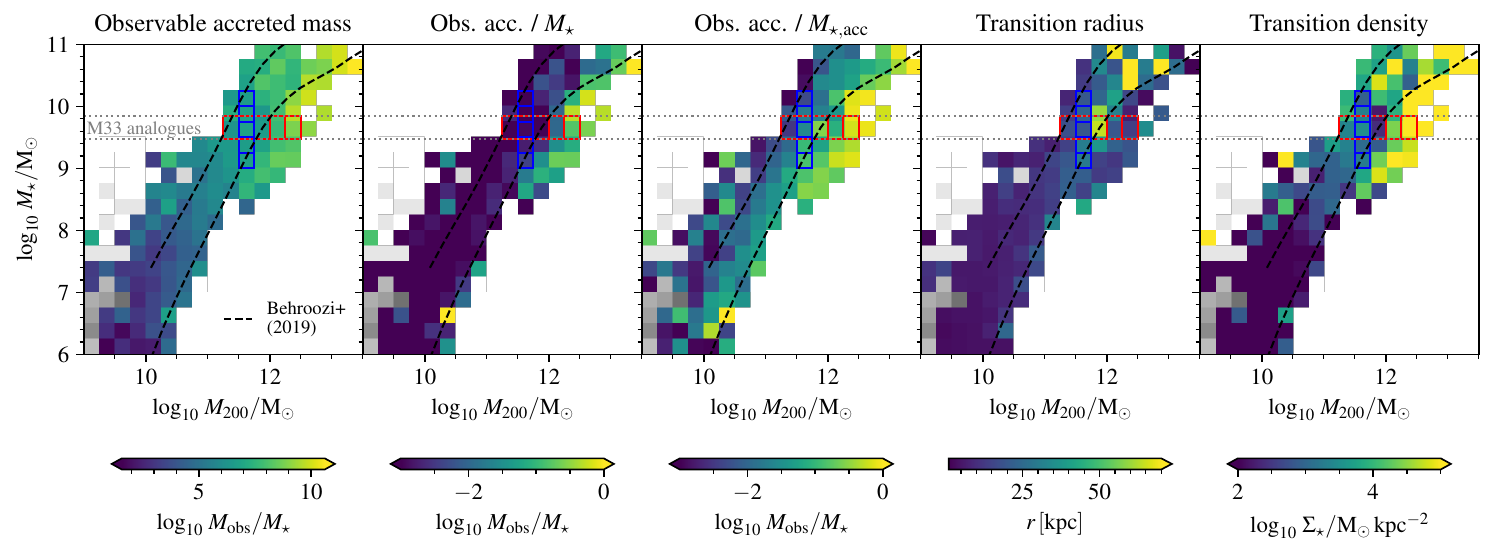}
  \caption{These `heatmaps' show five statistics of the `observable' stellar haloes around galaxies in our simulations, averaged in bins of stellar mass and virial mass. From left to right respectively, panels show the total mass of accreted stars beyond the transition radius; the same quantity as a fraction of total galaxy stellar mass (including the \insitu{} component); the same quantity as a fraction of accreted stellar mass only; the transition radius itself, and the total stellar mass density at the transition radius. Our fiducial M33 analogue mass range is indicated by horizontal dotted lines; the red boxes correspond to the panels in Fig.~\ref{fig:m33_profiles_by_mhalo} and the blue boxes to panels in Fig.~\ref{fig:dwarf_profiles_fix_mstar}.}
\label{fig:mstar_mhalo_accreted_properties}
\end{figure*}

Fig.~\ref{fig:m33_profiles_by_mhalo} shows the median stellar mass surface density profiles of accreted and \insitu{} stars for our M33 stellar mass analogues, divided into equally-spaced logarithmic bins of virial mass. All our M33 analogues have a non-negligible fraction of accreted stars, as indicated in previous figures. The increasing trend of $f_{\mathrm{acc}}$ with virial mass for M33 analogues is apparent in Fig.~\ref{fig:m33_profiles_by_mhalo}, with the exception of one bin, $11.75 < \log_{10} M_{200}/\mathrm{M_{\odot}} < 12$. Regardless of their virial mass, \insitu{} stars always dominate the central regions of galaxies in this sample.  We see a modest increase in the characteristic radius of the \insitu{} component with virial mass. Nevertheless, accreted stars always dominate the profile at sufficiently large radius. Blue squares mark the radius at which the density of accreted stars is an order of magnitude larger than that of \insitu{} stars. We call this the \textit{transition radius}. We refer to the region beyond the transition radius as the `observable' stellar halo, because the mass and other properties of accreted stars in that region could, in principle, be recovered from observations, \apc{with less} uncertainty \apc{regarding} the contribution of \insitu{} stars\footnote{Recent hydrodynamical simulations suggest that \insitu{} halo stars could make a substantial contribution at large distances, e.g. \citet{Monachesi:2016aa}, \citet{Font:2020aa}; see however \citetalias{Cooper:2015ab}.}. \citet{Proctor:2024aa} measure a similar transition radius for galaxies in the \textsc{Eagle} simulation, although they consider mostly 
more massive galaxies. They define the transition as the radius where the density of accreted stars exceeds that of \insitu{} stars, which by construction must be smaller than our definition. In Fig.~\ref{fig:m33_profiles_by_mhalo}, we indicate the ranges of transition radii reported in \citet{Proctor:2024aa} for discs and spheroids, in the range of $12<\log_{10} M_{200}/\mathrm{M_\odot}<12.5$; these results are broadly consistent with ours, especially given that the sample of \citeauthor{Proctor:2024aa} is not limited in stellar mass, and hence will include many galaxies with higher-amplitude \insitu{} profiles (see also Fig.~\ref{fig:mstar_mhalo_accreted_properties}).

As discussed above, it is extremely difficult to identify this transition radius in practice. Various approximations have been proposed to standardize comparisons between observations and models. For example, in their comparison of galaxies observed by the Dragonfly survey to mock observations from Illustris TNG, \citet{Merritt:2020aa} considered separations between \insitu{} and accreted stars at a threshold radii of $20$~kpc and $2R_{50}$, and a threshold stellar mass surface density $10^{6}\,\mathrm{M_{\odot}\,kpc^{-2}}$ (the latter following \citetalias{Cooper:2013aa}). \citet{Gilhuly:2022aa} further consider $5R_{50}$ and $5R_\mathrm{d}$, where $R_\mathrm{d}$ is the exponential scale-length of the galactic disc \citep[following][]{Pillepich:2014aa}. Some of these definitions scale with the central galaxy, others do not. As discussed by \citet{Gilhuly:2022aa}, the use of different definitions complicates comparisons of `stellar halo' masses or mass fractions between different studies.

A significant problem with such definitions \citep[as noted by][]{Gilhuly:2022aa} is that they are mostly `tuned' to the surface brightness profiles of central galaxies at a particular (MW-like) mass.
Fig.~\ref{fig:m33_profiles_by_mhalo} confirms that, in our M33 analogue range of stellar mass, any approximation of the transition radius by a fixed physical radius or radius at a fixed density can only apply over a narrow range of both stellar mass and virial mass. %
An `adaptive' definition, for example $5R_{50}$, is not an obvious improvement over a fixed radius. The problem of how to identify the transition radius reliably in real galaxies is clearly a difficult one, which we do not address here\footnote{Fig.~\ref{fig:m33_profiles_by_mhalo} (and our results in general) suggest that the azimuthally-averaged transition radius could be inferred from deviations relative to a Sersic profile fit to the inner (overwhelmingly \insitu{}) region of the galaxy, as in \citet{DSouza:2014aa}. This clearly requires further study and care in accounting for inclination effects.}.

Fig.~\ref{fig:mstar_mhalo_accreted_properties} shows how the properties of the `observable stellar halo' vary across the SMHM diagram. The M33 analog bins used in Fig.~\ref{fig:m33_profiles_by_mhalo} are shown as red boxes. For these galaxies, the observable stellar halo mass increases with virial mass, accounting for $\gtrsim10$ per cent of the total stellar mass at the highest virial masses. It can be seen that the range of M33 analogues by stellar mass includes the subset of MW analogues \textit{by virial mass} with the lowest \insitu{} formation efficiency - the `failed Milky Ways'. The central panel of Fig.~\ref{fig:mstar_mhalo_accreted_properties} shows that the \textit{total} accreted mass fraction scales roughly in proportion to the observable accreted mass fraction, reaching $\gtrsim30$~per cent at the highest virial masses. 

The average transition radius peaks in our central virial mass bin, $11.75<\log_{10} M_{200}/\mathrm{M_{\odot}}<12$ at $r\approx63\,\mathrm{kpc}$, falling to $\approx10-15\,\mathrm{kpc}$ at the highest and lowest viral masses of the M33 analogue sample. 
As shown in Fig.~\ref{fig:m33_profiles_by_mhalo}, the amplitude of the \insitu{} density profile is roughly constant with virial mass (by construction, because these galaxies were chosen in a narrow range of stellar mass). However, the Sersic index of the \insitu{} profile increases from $n\lesssim1$ from the lowest virial masses to \refrep{$n\sim1.6$ for the highest}. Massive mergers (associated with the growth to higher virial mass) reduce the concentration of the \insitu{} profile, which is not restored by the formation of significantly more \insitu{} stars (again by construction, for this sample). Conversely, the Sersic index of the accreted profile \refrep{does not show a trend with} virial mass \refrep{for this subset of galaxies}, while its amplitude increases steadily. The combination of these trends explains the behaviour of the \apc{transition radius} in Fig.~\ref{fig:mstar_mhalo_accreted_properties}.

\begin{figure*}
  \includegraphics[width=168mm, trim=0.0cm 0cm 0.0cm 0cm,clip=True]{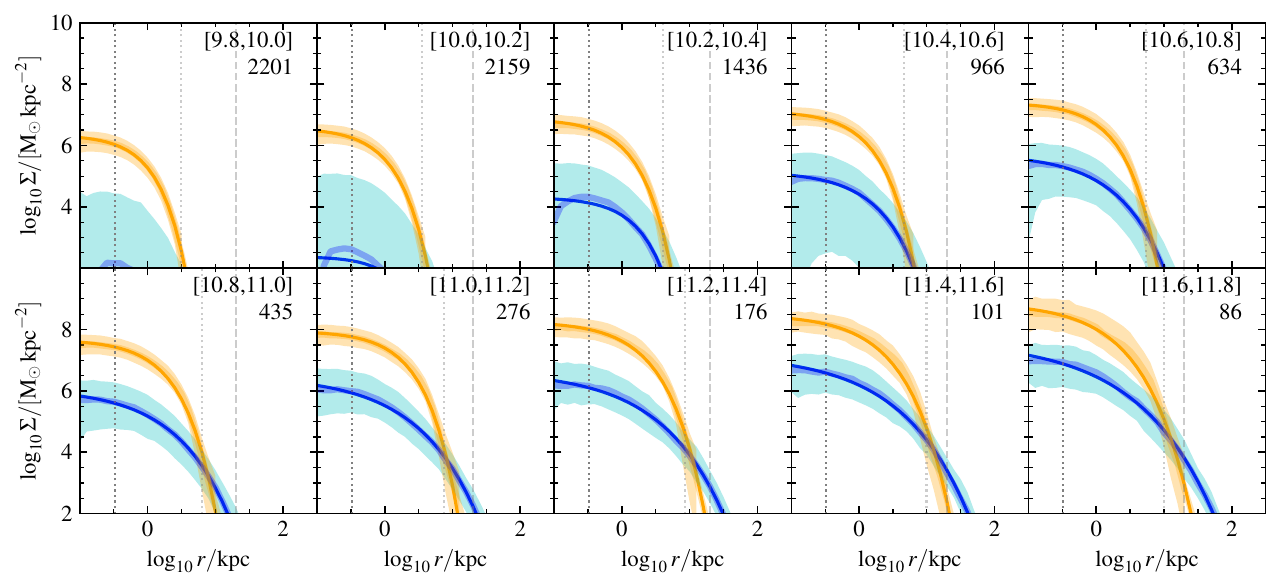}
  \caption{Decomposition of average projected stellar mass surface density profiles for low mass galaxies in our model, in bins of virial mass. Fainter, thicker lines show the median density of \insitu{} stars (orange) and accreted stars (blue). Darker lines of the same colour show single-Sersic fits to the median profiles.  Shaded regions show the corresponding 16-84 percentile range. In each panel, we show the range of $\log_{10} M_{200}$ and the corresponding number of galaxies. Vertical lines, from left to right in each panel, indicate the force softening scale (dark dotted), $5R_{50}$ (light dotted) and $20\,\kpc$ (dashed), respectively. \refrep{As in Fig.~\ref{fig:macc_mhalo}, galaxies with a total stellar mass $M_{\star}<10^{5.5}\Msol$ are excluded from these averages; this makes only a very small difference to the average accreted profiles in the two lowest virial mass bins.}}
\label{fig:density_profiles_by_halo_mass}
\end{figure*}

\apc{Figs.~\ref{fig:m33_profiles_by_mhalo} and \ref{fig:mstar_mhalo_accreted_properties} suggest how failed Milky Ways (should they exist) might stand out from the general population of low-mass galaxies. Their \insitu{} components remain relatively concentrated, despite the substantial mass fraction of accreted stars, suggesting either M33-like discs embedded in bright haloes (like M91, for example) or else a `two-component' elliptical, with a diffuse halo around a compact central core.}

The central stellar mass bin in our M33 analogue range is also peculiar, for a different reason. It is dominated by galaxies in which enough \insitu{} stars have scattered outwards to overwhelm the accreted component even at very large radii (i.e. $\gtrsim 10R_{50}$). In lower virial mass bins, the \insitu{} stars are more concentrated; in higher virial mass bins the accreted halo is more massive as described above. The result, in both cases, is a smaller transition radius.  Based on the profiles in this central mass bin, the corresponding galaxies may be early or intermediate-types (S0-like), although such galaxies are rare in observations at these masses (at least in isolated environments).

\subsubsection{Accreted bulges}
\label{sec:accreted_bulges}

\begin{figure}
\centering
  \includegraphics[height=110mm, trim=0.0cm 0cm 0.0cm 0cm,clip=True]{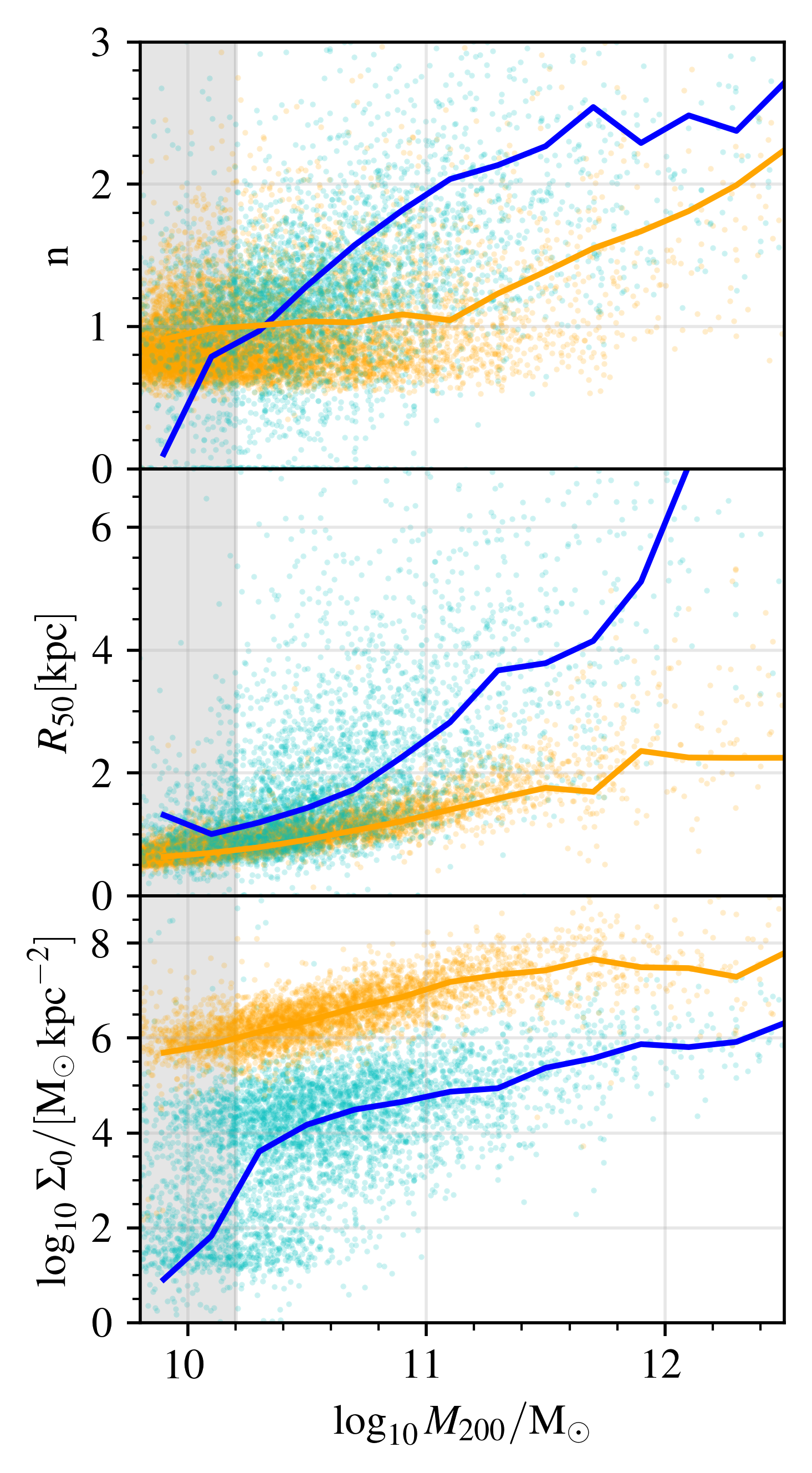}
  \caption{The parameters of single-Sersic profile fits to the \insitu{} stars (orange points) and accreted stars (blue points) in our simulation. From top to bottom, panels plot virial mass against Sersic index, $n$; half-mass radius,  $R_{50}$; and central surface density, $\Sigma_{0}$. Lines show medians of the full distribution. The shaded region indicates the range of virial mass for which the extremely low mass of accreted stars makes profile fits unreliable for typical galaxies; the blue lines are highly uncertain in this regime.}
\label{fig:density_parameters_by_halo_mass}
\end{figure}

For M33 analogues, the discussion in the previous section regarding the transition between \insitu{} and accretion-dominated regions of the stellar mass density profile, concerns densities far below those typically observable. The regions of these galaxies \apc{that can be observed with current wide-field surveys} are \apc{almost certainly} overwhelmingly dominated by \insitu{} stars. This implies that, in this \apc{sub-Milky Way} mass range, `classical' bulges do not form by accretion in our model. Accreted stars in these galaxies never reach a central surface density comparable to that of the \insitu{} component, unless they also comprise a large fraction of the total stellar mass -- in which case they dominate at all radii, not just at the centre. 

Taking this result at face value, the absence of classical (i.e. accretion-dominated, `mini-elliptical') bulges in low mass galaxies appears natural in $\Lambda$CDM. This is simply because, at lower masses, there are many galaxies that do not accrete significant amounts of stellar mass (although there are also plenty with significant accretion). Any bulge-like structures that are observed in galaxies without significant accretion must be associated with \insitu{} star formation (i.e. pseudo-bulges). This diversity holds up to galaxies of luminosity $\sim 10 L_{\star}$. Therefore, our model does not conflict with observations of $\lesssim L_{\star}$ galaxies with little or no mass in a kinematically hot spheroidal component, as discussed by \citet{Peebles:2020aa}.

However, our model of the \insitu{} stellar structure \apc{may be} too simplistic to draw a firm conclusion on this point. A particle tagging approach like ours, in which the newly formed stars are distributed uniformly within a fixed (fractional) binding energy threshold, generically produces an \insitu{} component with a spherical exponential profile, rather than a disc\footnote{As shown in \citetalias{Cooper:2017wv} and discussed above, concentrated \insitu{} `starbursts', driven for example by instabilities, interactions and dissipative mergers, are likely to have a different binding energy distribution; this is not currently accounted for in our approach. If it were, it would make \insitu{} bulges more prominent.
}. If the \insitu{} stars were concentrated into thin discs, the innermost accreted stars might be dense enough to be detectable as a central spheroidal component when galaxies are projected edge-on, (i.e. to `pop up above the disc' in the innermost part of the surface brightness profile) while still being fainter than the \insitu{} stars at larger radii or in a face-on projection . This seems most likely in cases similar to that of our $11.75<\log_{10} M_{200}/\mathrm{M_{\odot}}<12$ bin for M33 analogues, in which we see less than an order of magnitude difference between the central densities of the two components. It seems less likely in galaxies that lie on or above the average SMHM relation. For those kinds of galaxies, we find the central density of accreted stars is $\lesssim1$~per cent of the stars formed \insitu{}. 

\subsection{Galaxies less massive than M33}
\label{sec:below_m33}

\begin{figure*}
  \includegraphics[width=\linewidth,trim=0.0cm 0cm 0.0cm 0cm,clip=True]{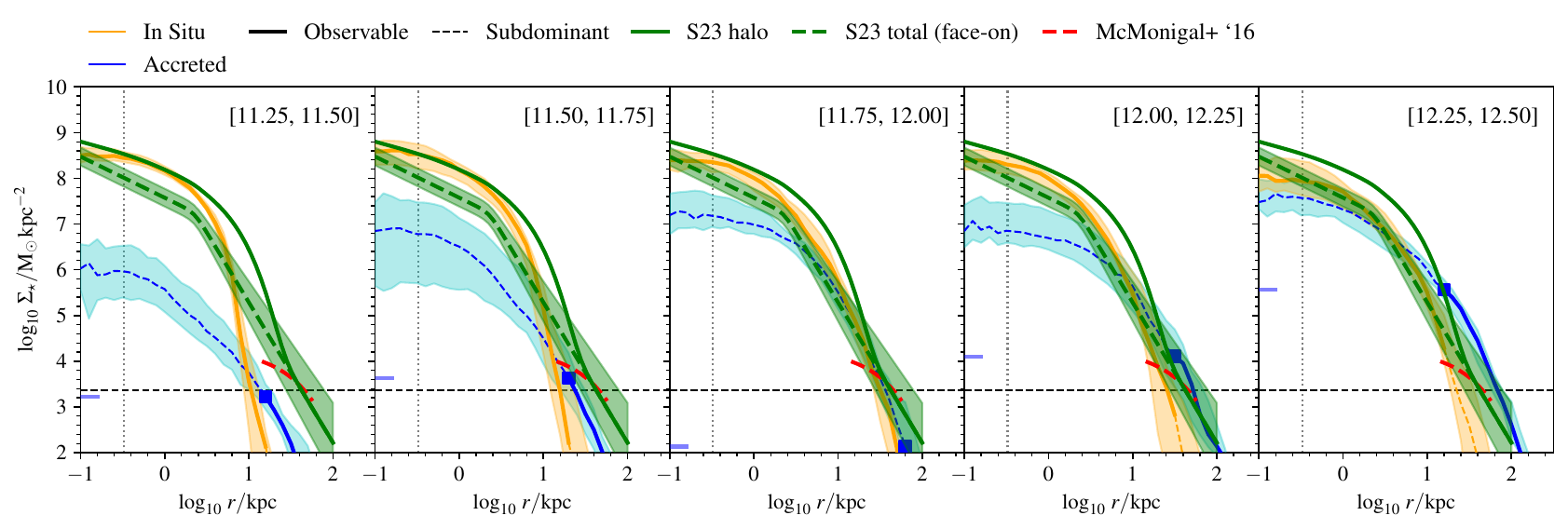}
  \caption{Stellar mass surface density profiles of M33 analogues (defined by stellar mass) in bins of virial mass, identical to Fig.~\ref{fig:m33_profiles_by_mhalo}, but here overlaying observational results for M33. The solid green line shows the total surface density profile obtained by \citet{Smercina:2023aa}; the dashed line and shaded of the same colour shows the inferred contribution of a broken power-law stellar halo. The short dashed red line shows the M33 halo profile reported by \citet{McMonigal:2016aa}. The solid pink line shows the RR Lyrae profile from \citet{Tanakul:2017aa}, with solid purple lines showing the (approximate) range of corresponding uncertainty. The scaling of these three data sets to stellar mass surface density is uncertain (see the text). Black symbols (crosses, circles and squares) show selected density profiles observed by \citet[][NGC 3432 and NGC 4565]{Gilhuly:2022aa} and \citet[][NGC 1042]{Trujillo:2021aa}. Note that the simulated \insitu{} profiles in these figures (orange lines) are those of exponential spheroids; they do not account for the combined effect of disc-like morphology and projection.}
\label{fig:dwarf_profiles_fix_mstar}
\end{figure*}

\begin{figure*}
  \includegraphics[width=\linewidth,trim=0.0cm 0cm 0.0cm 0cm,clip=True]{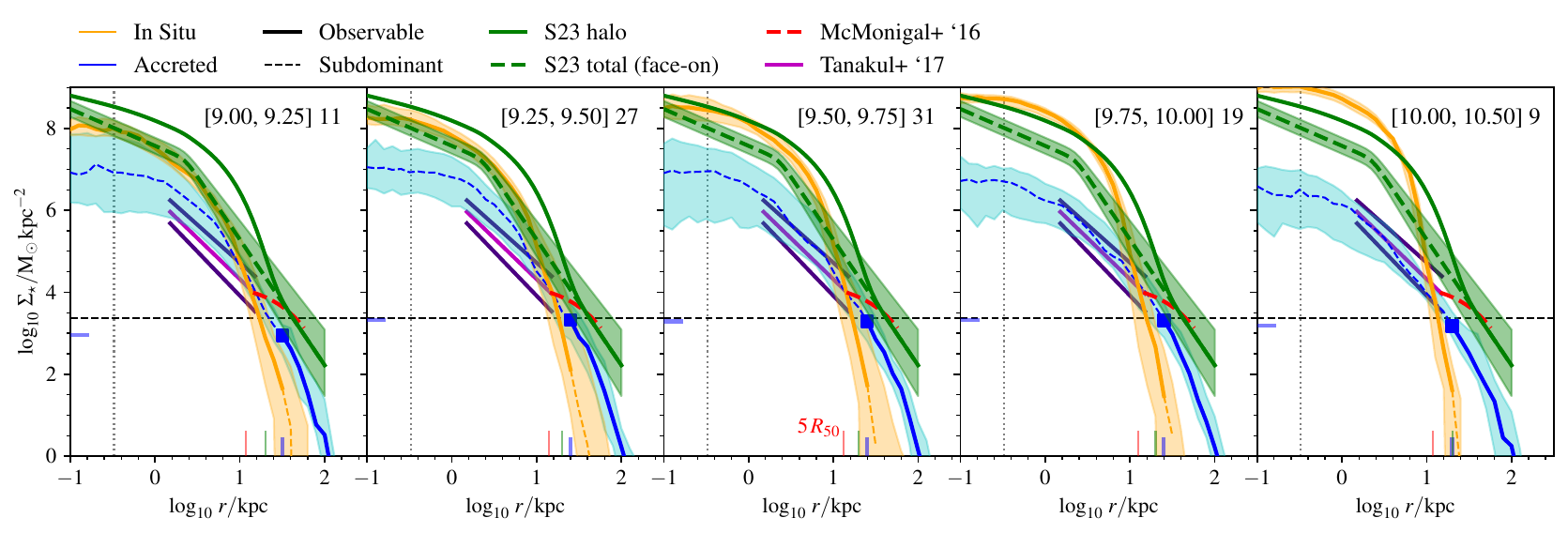}
  \caption{As Fig.~\ref{fig:dwarf_profiles_fix_mstar}, but for logarithmic bins of \textit{stellar} mass (indicated in the top right of each panel, along with the number of galaxies in the bin), across a fixed range of \textit{virial} mass, $11.5<\log_{10}\,M_{200}/\Msol<11.75$. The vertical range of this figure is less than that of Fig.~\ref{fig:dwarf_profiles_fix_mstar}.}
\label{fig:dwarf_profiles_fix_mhalo}
\end{figure*}

Much of the complexity discussed in the previous subsection is caused by the change in the dominant mode of stellar mass growth around the knee of the SMHM relation. Fig.~\ref{fig:mstar_mhalo_accreted_properties} suggests that this becomes less relevant at lower stellar mass and virial mass, and unimportant below $M_{200}\sim10^{11}\,\Msol$ or $M_{\star}\sim10^{8}\,\Msol$. Dwarf galaxies with stellar masses below the average for their virial mass, or equivalently, virial masses higher than the average for their stellar mass, have typical accreted fractions $\lesssim10$~per cent, with no strong dependence on their stellar mass. Although accreted stellar mass fractions can be high, the stars are typically centrally concentrated, hence accreted stars make effectively no observable contribution to the overall surface density profiles of these galaxies\footnote{Again, our model does not account for the 3d morphology of the galaxies; the accreted component may well be detectable if the \insitu{} component corresponds to a thin disc and the galaxy is viewed edge-on.}.

Fig.~\ref{fig:density_profiles_by_halo_mass} shows how the median projected stellar mass surface density profiles of \insitu{} and accreted stars change with virial mass. Sersic fits to the median profiles in Fig.~\ref{fig:density_profiles_by_halo_mass} are shown with darker lines. The fits to the accreted component are not reliable in haloes where the accreted mass is close to zero, but are representative otherwise. These are the `average' profiles that would be recovered, in principle, from stacked images of many such galaxies. Individual galaxies scatter around these averages. The scatter (shown here by the 16-84$^{\mathrm{th}}$ \apc{percentile} range) is larger for the accreted component, in which a single bright stream can boost the density by several orders of magnitude over a narrow range of radius. 

Below $M_{200}\sim10^{10}\,\Msol$, the accreted mass is effectively zero in many haloes. These are cases in which only the main branch of the halo merger tree has a potential well with virial temperature high enough to allow efficient accretion and cooling of atomic hydrogen from the intergalactic medium \citep[the critical virial mass at $z=0$ in this model is $M_{200}\sim10^{10}\,\Msol$; see e.g.][]{Benitez-Llambay:2020uy}. Their stellar halo masses would increase if the suppression of baryon accretion and cooling by the UV background after reionization were reduced, or if reionization occurred later - \apc{such changes} would favour a larger number of star-forming progenitors \citep{Font:2011aa,Benitez-Llambay:2020uy,Kim:2024aa}. More generally, the stellar halo masses of these galaxies are potentially sensitive to \apc{any factors that affect the halo occupation fraction and slope of the SMHM relation at low masses, including the details of} star formation before reionization. \apc{Currently there are few strong constraints on the calibration of cosmological simulations in this regime
\citep[mostly from Milky Way satellites, e.g.][]{Benson:2003aa,Bose:2018aa}.} \apc{\citet{Deason:2022aa} use a simple particle tagging method (also with \coco{}) and empirical SMHM models to demonstrate the significant impact of these uncertainties on predictions for accreted stellar masses in dwarf galaxies. \citet{Wang:2023aa} consider a  related phenomenon in which a subset of halo assembly histories at low mass may give rise to faint galaxies consisting almost entirely of accreted stars.}

In our model, where haloes with $M_{200}\lesssim10^{10.6}\,\Msol$ do have a substantial mass of accreted stars, they are more compact than the \insitu{} component. This is because star-forming (minor) branches in the merger trees of these systems are typically associated with relatively dense and early-collapsing progenitors that merge at early epochs (in these systems, less massive \apc{progenitor} branches and branches that form later \apc{may} not exceed the cooling threshold). Conversely, the \insitu{} component is dominated by stars formed closer to $z=0$ when the \apc{main branch of the merger tree has a much higher virial mass, and the characteristic scale of the star-forming gas is larger due to its high angular momentum}. In situ stars may also be scattered outwards during mergers. This scattering may produce an extended in situ component with halo-like kinematics; \citet{Tau:2024aa} find this mechanism to be the dominant contributor to the outskirts of low-mass galaxies in the Auriga simulations. \refrep{We caution that, in particle-based simulations such as ours, the scale length of the \insitu{} component may be overestimated in galaxies with sizes close to the resolution limit (see section~\ref{sec:mass_size} and Fig.~\ref{fig:sizemass_central}). This could, in turn, give rise to a more diffuse accreted component for low-mass galaxies.}

Fig.~\ref{fig:density_parameters_by_halo_mass} shows the run of the Sersic index $n$, stellar half-mass scale $R_{50}$ and amplitude $\Sigma_{0}$, for the fits to the median stellar mass surface density profiles shown in Fig.~\ref{fig:density_profiles_by_halo_mass}. Clouds of points show the corresponding parameters of Sersic fits to the \insitu{} and accreted profiles of individual galaxies\footnote{The averages of these points need not correspond to the parameters of the median profile, although they are similar in practice.}. The $\Sigma_{0}$ panel shows clearly the mass range where the population of haloes \apc{that have} near-zero accreted mass create a bimodality, which makes the average trends of the Sersic index and scale for the accreted component less reliable (this region is indicated by the grey band). The exponential ($n\simeq1$) shapes of the \insitu{} profiles are a consequence of our particle tagging prescription (see Section~\ref{sec:fmb}). The accreted component also has a roughly exponential profile in the least massive star-forming dark matter haloes. In that case, it could be hard to distinguish the accreted component from the \insitu{} component by Sersic index, if the latter is also spheroidal (although the \insitu{} component has fractionally larger $R_{50}$). The \apc{average} Sersic index of the accreted component increases steadily towards $n\sim3$ at the upper limit of our virial mass range, with a correspondingly rapid increase in $R_{50}$. These predictions can be tested by future observations that distinguish between  \insitu{} and accreted components.

\subsection{Comparisons with observations}
\label{sec:low_mass_observations}

\apc{M33 can be studied in comparatively greater detail than more distant low-mass galaxies of its type. Observations of possible accreted and \insitu{} stellar haloes in M33 therefore serve to illustrate the present state of data in this regime, and the possible insights offered by comparison to simulations such as ours}.

\citet{Gilbert:2022aa} provide a recent summary of work on the accreted halo of M33, in the context of their detection of a centrally concentrated population of RGB stars that have high velocity dispersion ($\sigma\approx60\,\kms$) and low bulk rotation, consistent with an accreted component. The mass fraction they attribute to this component in a kinematic decomposition \textit{decreases} with radius, which they interpret as evidence of a compact stellar halo formed by \insitu{} processes rather than through accretion \citep[similar to the scenario discussed in the context of the Milky Way by][]{Belokurov:2018aa}. The implications of this detection for the overall properties of M33 stellar halo have been explored by \citet[][hereafter S23]{Smercina:2023aa} using multiwavelength stellar density maps of the inner region of M33 from the PHATTER survey \citep{Williams:2021aa}. S23 argue for a stellar halo of mass $\sim5\times10^{8}\,\Msol$, following a broken power-law density profile with inner and outer logarithmic slopes of $\sim-1$ and $-3$, respectively, changing slope in the range $1<r<3~\kpc$. They note that this stellar halo mass would be consistent with theoretical and empirical relations between stellar halo mass and metallicity \citep[e.g.][]{DSouza:2018aa}, given the value of $\mathrm{[Fe/H]} \approx -1.3$ measured for the high velocity dispersion component identified by \citet{Gilbert:2022aa}. S23 remark that, in light of the high concentration of their inferred halo, the wide range of previous estimates based on star counts at larger radii \citep{Cockcroft:2013aa, McMonigal:2016aa} could be due to systematic uncertainties in background subtraction. Recent observations of RGB and red clump stars by \citet{Ogami:2024aa} support the existence of an extended halo to $15\lesssim r\lesssim30\,\mathrm{kpc}$ with logarithmic slope $\gtrsim-3$. Meanwhile, \citet{Tanakul:2017aa} have studied the surface density profile of RR Lyrae stars in M33, finding a logarithmic slope of $-2$ over the range $1<r<10~\kpc$.

Assuming a stellar mass of $3\times10^{9}\,\Msol$ \citep{van-der-marel:2019aa}, S23 determine that their broken power law halo model corresponds to a total stellar halo mass fraction of $\sim16$ per cent. In our models, such a high fraction would be unusual if all of the mass ascribed to the halo component was accreted; see, for example, Fig.~\ref{fig:mstar_mhalo}, in which we show that the S23 halo mass fraction puts M33 in the region of that diagram populated by haloes with the least efficient \insitu{} star formation (see discussion in Section~\ref{sec:mstar_mhalo}). A high accreted fraction would therefore be in tension with the idea that M33 is a prototypical low-mass, bulgeless disc galaxy:  galaxies with high accreted mass fractions are more likely to show either signs of disturbance (i.e. perturbations to the \insitu{} component) or elliptical morphology (an accreted component that dominates the surface brightness profile over a wide range of radius). There is considerable uncertainty in that expectation, as also discussed by S23. Nevertheless, our models do not predict such high accreted mass fractions for galaxies like M33, unless their total virial mass is also relatively high. Our models therefore suggest that the S23 star count-based halo model may be dominated by \insitu{} halo material in the central region, rather than accreted stars. Although the argument for a high central density in S23 is based on the high dispersion component described by \citet{Gilbert:2022aa}, this does not necessarily mean those stars were formed \insitu{}; only that, if they are accreted, their overall contribution to the total mass may be lower than assumed by S23.

To illustrate the way in which the \insitu{} and accreted profiles separately constrain the likely stellar mass and virial mass (at least in our model) , Fig.~\ref{fig:dwarf_profiles_fix_mstar} overlays data from \citet{McMonigal:2016ab}, \citet{Tanakul:2017aa}, and S23 on our surface density profiles for stellar-mass-selected M33 analogues in bins of virial mass. This is the same selection of model profiles shown in Fig.~\ref{fig:m33_profiles_by_mhalo}. For reference, we also show selected surface density profiles from the Dragonfly Survey \citep{Merritt:2016aa,Gilhuly:2022aa} and the LIGHTS survey \citep{Trujillo:2021aa, Zaritsky:2024aa}. In the highest virial mass bin ($12.25<\log_{10}\,M_{200}/\Msol < 12.5$),  the S23 profile agrees approximately with our prediction for the density of accreted stars in the inner galaxy, but it falls below the average density of our model in the `directly observable' stellar halo region ($\gtrsim10\,\kpc$). At lower virial mass, the S23 profile agrees with our predicted density in the outskirts, \apc{but the central density of accreted stars in the S23 profile is significantly higher than our model}. In the range of virial mass for M33 estimated by \citet{Patel:2018aa}, $1\lesssim M_{200} \lesssim 3\times10^{11}\,\Msol$, the S23 profile is \apc{higher than our model average} accreted mass profile at all radii. In general terms, this implies that the detection of any significant stellar halo mass in M33 at $\sim10$--$20\,\kpc$ favours a relatively high virial mass for M33. M33 does not seem to be unusual in this respect; the galaxies NGC 3432 \citep{Gilhuly:2022aa} and NGC 1042 \citep{Merritt:2016aa,Trujillo:2021aa} show similar profiles. Stellar mass estimates for these galaxies are $M_{\star}\approx4.8\times10^{9}$ \citep{Gilhuly:2022aa} and $M_{\star}=6.5\pm1.6\times10^{9}\,\Msol$ \citep{Trujillo:2021aa}, respectively. For comparison \apc{to these low-mass galaxies,} we also show the profile of NGC 4565 (a MW analogue, $M_\star \approx 7.5\times10^{10}\,\Msol$) from \citet{Gilhuly:2022aa}.

Fig.~\ref{fig:dwarf_profiles_fix_mhalo} compares the same M33 data to a different sample of M33 analogues from our model, \apc{this time} selected by \textit{virial mass}, $11.5<\log_{10}\,M_{200}/\Msol < 11.75$. The profiles are averaged in logarithmic bins of stellar mass (see the blue boxes in Fig.~\ref{fig:mstar_mhalo_accreted_properties}). For such a selection, as explained above, the profile of accreted stars is essentially independent of the stellar mass bin, whereas the amplitude of the \insitu{} profile increases from low-to-high total stellar mass.

\begin{figure*}
  \includegraphics[width=168mm, trim=0.0cm 0cm 0.0cm 0cm,clip=True]{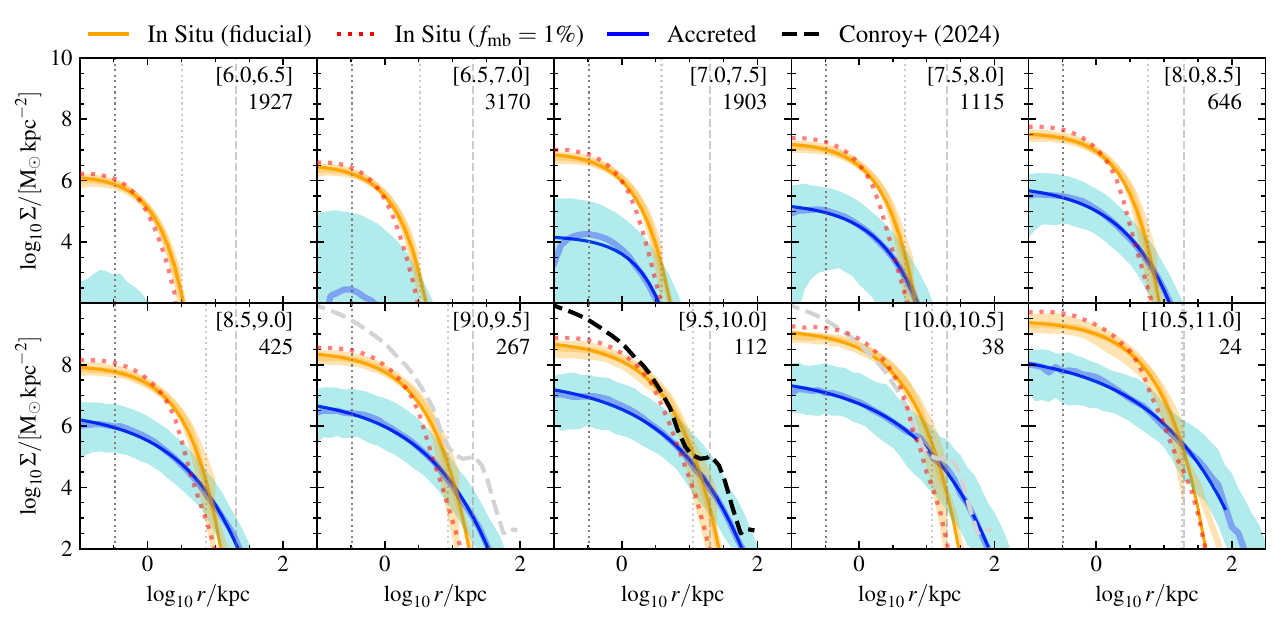}
  \caption{Decomposition of average projected stellar mass surface density profiles for low mass galaxies in our model, as Fig.~\ref{fig:density_profiles_by_halo_mass}, but here in bins of stellar mass. Fainter lines show the median density of \insitu{} stars (orange) and accreted stars (blue). Darker lines of the same colour show single-Sersic fits to the median profiles. Shaded regions show the corresponding 16-84 percentile range. \refrep{Red dotted lines show the corresponding \insitu{} profiles for a model with $\fmb=1\%$.} In each panel, we show the range of $\log_{10} M_{200}$ and the corresponding number of galaxies. Vertical lines, from left to right, indicate the force softening scale (dark dotted), $5R_{50}$ (light dotted) and $20\,\kpc$ (dashed), respectively. The dashed curve in the $[9.5,10]$ panel shows the density of Ark 227 reported by \citet{Conroy:2024aa}; the same profile is repeated in panels $[9.0,9,5]$ and $[10,10.5]$.}
\label{fig:density_profiles_by_stellar_mass}
\end{figure*}

The RR Lyrae profile from \citet{Tanakul:2017aa} hints at an accreted component that would be broadly consistent with a virial mass in this range, in line with most estimates for M33. However, the amplitude of this profile is highly uncertain. To plot these data in  Figs.~\ref{fig:dwarf_profiles_fix_mstar} and \ref{fig:dwarf_profiles_fix_mhalo}, we make a very rough estimate of the ratio between RR Lyrae counts and the total mass of the accreted halo population as follows. We assume that the underlying specific frequency of RR Lyrae in their (very metal poor and old) parent stellar population is comparable to that found for the Milky Way dwarf spheroids and globular clusters: $S_\mathrm{RR}\sim50-100$ for a population of absolute magnitude $M_{V}=-7.5$ \citep[e.g.][]{Mackey:2003aa,Tanakul:2017aa}. Taking $S_\mathrm{RR}=50$ and assuming a mass-to-light ratio for the parent population of $M_\star/L_V\sim3$ then gives a ratio of one RR Lyrae per $\sim5226\,\Msol$. The progenitors of the accreted halo at this scale are likely to contain stellar populations somewhat more metal rich than these systems (see later sections), which may have lower $S_\mathrm{RR}$ and hence lower the overall ratio for the progenitor \citep{Tanakul:2017aa}. Ignoring this (i.e. assuming that the RR Lyrae-generating populations dominate the progenitor) and converting the \citet{Tanakul:2017aa} RR Lyrae surface density to mass, following the simple prescription above, results in a stellar mass surface density estimate that matches reasonably well the inward extrapolation of 
the 
\citet{McMonigal:2016ab} profile and (to a lesser degree) the S23 profile. The slope of the \citet{Tanakul:2017aa} data (assuming a constant ratio of RR Lyrae to underlying stellar halo mass) is in better agreement with the accreted star density slope in our models than the S23 fit. Both the slope and amplitude we have estimated from the \citet{Tanakul:2017aa} data could be adjusted to better match the S23 data by assuming (plausible) radial variation in the RR Lyrae-to-total stellar mass ratio.

We conclude that the predictions from our simulation are at least qualitatively consistent with the current understanding of M33, and provide a useful way to combine inferences about the stellar haloes in low mass galaxies with other constraints on the total stellar mass and virial mass. In simple terms, the central surface density (dominated by \insitu{} stars) effectively constrains the total stellar mass, while the amplitude of the accreted profile, and hence the of the overall profile overall at sufficiently large radius, correlates with virial mass. This has been shown before (e.g. \citetalias[e.g.][]{Cooper:2013aa}; \citealt{Pillepich:2014aa}) and discussed in the context of density profile slopes and Sersic indices fit to the outskirts of brighter galaxies, but the potential usefulness of these two separate constraints is particularly clear (and worth emphasizing) in the low-mass regime. It would be interesting to compare our models with more photometric data sets for other low mass galaxies \citep{Harmsen:2017aa,Merritt:2020aa,Gilhuly:2022aa}. In further work, we intend to extend these comparisons, to explore the uncertainty due to the semi-analytic component of our model, and to incorporate more `observational realism' with mock images. 

\subsubsection{Very low-mass galaxies}

There are very few stellar halo data in the literature with which to compare for isolated galaxies less massive than M33 analogues. Mostly these are for galaxies in the Local Group. There are several tentative detections of extended components around the dwarf satellites of the Milky Way \citep[e.g.][]{Irwin:1995aa, Chiti:2021aa, Jensen:2024aa}, but these are difficulty to disentangle from the effects of tidal interaction with the Galaxy. \citet[][see references therein]{Kang:2019aa} summarize data from six relatively isolated galaxies in the Local group: Leo T, Leo A, WLM, IC 1613 and NGC 6822 
These galaxies have stellar masses $\sim10^7\,\Msol$; by fitting profiles to these data, \citet{Kang:2019aa} obtain accreted halo mass fractions $\gtrsim10$~per cent \citep[see also][]{Ricotti:2022aa}. This may appear high at first glance, but would nevertheless be consistent with our models. \citet{Higgs:2021aa} examine other nearby isolated galaxies and find evidence for an extended component in UGC 4879 (of comparable, perhaps lower mass), but no evidence of substructure. At slightly greater distances, \citet{Annibali:2020aa} find extended components and tentative evidence of streams and other features around NGC 2366, UGC 685, NGC 5477, UGC 4426 and (most strikingly) UGC 12613. The galaxies in their sample (not all of which show extended components) have magnitudes $-12 \lesssim M_{B} \lesssim -18$. \apc{\citet{Hunt:2024aa} demonstrate the potential for Euclid to greatly expand the number of deep surface brightness profiles in the dwarf regime.} One difficulty of interpreting extended components in dwarfs as evidence for stellar haloes is that \insitu{} star formation appears to be sustained for longer in the inner regions of low mass galaxies compared to their outskirts \cite[apparent in their blue-to-red colour gradients, e.g.][]{Liao:2023aa}, leading to the expectation \apc{that faint dwarf galaxies typically have} a diffuse, older component formed through \insitu{} processes alone \citep{Kado-Fong:2022aa}. 

Recently, \citet{Conroy:2024aa} reported a serendipitous observation with \textit{JWST} of an extended component around Ark 227 (distance $\sim35$~Mpc, $M_{\star}\sim5\times10^9\,\Msol$). This provides a very useful point of comparison in the regime intermediate between M33-like objects and the nearby dwarfs listed above. Fig.~\ref{fig:density_profiles_by_stellar_mass} compares the stellar mass surface density profile derived by \citet{Conroy:2024aa} to our simulations in a similar range of stellar mass, alongside comparable predictions for other stellar \refrep{mass ranges}. The profile is close to the model prediction overall; differences in the centre of the profile are likely due to the limitations of the particle tagging approach and resolution effects. In the outskirts, the observed profile shows a `shelf'-like inflection (discussed by \citeauthor{Conroy:2024aa}) but is close on average to the expected profile of accreted stellar mass. The inflection occurs close to the radius and surface brightness where, following our models, we would expect the transition between \insitu{} and accretion-dominated regions of the profile. \refrep{We note that this transition radius depends weakly on our choice of the particle tagging parameter, $\fmb$, which sets the scale of the \insitu{} component as described in Sections~\ref{sec:fmb} and \ref{sec:mass_size}. To illustrate this, we show \insitu{} profiles for an $\fmb=1\%$ model in Fig.~\ref{fig:density_profiles_by_stellar_mass}. Variation of $\fmb$ within a range consistent with the size--mass relation does not change the average profile of the accreted component (in our model, this is determined mostly by the orbital dynamics of the accretion events, and not by the internal structure of the progenitors)}.

\section{Progenitors and stellar populations}
\label{sec:progenitors_and_populations}

Here we summarize some of the properties that have \apc{previously} been the focus of \apc{studies of Milky Way-like stellar haloes}, in the context of lower mass systems. \apc{These results illustrate the capabilities of the data we release along with this paper.}

\subsection{Progenitor diversity}
\label{sec:diversity}

\begin{figure}
  \includegraphics[width=\linewidth,trim=0.0cm 0cm 0.0cm 0cm,clip=True]{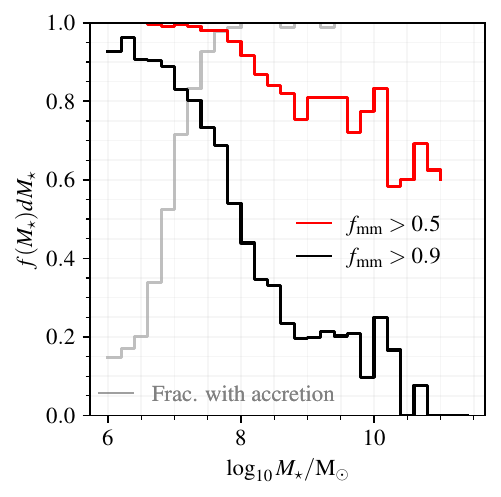}
  \caption{The fraction of galaxies of a given stellar mass in which more than 90 per cent (black line) or more than 50 per cent (red line) of the accreted mass is contributed by the most massive progenitor. These lines only count galaxies with progenitors, and only stellar mass bins with more than three galaxies are included. The fraction of galaxies of a given mass \apc{that have accreted at least one progenitor} is shown by the light grey line.}
\label{fig:mm_prog_fractions}
\end{figure}

\begin{figure}
  \includegraphics[width=\linewidth,trim=0.0cm 0cm 0.0cm 0cm,clip=True]{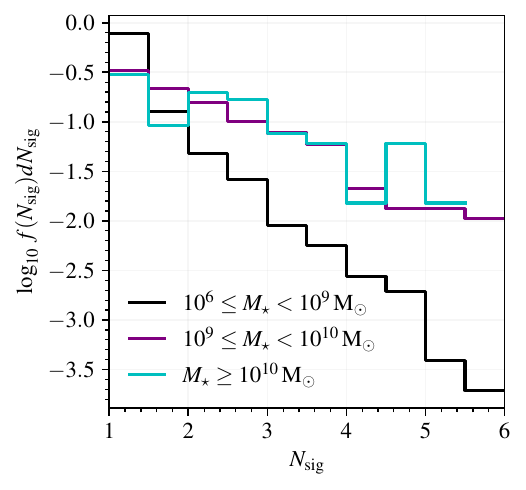}
  \caption{The distribution of $N_\mathrm{sig}$, a measure of the number of significant progenitors of the accreted stellar halo (defined in the text), for galaxies in three ranges of total stellar mass. The vertical scale shows the fraction of galaxies from each mass range in a given bin (width $0.5$).}
\label{fig:nsig_by_mstar}
\end{figure}

The stochastic nature of hierarchical clustering results in a broad range of progenitor accretion times, masses, and star formation histories among central galaxies, even at a fixed combination of stellar mass and virial mass. For this reason, the properties of present-day satellite systems and stellar haloes can vary substantially among otherwise similar galaxies, particularly \apc{because} they are disproportionately sensitive to the fate of the few most massive progenitors \citep{Bullock:2005aa,Amorisco:2017aa}. The Milky Way provides a clear example: if the ongoing merger of the LMC had proceeded more rapidly, the observable features of the Milky Way's accreted stellar halo (and to some extent, the interpretation of its satellite system) would be very different, even though the merger would be "minor" in terms of the LMC's contribution of stellar mass and virial mass.

Our simulation allows us to investigate the typical diversity of stellar haloes in low mass galaxies. In Fig.~\ref{fig:mm_prog_fractions} we show the fraction of dwarf galaxies that have \apc{an} accreted halo component (the grey line in the figure) rises rapidly to 100 per~cent around a total stellar mass of $M_{\star}\sim 10^{8}\,\mathrm{\Msol}$. This is the straightforward consequence of the steep decline of galaxy formation efficiency in low mass haloes. At the lowest virial masses, \apc{any stars that the galaxy accretes are}  invariably contributed by a single accretion event. At higher masses, the fraction of galaxies in which a single progenitor accounts for $\gtrsim90$~per~cent of the accreted stellar mass (black line) falls from 60~per~cent at $M_{\star}\sim 10^{8}\,\mathrm{\Msol}$ to $\lesssim20$~per~cent at $M_{\star}\sim 10^{10}\,\mathrm{\Msol}$. Milky Way-mass galaxies that have stellar haloes dominated by a single `LMC-like' merger (in terms of fraction rather than absolute mass) are therefore relatively rare in our model \citep[see also the discussion in][]{Bose:2023aa}. Systems in which the most massive progenitor accounts for $\gtrsim50$~per~cent of the accreted stellar mass are more common, $\sim60$~per~cent \apc{of the population} at the Milky Way mass scale. These results are broadly in line with previous work on the diversity of haloes in Milky Way analogues \citep{Monachesi:2019aa}. Our results suggest that this picture extends to galaxies of mass $M_{\star}\gtrsim  10^{9}\,\mathrm{\Msol}$ (for example, to M33 analogues), but that dwarf galaxies \apc{with $M_{\star}\lesssim  10^{8}\,\mathrm{\Msol}$}  are increasingly dominated by their most massive progenitors.

\begin{figure*}
  \includegraphics[width=\linewidth,trim=0.0cm 0cm 0.0cm 0cm,clip=True]{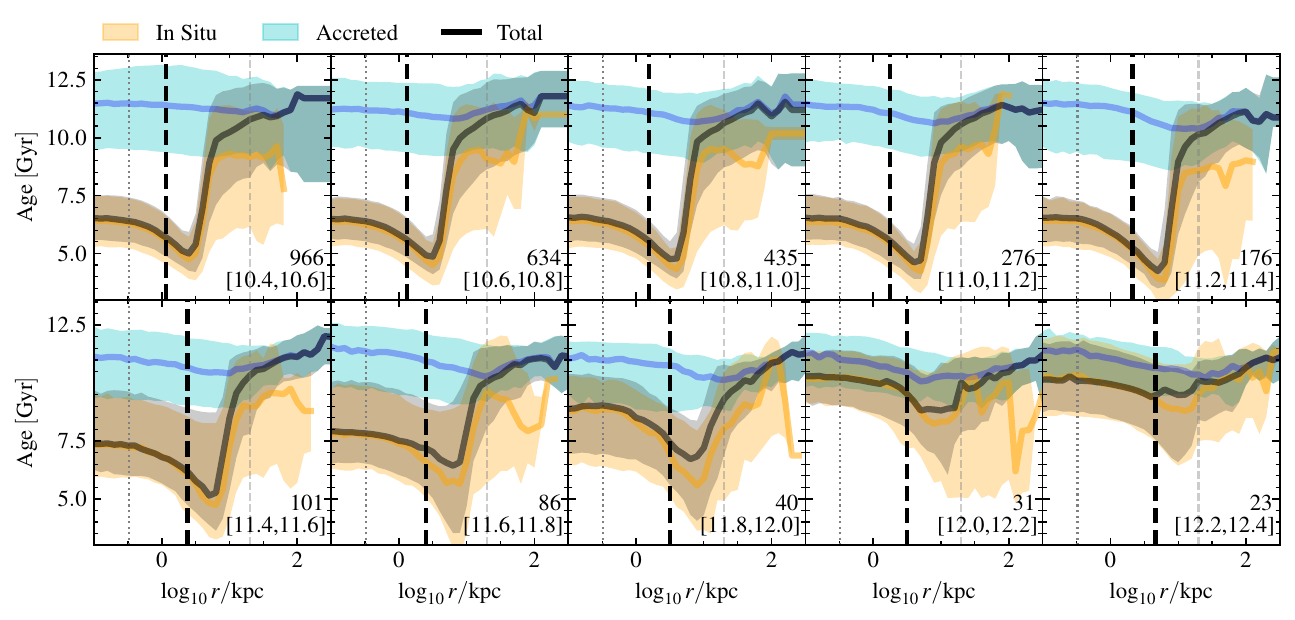}
  \caption{Projected age profiles in bins of virial mass ($\log_{10} M_{200}/\Msol$, shown in each panel along with the number of galaxies in each bin). Lines show the median total (black), accreted (blue) and \insitu{} (orange) age at a given radius; blue and orange ranges show the  16-84$^\mathrm{th}$ percentile ranges of for accreted and \insitu{} stars, respectively. The bold vertical dashed line indicates the average projected half-light radius, $R_{50}$, in each bin, and the lighter vertical dashed line a fixed radius of $20$~kpc.}
\label{fig:ages_mhalo}
\end{figure*}

Fig.~\ref{fig:nsig_by_mstar} quantifies the above in terms of a measure of overall `progenitor diversity', $N_\mathrm{sig} $ (\citetalias{Cooper:2010aa}; \citealt{Pu:2025aa}). This is defined as the second moment of the progenitor mass distribution, $N_\mathrm{sig} = [\sum_i m_i]^2\ /\sum_i m_i^2$, such that $N_\mathrm{sig}\sim1$ for accreted haloes dominated by a single progenitor and $N_\mathrm{sig}\sim N$ for accreted haloes in which $N$ progenitors contribute equal fractions of the mass. \apc{Diverse accreted haloes ($N_\mathrm{sig}>1$) are substantially less likely for galaxies with $M_{\star}\lesssim 10^{9}\,\mathrm{\Msol}$, compared to galaxies in the M33 or Milky Way mass range}\footnote{The average $N_\mathrm{sig}$ will of course increase rapidly with virial mass above the break in the SMHM relation, although there are few such galaxies in our simulation volume.}.

\subsection{Stellar populations}
\label{sec:populations}

\begin{figure*}
  \includegraphics[width=\linewidth,trim=0.0cm 0cm 0.0cm 0cm,clip=True]{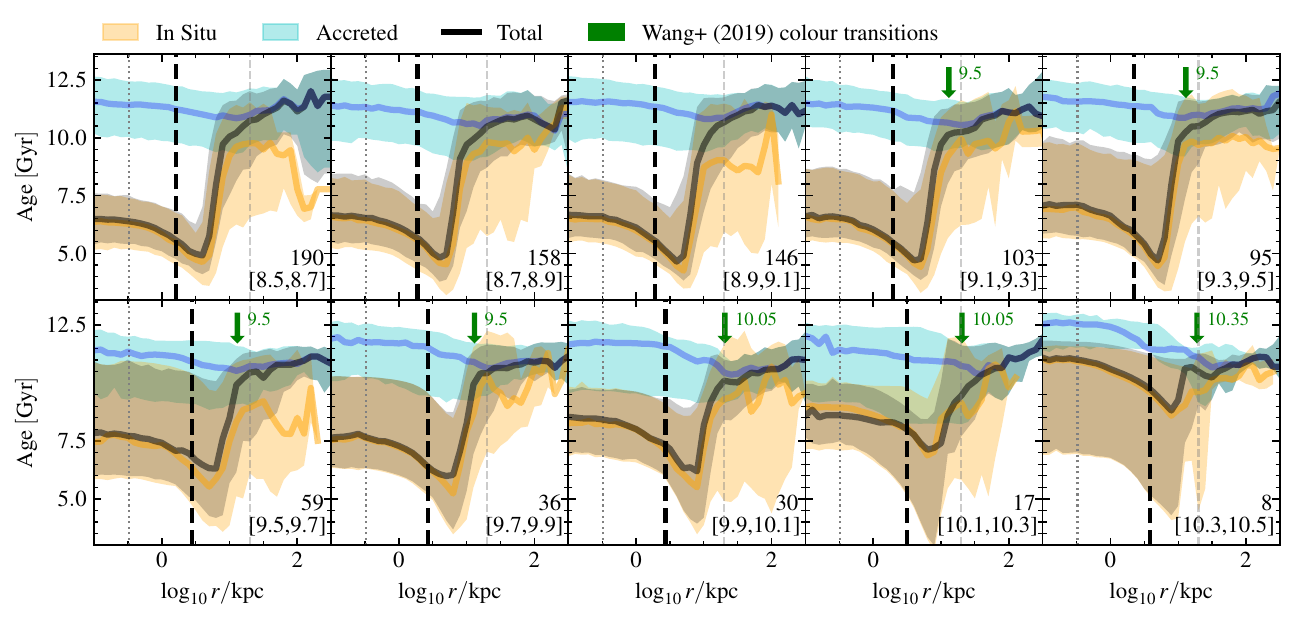}
  \caption{Age profiles in bins of galaxy stellar mass. 
 Green arrows mark the inflection point of $g-r$ with radius found for galaxies stacked in similar ranges of stellar mass by \citet{Wang:2019aa}. The number next to each arrow gives the approximate mean $\log_{10}\,M_{\star}/\Msol$ of the stack from which it is obtained. Other colours and labels are as in Fig.~\ref{fig:ages_mhalo}.}
\label{fig:ages_mstar}
\end{figure*}

\begin{figure*}
  \includegraphics[width=\linewidth,trim=0.0cm 0cm 0.0cm 0cm,clip=True]{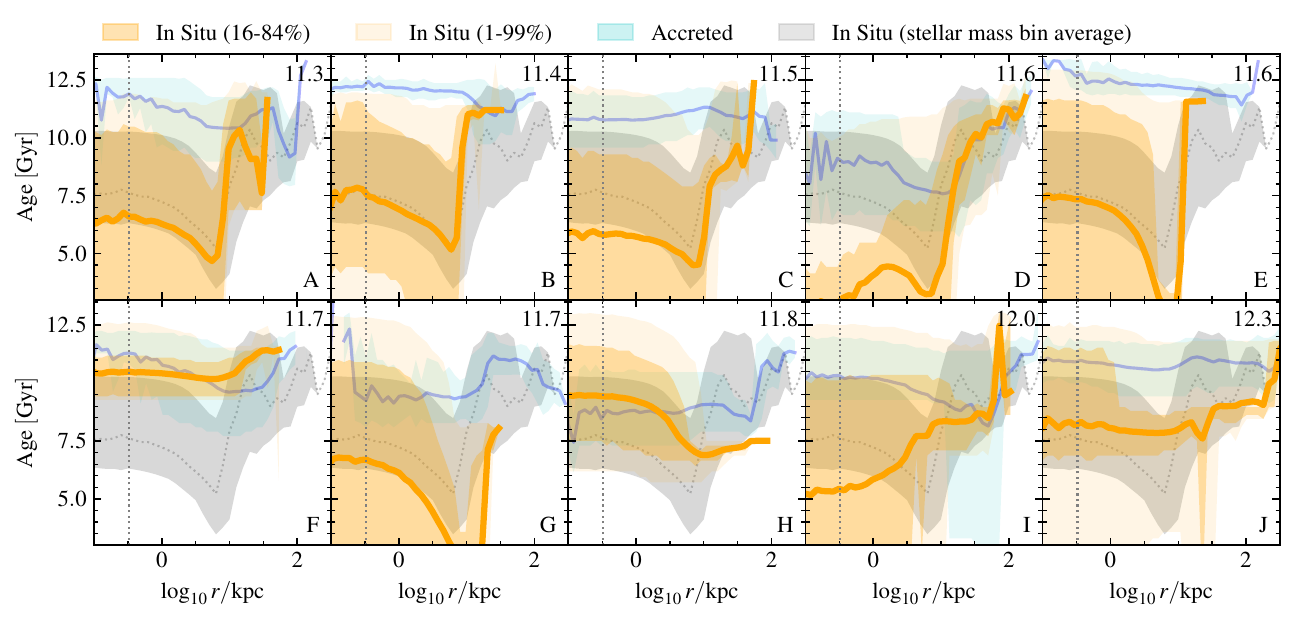}
  \caption{Age profiles for individual galaxies selected in a range of stellar mass $3 \times 10^{9} < M_{\star} < 7 \times 10^{9}\,\Msol$. For each galaxy, $\log_{10}\,M_{200}/\mathrm{M_{\sun}}$ is given in the upper right of the panel. Letters are labels referred to in the text. The shaded orange and blue regions indicate the range of age in the galaxy at a particular radius about the mean. The grey region, reproduced from previous figures, indicates the range of the mean profile in this bin of stellar mass.}
\label{fig:ages_mstar_examples_m33}
\end{figure*}

\begin{figure*}
  \includegraphics[width=\linewidth,trim=0.0cm 0cm 0.0cm 0cm,clip=True]{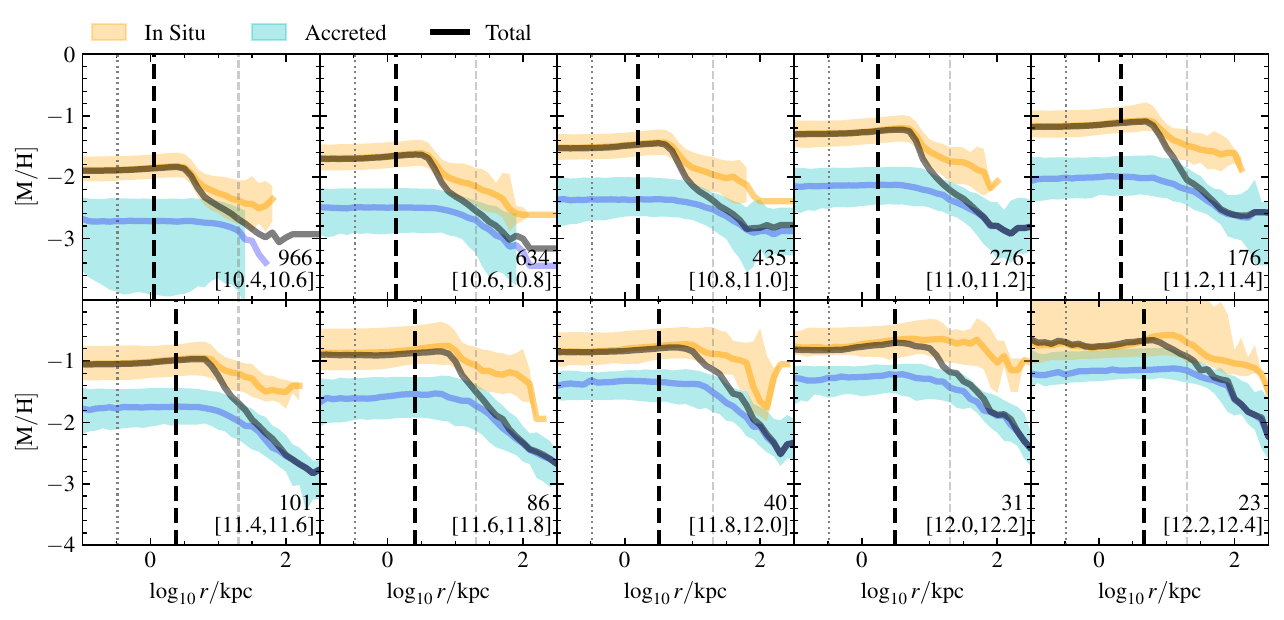}
  \caption{Metallicity profiles in bins of virial mass. Metallicity, $\mathrm{[M/H]}$, is defined as the ratio of metal mass to total mass, as defined in the text. Colours and labels are as in Fig.~\ref{fig:ages_mhalo}.}
\label{fig:zmet_mhalo}
\end{figure*}

Figs.~\ref{fig:ages_mhalo}--\ref{fig:zmet_mhalo} show \apc{average} age and metallicity profiles of accreted and \insitu{} stars, in bins of virial mass and stellar mass, respectively. 

\subsubsection{Stellar ages}
\label{sec:ages}

We first consider profiles of stellar age. Fig.~\ref{fig:ages_mhalo} shows that accreted stars are typically old (formed at lookback times $\gtrsim 10$~Gyr) across the entire range of virial mass we consider. \apc{We find that accreted stars are slightly older on average near the galactic centre than in the outskirts of the galaxy}. This trend with radius is weak, and does not vary significantly with virial mass. However, with increasing virial mass, the scatter in the age of accreted stars at a fixed radius decreases. 

The age gradient of the \insitu{} component is more obvious, with older (but not ancient) stars in the centre, a minimum age slightly beyond the half-mass radius, and much older ages in the outskirts. At the largest radii, \insitu{} stars have (average) age comparable to the youngest accreted stars. They also show a similar `young-to-old' trend as the accreted component at larger radii. These \apc{ancient} \insitu{} stars in the outskirts of the simulated galaxies represent only a very small fraction of the \insitu{} mass, particularly in low-mass galaxies. Nevertheless, at fixed virial mass, Fig.~\ref{fig:ages_mhalo} implies that 
accreted and \insitu{} stars \apc{have similar ages} \apc{around the `transition radius' we define in  section~\ref{sec:m33_stellar_haloes} (at which they contribute roughly equal fractions of the projected stellar density)}. Notably, in the age profile of the total stellar mass (black line), the step-change from young to old age outside the transition radius is due to the trend of the \insitu{} component alone, rather than the increasing contribution of accreted stars.

In the most massive haloes, there is little difference in the \apc{average} age distribution of the two components at any given radius (e.g. lower right panel). \apc{This contrasts with} the least massive haloes, in which the \insitu{} component is clearly much younger within the half-mass radius. This is not surprising \apc{because} low mass haloes are more likely to have formed recently and to host ongoing, \apc{centrally concentrated} star formation at the present day. \apc{At this level of} detail, \apc{the baryonic physics that determines the radial distribution of star formation may not be well represented} by $f_\mathrm{mb}$ tagging \citepalias[see][]{Cooper:2017wv}.

Similar trends are apparent in Fig.~\ref{fig:ages_mstar}, where we show average age profiles in bins of stellar mass. The scatter at fixed stellar mass is somewhat larger than in Fig.~\ref{fig:ages_mhalo}. \insitu{} and accreted stars in the central regions of galaxies with the largest stellar masses in our sample (i.e. Milky Way analogues by stellar mass) are older on average, compared to those of Milky Way analogues selected by virial mass. Milky Way stellar mass analogues also have more diversity among their individual profiles at fixed radius.

\citet{Wang:2019aa} report $g-r$ colour profiles for galaxies $M_\star \gtrsim 10^{9}\,\Msol$ measured from stacked Hyper-SuprimeCam images (in circular apertures; e.g.\ their fig.~11). Their profiles show red-to blue gradients that appear to invert at a radii $\sim15-20$~kpc, with the inflection occurring at larger radius for galaxies with greater stellar mass. These inflection radii are marked on Fig.~\ref{fig:ages_mstar} with green arrows. Stellar mass-dependent changes in colour (and hence presumably in age and metallicity), \apc{are qualitatively consistent} with the results we show in Section~\ref{sec:low_mass_haloes} for the transition between \insitu{} and accretion-dominated regions of the surface brightness profile. However, Fig.~\ref{fig:ages_mstar} suggests that, in dwarf galaxies, our models might predict a rapid inversion of the profile towards redder colour at somewhat at somewhat smaller radii ($\sim10$--$15$~kpc), driven by the older ages of \insitu{} stars alone. 

\apc{It is not clear that the trend of increasing \apc{average} \insitu{} star age with total stellar mass in the outskirts of our simulated galaxies is robust}. It may be an artefact, either of the simulation technique, or of how the profiles are averaged (stacked) across the sub-samples. For example, `glitches' in \apc{merger tree} links between progenitor and descendant haloes could lead to particular accreted stellar populations being wrongly classified as \insitu{}, or vice versa. \refrep{\citet{Chandro-Gomez:2025aa} discuss the nature and origin of such `glitches' in detail, and examine their effects on semi-analytic models, including \galform{}}.  \refrep{Spurious star formation events associated with merger tree} `glitches' could have a disproportionate effect the stellar mass budget in the outskirts, \apc{and may be the dominant source of \insitu{} stars in those regions}. On the other hand, it \refrep{is also the case that} \apc{a fraction of early-forming \insitu{}} stars are scattered to \refrep{significantly} \apc{lower binding energy} \refrep{during low mass ratio mergers} and hence are more extended than  \refrep{stars that form} after the relaxation of the central potential \citep[e.g.][]{Zolotov:2009aa, Font:2011uz, Tau:2024aa}. \refrep{This is thought to be the principle reason for the higher Sersic indices of early-type galaxies at a fixed mass.}

To explore this further, in Fig.~\ref{fig:ages_mstar_examples_m33} we show 10 examples of the age profiles for individual galaxies (labeled A-J in the figure). These were chosen to span the range of virial mass for systems in the M33 stellar mass range, $3 \times 10^{9} < M_{\star} < 7 \times 10^{9}\,\Msol$.
\begin{figure*}
  \includegraphics[width=\linewidth,trim=0.0cm 0cm 0.0cm 0cm,clip=True]{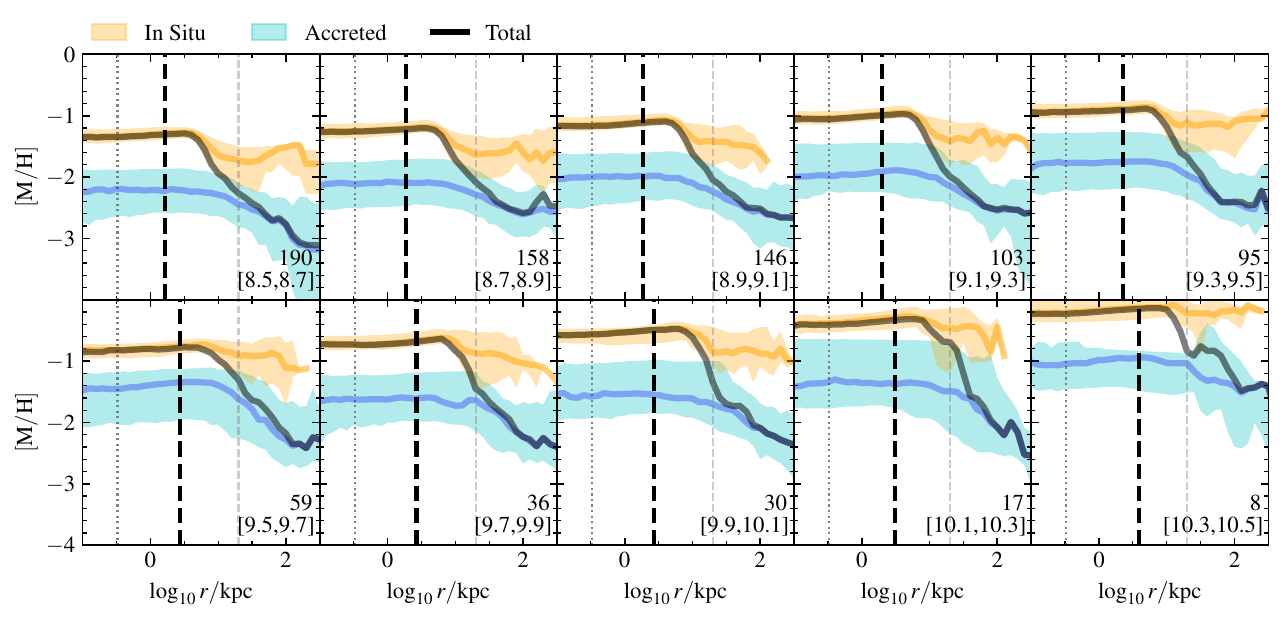}
  \caption{Metallicity profiles in bins of stellar mass. Colours and labels are as in Fig.~\ref{fig:ages_mhalo}.}
\label{fig:zmet_mstar}
\end{figure*}
The individual median \insitu{} age profiles in this figure (thick lines) show a wide scatter around the behaviour of the sample average (the grey region shows the average profile for the \insitu{} stars for the [9.5, 9.7] bin in Fig.\ref{fig:ages_mstar}). The distribution of ages at a fixed radius is broad, particularly for \insitu{} stars \apc{at smaller radii}. Galaxies F, H and J have relatively flat age profiles, in some cases significantly older than the average; \refrep{these are consistent with the expectation for early-type galaxies}. Others (D, I) have notably young centres. The average behaviour is represented by A, B, C, E and G; in these individual cases it is more obvious that the inflection in age is due to a very small mass of \insitu{} stars (suggested by the small scatter associated with those parts of the profile). If these regions \apc{have no} very young stars (for central galaxies, this almost guaranteed in an $f_\mathrm{mb}$ particle tagging approach) then a handful of old \insitu{} stars scattered into this region (or misclassified \apc{as} accreted stars) can have a disproportionate effect. Those effects could accumulate in the average, resulting in a rising trend. Nevertheless, there are cases (A, D, I) where a trend towards very old ages in the outskirts seems more robust \refrep{and may correspond to systems that re-grow a concentrated \insitu{} component after a major merger at intermediate redshift}.

\subsubsection{Metallicities}
\label{sec:metals}

\begin{figure*}
  \includegraphics[width=\linewidth,trim=0.0cm 0cm 0.0cm 0cm,clip=True]{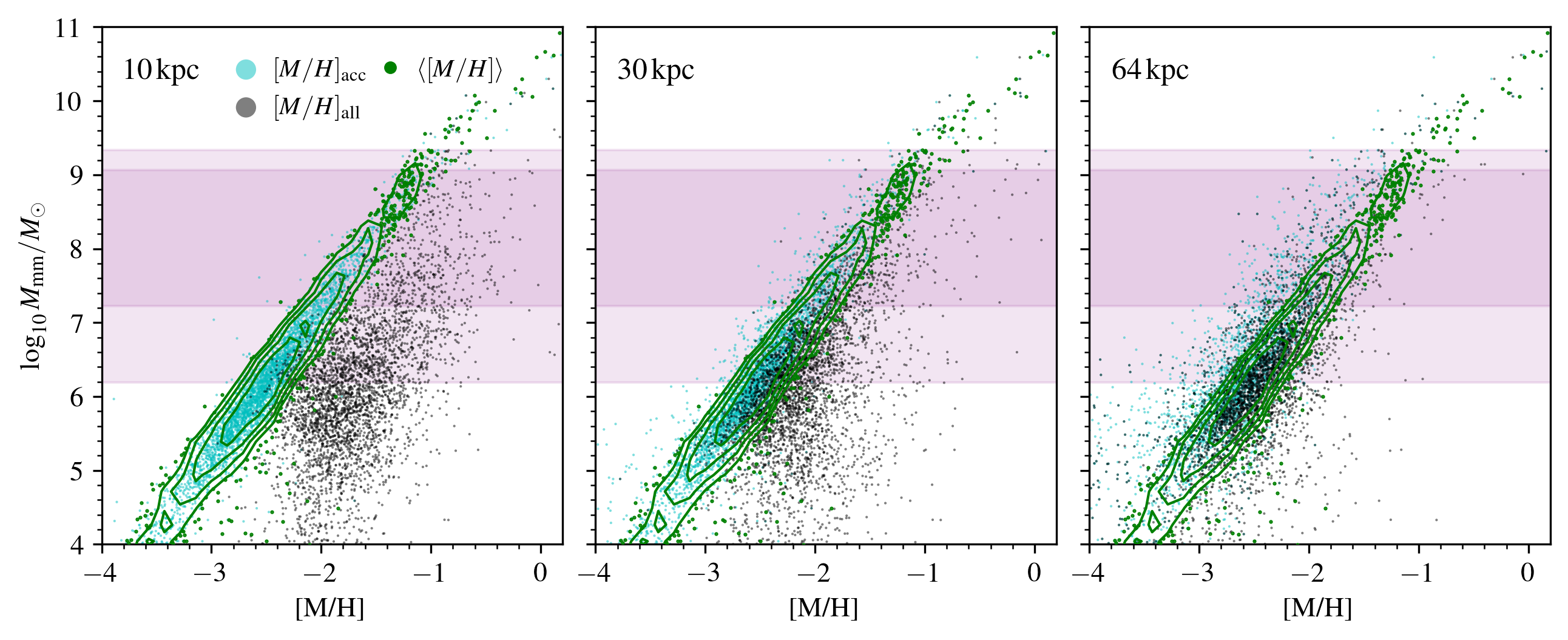}
  \caption{Relation between progenitor mass and mean metallicity of accreted stars (blue) and all stars (black) measured at three different radii (panel labels). Green contours and points show the overall mass--metallicity relation for progenitor galaxies. Purple bands indicate the 5-95 and 16-84 percentile ranges (light and dark shading, respectively) of the mass of the most-massive progenitor for M33 analogues (selected by stellar mass). }
\label{fig:metals_at_radius}
\end{figure*}

Almost all galaxy formation models predict a very tight correlation between the total stellar mass of a galaxy and its average metallicity. \apc{In principle, this can} be exploited to determine the masses of accreted progenitors from observations of metallicity or colour, provided a representative sample of accreted stars can be identified \citep[e.g.][]{DSouza:2018aa, Monachesi:2019aa}. Figs.~\ref{fig:zmet_mhalo} and \ref{fig:zmet_mstar} show average metallicity profiles for our simulated galaxies in bins of virial mass and stellar mass respectively, analogous to Figs.~\ref{fig:ages_mhalo} and \ref{fig:ages_mstar}. \apc{Average} metallicity gradients \apc{in our model} are relatively flat to $\sim2R_{50}$ for both \insitu{} and accreted stars, with a steeper decline in total metallicity driven by the transition between the two components (e.g.\ \citealt{Font:2006aa}; \citetalias{Cooper:2010aa}). \apc{This is broadly in line with observations, although the relatively small sample of galaxies with sufficiently extensive halo metallicity gradients shows notable diversity \citep[e.g.][]{Monachesi:2016aa, Monachesi:2016ab}.We find} the average metallicity of the accreted component increases more rapidly with virial mass than the metallicity of the \insitu{} component, because the declining efficiency of galaxy formation with virial mass reduces the mass ratio between galaxies and their most massive progenitors. 
The metallicity difference between accreted and \insitu{} stars for Milky Way-like systems is more pronounced when galaxies are grouped by stellar mass, because a stellar mass selection for such galaxies is dominated by efficient systems in lower mass haloes (resembling the Milky Way). These almost inevitably have higher \insitu{} metallicity and lower accreted metallicity, due to the absence of massive progenitors (see sec.~\ref{sec:accreted_fraction_trends}). 

Despite the generally flat gradients in the \apc{highest density} regions of both \apc{\insitu{} and accreted} components, stars in the outskirts of galaxies are more metal poor on average. 
This is mainly because accreted debris from weakly-bound, low-mass progenitors typically extends to larger distances than debris from massive progenitors, which sink rapidly due to dynamical friction before the majority of their stars are unbound (\citealt{Bullock:2005aa}; \citetalias{Cooper:2010aa}; \citealt{Amorisco:2017aa}). In our models, we also see that the handful of (old) \insitu{} stars that contribute to the average \insitu{} profile at larger distances are also more metal-poor than the typical \insitu{} metallicity for their virial or stellar mass. This has the effect of blurring the transition between the \insitu{} and accreted \apc{stellar populations}.

Fig.~\ref{fig:metals_at_radius} shows the average stellar metallicity\footnote{These metallicities correspond to the total mass fraction of metals in stars, rather than more directly observed quantities such as [Fe/H]. Although the simulated relation is as tight as observations imply, its zero-point is model-dependent. We caution against using this figure
to read off absolute mass estimates from observed metallicities.} in circular annuli drawn at 10, 30, and 65~kpc, plotted against the mass of the most massive progenitor of the galaxy. We plot an individual point for each galaxy in our sample (black; note that metallicity is on the horizontal axis in this figure). Cyan points show the average metallicity of accreted stars only, in the same annuli. We overplot contours showing the galaxy-average mass--metallicity relation in our model. An annulus at 30~kpc might be assumed to be \apc{dominated by accreted stars} in the majority of galaxies, \apc{although, in a detailed study of the Auriga simulations, \citet{Monachesi:2019aa} find a somewhat more complex picture}. \apc{At this radius in our models, we find that an estimate of the mass} of the most massive progenitor based on a measurement of average metallicity, combined with the mass--metallicity relation, would be biased \apc{high{}} by $\lesssim1$~dex. 

Towards lower masses, the scatter increases, such that the progenitor mass would be overestimated by several orders of magnitude in some cases. This is the consequence of the metallicity at this seemingly large radius not always being dominated by the most massive progenitor. In many cases, in our model, \insitu{} stars make a significant contribution, increasing the measured metallicity for a given most-massive progenitor mass. This issue is substantially worse for metallicity measured at 10~kpc, even in the least massive galaxies for which $\sim10$~kpc is $\gtrsim 5 r_{50}$. In an annulus of radius $\sim65$~kpc, mapping the average metallicity to the mass metallicity relation gives an almost unbiased estimate of the mass of the most massive progenitor, albeit with a slight increase in the scatter, and a tendency towards underestimates at higher mass. However, even for massive galaxies,  measurements of colour (let alone spectroscopic metallicity) at such distances will be challenging. In this sense, it is encouraging that the bias at $\sim30$~kpc appears quite small, especially for massive galaxies. Our model suggests caution when this approach is extended towards lower masses. For illustration, the purple regions on Fig.~\ref{fig:metals_at_radius} indicate the  $1\sigma$ and $2\sigma$ mass range for the most-massive progenitors of M33 analogues in our model. In such systems, our results imply that measurements at $\sim 30$~kpc should have the least bias and uncertainty $\lesssim1$~dex, although with the potential for outliers, particularly if the metallicity is low.

\section{Conclusions}
\label{sec:conclusions}

We have examined the masses and density profiles of stellar haloes around field dwarf galaxies in the \coco{} $N$-body simulation by applying the \stings{} particle tagging methodology \citepalias{Cooper:2010aa,Cooper:2017wv} together with the semi-analytic galaxy formation model, \galform{}, of \citet{Lacey:2016aa}. This study extends previous work on the stellar haloes of systems with $M_{200} \gtrsim 10^{12}\,\mathrm{M_{\odot}}$ by \citetalias{Cooper:2013aa} to much lower virial masses, $M_{200} \gtrsim 10^{8}\,\mathrm{M_{\odot}}$, accounting for effectively all  haloes in the field capable of forming stars at some point in their history. 

We have validated our baseline model through comparisons to the SMHM relation (Fig.~\ref{fig:mstar_mhalo}) and observed size--mass relations (Fig.~\ref{fig:sizemass_central}, see also Appendix \ref{fig:appendix:m33_sizemass}). We find our model reproduces these relations to a level of accuracy similar to current state-of-the-art hydrodynamical simulations \apc{in cosmological volumes}. \apc{We make a portion of our fiducial model publicly available and plan further data releases in future (see Appendix~\ref{appendix:datarelease}).}

Across the \apc{present day} galaxy population, we find a distribution of accreted stellar mass fraction as a function of total stellar mass that is consistent with our previous work and with observational data. Our approach is  different to the hydrodynamical methods used by most current cosmological simulations, so this consistency suggests that the overall statistical properties of accreted haloes are not strongly model dependent, provided models are reasonably well calibrated to the present-day stellar mass function. Differences in detail may be due to resolution limitations in cosmological hydrodynamical simulations, the difficulty of calibrating zoom simulations, uncertainties associated \apc{with } the formation of \insitu{} haloes in hydrodynamical models, and
the dynamical limitations of particle tagging. These factors are coupled. In particular, the presence (and in detail, mass and evolution) of a concentrated baryonic contribution to the central potential (e.g.\ a disc) may have a substantial effect on the shape and structure of the stellar halo, especially for Milky Way analogues where the mass fraction of baryons in a central disc is greatest \citep[e.g][]{Errani:2017aa,Richings:2020aa}. The growth of galactic discs and their effects on satellite disruption remain highly uncertain. For example, there is some evidence that the Milky Way satellite population has not been depleted to the extent typically predicted by hydrodynamical simulations of galaxies with comparable discs \citep[e.g.][]{Kim:2021aa, Shipp:2024aa}. Although more detailed comparisons with particle tagging methods in cosmological simulations would be worthwhile, we suspect \citepalias[following][]{Cooper:2017wv} that different implementations of baryonic physics overwhelmingly dominate differences between the predictions for stellar halo properties from different models. 

The wide ranges of stellar mass and virial mass in our simulation help to emphasize a straightforward but important point concerning statistics such as accreted mass fractions for samples of Milky Way analogues. This point was discussed at length in Section~\ref{sec:accreted_fraction_trends} and illustrated in Figs.~\ref{fig:macc_mstar_g13_3}--\ref{fig:loweff_smhm_breakdown}. If the stellar mass of the Milky Way is \apc{exceptionally} high for its virial mass (as suggested by abundance matching), it follows that, in a sample of analogues selected at fixed \textit{stellar mass}, galaxies with similar \textit{virial} mass to the Milky Way will necessarily have  \apc{exceptionally} \textit{low} accreted mass fractions. Correspondingly, in a sample selected by \textit{virial mass} (e.g. in a simulation, or \apc{based on} an observable proxy such as circular velocity or weak lensing), galaxies with the \textit{stellar mass} of the Milky Way are expected to have \apc{exceptionally} \textit{high} accreted mass fractions. The Milky Way itself appears to be consistent with this logic  since it \apc{is thought to have} a relatively underweight \apc{stellar halo} \apc{compared to other galaxies of similar} stellar mass.

Our main results concern the distribution of accreted stars around field galaxies less massive than the Milky Way. The population statistics of the accreted haloes in such galaxies have not been studied extensively in the literature. In this regime, we find the following:

\begin{enumerate}

  \item Accreted stars contribute only a small fraction of the total mass ($\lesssim 10$~per~cent) and \apc{typically} make no significant contribution to the total stellar mass surface density  (Figs.~\ref{fig:m33_profiles_by_mhalo}, \ref{fig:mstar_mhalo_accreted_properties}, \ref{fig:density_profiles_by_halo_mass}, and \ref{fig:density_profiles_by_stellar_mass}). 
  
  \item In haloes with $M_{200} \lesssim 10^{11}\,\mathrm{M_{\odot}}$, the density of accreted stars is significantly lower than that of \insitu{} stars at all radii (in particular at radii with surface mass densities that are potentially observable with integrated light images, $\Sigma \gtrsim1000\,\mathrm{M_{\odot}\,kpc^{-2}}$ (Figs.~\ref{fig:density_profiles_by_halo_mass} and \ref{fig:density_parameters_by_halo_mass}).
  \item The \apc{scale lengths of accreted stellar haloes} around low mass galaxies are comparable to those of their \insitu{} components: this contrasts with the \apc{accreted} stellar haloes of Milky Way analogues, which are typically much more extended than their \insitu{} stars (Figs.~\ref{fig:density_profiles_by_halo_mass} and \ref{fig:density_parameters_by_halo_mass}).
  \item At the lowest virial masses, the typical density profiles of accreted and \insitu{} stars \apc{(measured in circular annuli)} have almost identical Sersic index and scale radius, differing only in their amplitude (Fig.~\ref{fig:density_parameters_by_halo_mass}). 
  \item Our results are in reasonable agreement with data on M33, with the best agreement for \coco{} haloes in the range $11.5<\log_{10}\,M_{200}/\Msol<11.75$ (Figs.~\ref{fig:dwarf_profiles_fix_mstar} and \ref{fig:dwarf_profiles_fix_mhalo}). At this stellar mass scale, the surface density of accreted stars exceeds that of the \insitu{} component at  $\approx15\,\mathrm{kpc}$, corresponding to surface densities $3.5\lesssim \log_{10}\,M_\star/\Msol\,\mathrm{kpc^{-2}}\lesssim6$, varying systematically with virial mass. 
  \item At stellar masses, $M_\mathrm{\star} \sim 10^{9}\,\Msol$, accreted components comprising multiple significant progenitors become increasingly rare; below $M_\mathrm{\star} \sim 10^{8}\,\Msol$, the fraction of galaxies with no significant accreted component increases, reaching 50~per~cent at $M_\mathrm{\star} \sim 10^{7}\,\Msol$ (Figs.~\ref{fig:mm_prog_fractions} and \ref{fig:nsig_by_mstar}). However, since accreted and \insitu{} material in galaxies are hard to distinguish at these masses, this does not imply that 50 per cent of galaxies of  $M_\mathrm{\star} \sim 10^{7}\,\Msol$ will be observed to have a `stellar halo' in the classical sense of metal-poor stars on eccentric orbits. 
  \item \apc{The radius at which the surface density of accreted stars exceeds that of \insitu{} stars does not correspond to a sharp transition in age or metallicity; rather, we find a gradual variation of stellar populations across this transition, with steeper age and metallicity gradients in the outskirts of less massive galaxies. In the inner regions of low-mass galaxies, stellar population gradients in the accreted material are weak and driven mostly by outwardly decreasing age. Average stellar population profiles show clear trends with viral mass or stellar mass, but individual galaxies can show substantial variation around these averages. Figs.~\ref{fig:ages_mhalo} -- \ref{fig:zmet_mstar} summarize our predictions for the age and metallicity profiles of our simulated galaxies.}
  \item \apc{We find that the average metallicity at $\sim30\,\mathrm{kpc}$ provides a reasonable estimate of the mass of the most massive progenitor, with a bias of $\lesssim1$~dex and increasing scatter towards lower mass, both due primarily to the contribution of \insitu{} stars (Fig.~\ref{fig:metals_at_radius})}.
  
\end{enumerate}

In detail, our conclusions likely depend somewhat on our choice of galaxy formation model. For example, the inefficient tail of low stellar mass galaxies in the Milky Way mass range may be a peculiarity of the \citetalias{Lacey:2016aa}  model (see Appendix~\ref{appendix:mstar_mhalo}). %
Since we predict that \apc{a characteristic of such systems is that they will have relatively bright accreted haloes} (in comparison to fainter galaxies), \apc{this provides potentially observable evidence for their existence}. \refrep{Our predictions for accreted stellar halo properties in low-mass systems also likely depend on resolution; at lower virial mass, the accreted halo is assembled from progenitors that are themselves closer to the resolution limit and likely subject to greater uncertainty in the treatment of star formation in \galform{}. While the \citetalias{Lacey:2016aa} \galform{} model is capable of reproducing the Milky Way satellite luminosity function in haloes of similar virial mass \citep[e.g][]{Bose:2018aa}, there is room to improve its treatment of star formation in low mass galaxies \citep[e.g.][]{Benson:2010ab}}. \apc{More generally, sufficiently deep observations may provide an alternative means of testing the predictions of our model regarding the correspondence between stellar mass, virial mass (inferred, for example, from lensing, rotation curves or satellite / globular cluster counts and kinematics), and the surface brightness profile of diffuse light.}. \apc{Meanwhile,} the particle-tagging approach is well suited to \apc{further exploration of the} differences between the predictions of semi-analytic galaxy formation models in the regime of stellar haloes and low surface brightness features, and we expect to do so in future work.

\section*{Data Availability Statement }

A large subset of the `star particle' data described in this work is made available as described in Appendix~\ref{appendix:datarelease}. Further data sets will be released incrementally, documented at \url{http://github.com/nthu-ga/stings-data}. The underlying dark matter simulation is available on request to the authors of \citet{Hellwing:2016aa}.

\section*{Acknowledgments}

APC acknowledges support from the Taiwanese Ministry of Education Yushan Fellowship, MOE-113-YSFMS-0002-001-P2, and grants 109-2112-M-007-011-MY3, 112-2112-M-007-017, and 112-2112-M-007-009 awarded by the Taiwanese National Science and Technology Council. Support for APC in the early stages of this work was provided by a COFUND/Durham Junior Research Fellowship under EU grant [267209]. CSF acknowledges STFC Consolidated Grant ST/X001075/1 and support from the European Research Council through ERC Advanced Investigator grant, DMIDAS [GA 786910] to CSF. SB is supported by the UK Research and Innovation (UKRI) Future Leaders Fellowship [grant number MR/V023381/1]. This work is supported via the research project `COLAB' funded by the National Science Center, Poland, under agreement number UMO-2020/39/B/ST9/03494.
This work used the DiRAC@Durham facility managed by the Institute for Computational Cosmology on behalf of the STFC DiRAC HPC Facility (\url{www.dirac.ac.uk}). The equipment was funded by BEIS capital funding via STFC capital grants ST/K00042X/1, ST/P002293/1 and ST/R002371/1, Durham University and STFC operations grant ST/R000832/1. DiRAC is part of the National e-Infrastructure. This work also used high-performance computing facilities operated by the Center for Informatics and Computation in Astronomy (CICA) at National Tsing Hua University. This equipment was funded by the Ministry of Education of Taiwan, the National Science and Technology Council, and National Tsing Hua University.

\section*{Software}

This work made use of \texttt{astropy} \citep{astropy:2013, astropy:2018, astropy:2022}, \texttt{jupyter} \citep{2007CSE.....9c..21P, kluyver2016jupyter}, \texttt{matplotlib} \citep{Hunter:2007}, \texttt{numpy} \citep{numpy}, \texttt{python} \citep{python}, \texttt{scipy} \citep{2020SciPy-NMeth, scipy_13352243}, and \texttt{PyTables} \citep{pytables}. Software citation information aggregated using \texttt{\href{https://www.tomwagg.com/software-citation-station/}{The Software Citation Station}} \citep{software-citation-station-paper, software-citation-station-zenodo}.

\bibliographystyle{mnras} 
\bibliography{astro_master} 

\begin{thebibliography}{}
\makeatletter
\relax
\def\mn@urlcharsother{\let\do\@makeother \do\$\do\&\do\#\do\^\do\_\do\%\do\~}
\def\mn@doi{\begingroup\mn@urlcharsother \@ifnextchar [ {\mn@doi@}
  {\mn@doi@[]}}
\def\mn@doi@[#1]#2{\def\@tempa{#1}\ifx\@tempa\@empty \href
  {http://dx.doi.org/#2} {doi:#2}\else \href {http://dx.doi.org/#2} {#1}\fi
  \endgroup}
\def\mn@eprint#1#2{\mn@eprint@#1:#2::\@nil}
\def\mn@eprint@arXiv#1{\href {http://arxiv.org/abs/#1} {{\tt arXiv:#1}}}
\def\mn@eprint@dblp#1{\href {http://dblp.uni-trier.de/rec/bibtex/#1.xml}
  {dblp:#1}}
\def\mn@eprint@#1:#2:#3:#4\@nil{\def\@tempa {#1}\def\@tempb {#2}\def\@tempc
  {#3}\ifx \@tempc \@empty \let \@tempc \@tempb \let \@tempb \@tempa \fi \ifx
  \@tempb \@empty \def\@tempb {arXiv}\fi \@ifundefined
  {mn@eprint@\@tempb}{\@tempb:\@tempc}{\expandafter \expandafter \csname
  mn@eprint@\@tempb\endcsname \expandafter{\@tempc}}}

\bibitem[\protect\citeauthoryear{{Abadi}, {Navarro}, {Steinmetz}  \&
  {Eke}}{{Abadi} et~al.}{2003}]{Abadi:2003aa}
{Abadi} M.~G.,  {Navarro} J.~F.,  {Steinmetz} M.,   {Eke} V.~R.,  2003, \mn@doi
  [\apj] {10.1086/375512}, \href
  {https://ui.adsabs.harvard.edu/abs/2003ApJ...591..499A} {591, 499}

\bibitem[\protect\citeauthoryear{{Abraham} \& {van Dokkum}}{{Abraham} \& {van
  Dokkum}}{2014}]{Abraham:2014aa}
{Abraham} R.~G.,  {van Dokkum} P.~G.,  2014, \mn@doi [\pasp] {10.1086/674875},
  \href {http://adsabs.harvard.edu/abs/2014PASP..126...55A} {126, 55}

\bibitem[\protect\citeauthoryear{{Amorisco}}{{Amorisco}}{2017a}]{Amorisco:2017aa}
{Amorisco} N.~C.,  2017a, \mn@doi [\mnras] {10.1093/mnras/stw2229}, \href
  {http://adsabs.harvard.edu/abs/2017MNRAS.464.2882A} {464, 2882}

\bibitem[\protect\citeauthoryear{{Amorisco}}{{Amorisco}}{2017b}]{Amorisco:2017ac}
{Amorisco} N.~C.,  2017b, \mn@doi [\mnras] {10.1093/mnrasl/slx044}, \href
  {http://adsabs.harvard.edu/abs/2017MNRAS.469L..48A} {469, L48}

\bibitem[\protect\citeauthoryear{{Annibali} et~al.,}{{Annibali}
  et~al.}{2020}]{Annibali:2020aa}
{Annibali} F.,  et~al., 2020, \mn@doi [\mnras] {10.1093/mnras/stz3185}, \href
  {https://ui.adsabs.harvard.edu/abs/2020MNRAS.491.5101A} {491, 5101}

\bibitem[\protect\citeauthoryear{{Astropy Collaboration} et~al.,}{{Astropy
  Collaboration} et~al.}{2013}]{astropy:2013}
{Astropy Collaboration} et~al., 2013, \mn@doi [\aap]
  {10.1051/0004-6361/201322068}, \href
  {http://adsabs.harvard.edu/abs/2013A%26A...558A..33A} {558, A33}

\bibitem[\protect\citeauthoryear{{Astropy Collaboration} et~al.,}{{Astropy
  Collaboration} et~al.}{2018}]{astropy:2018}
{Astropy Collaboration} et~al., 2018, \mn@doi [\aj] {10.3847/1538-3881/aabc4f},
  \href {https://ui.adsabs.harvard.edu/abs/2018AJ....156..123A} {156, 123}

\bibitem[\protect\citeauthoryear{{Astropy Collaboration} et~al.,}{{Astropy
  Collaboration} et~al.}{2022}]{astropy:2022}
{Astropy Collaboration} et~al., 2022, \mn@doi [\apj]
  {10.3847/1538-4357/ac7c74}, \href
  {https://ui.adsabs.harvard.edu/abs/2022ApJ...935..167A} {935, 167}

\bibitem[\protect\citeauthoryear{{Bailin}, {Bell}, {Valluri}, {Stinson},
  {Debattista}, {Couchman}  \& {Wadsley}}{{Bailin}
  et~al.}{2014}]{Bailin:2014aa}
{Bailin} J.,  {Bell} E.~F.,  {Valluri} M.,  {Stinson} G.~S.,  {Debattista}
  V.~P.,  {Couchman} H.~M.~P.,   {Wadsley} J.,  2014, \mn@doi [\apj]
  {10.1088/0004-637X/783/2/95}, \href
  {http://adsabs.harvard.edu/abs/2014ApJ...783...95B} {783, 95}

\bibitem[\protect\citeauthoryear{{Behroozi}, {Wechsler}, {Hearin}  \&
  {Conroy}}{{Behroozi} et~al.}{2019}]{Behroozi:2019aa}
{Behroozi} P.,  {Wechsler} R.~H.,  {Hearin} A.~P.,   {Conroy} C.,  2019,
  \mn@doi [\mnras] {10.1093/mnras/stz1182}, \href
  {https://ui.adsabs.harvard.edu/abs/2019MNRAS.488.3143B} {488, 3143}

\bibitem[\protect\citeauthoryear{{Belokurov} \& {Kravtsov}}{{Belokurov} \&
  {Kravtsov}}{2023}]{Belokurov:2023aa}
{Belokurov} V.,  {Kravtsov} A.,  2023, \mn@doi [\mnras]
  {10.1093/mnras/stad2241}, \href
  {https://ui.adsabs.harvard.edu/abs/2023MNRAS.525.4456B} {525, 4456}

\bibitem[\protect\citeauthoryear{{Belokurov}, {Erkal}, {Evans}, {Koposov}  \&
  {Deason}}{{Belokurov} et~al.}{2018}]{Belokurov:2018aa}
{Belokurov} V.,  {Erkal} D.,  {Evans} N.~W.,  {Koposov} S.~E.,   {Deason}
  A.~J.,  2018, \mn@doi [\mnras] {10.1093/mnras/sty982}, \href
  {http://adsabs.harvard.edu/abs/2018MNRAS.478..611B} {478, 611}

\bibitem[\protect\citeauthoryear{{Benitez-Llambay} \&
  {Frenk}}{{Benitez-Llambay} \& {Frenk}}{2020}]{Benitez-Llambay:2020uy}
{Benitez-Llambay} A.,  {Frenk} C.,  2020, \mn@doi [\mnras]
  {10.1093/mnras/staa2698}, \href
  {https://ui.adsabs.harvard.edu/abs/2020MNRAS.498.4887B} {498, 4887}

\bibitem[\protect\citeauthoryear{{Ben{\'\i}tez-Llambay}, {Frenk}, {Ludlow}  \&
  {Navarro}}{{Ben{\'\i}tez-Llambay} et~al.}{2019}]{Benitez-Llambay:2019aa}
{Ben{\'\i}tez-Llambay} A.,  {Frenk} C.~S.,  {Ludlow} A.~D.,   {Navarro} J.~F.,
  2019, \mn@doi [\mnras] {10.1093/mnras/stz1890}, \href
  {https://ui.adsabs.harvard.edu/abs/2019MNRAS.488.2387B} {488, 2387}

\bibitem[\protect\citeauthoryear{{Benson} \& {Bower}}{{Benson} \&
  {Bower}}{2010}]{Benson:2010ab}
{Benson} A.~J.,  {Bower} R.,  2010, \mn@doi [\mnras]
  {10.1111/j.1365-2966.2010.16592.x}, \href
  {http://adsabs.harvard.edu/abs/2010MNRAS.405.1573B} {405, 1573}

\bibitem[\protect\citeauthoryear{{Benson}, {Frenk}, {Baugh}, {Cole}  \&
  {Lacey}}{{Benson} et~al.}{2003}]{Benson:2003aa}
{Benson} A.~J.,  {Frenk} C.~S.,  {Baugh} C.~M.,  {Cole} S.,   {Lacey} C.~G.,
  2003, \mn@doi [\mnras] {10.1046/j.1365-8711.2003.06709.x}, \href
  {https://ui.adsabs.harvard.edu/abs/2003MNRAS.343..679B} {343, 679}

\bibitem[\protect\citeauthoryear{{Bernardi}, {Meert}, {Sheth}, {Fischer},
  {Huertas-Company}, {Maraston}, {Shankar}  \& {Vikram}}{{Bernardi}
  et~al.}{2017}]{Bernardi:2017ab}
{Bernardi} M.,  {Meert} A.,  {Sheth} R.~K.,  {Fischer} J.~L.,
  {Huertas-Company} M.,  {Maraston} C.,  {Shankar} F.,   {Vikram} V.,  2017,
  \mn@doi [\mnras] {10.1093/mnras/stx176}, \href
  {https://ui.adsabs.harvard.edu/abs/2017MNRAS.467.2217B} {467, 2217}

\bibitem[\protect\citeauthoryear{{Boardman} et~al.,}{{Boardman}
  et~al.}{2020}]{Boardman:2020aa}
{Boardman} N.,  et~al., 2020, \mn@doi [\mnras] {10.1093/mnras/staa2731}, \href
  {https://ui.adsabs.harvard.edu/abs/2020MNRAS.498.4943B} {498, 4943}

\bibitem[\protect\citeauthoryear{{Bose} \& {Deason}}{{Bose} \&
  {Deason}}{2023}]{Bose:2023aa}
{Bose} S.,  {Deason} A.~J.,  2023, \mn@doi [\mnras] {10.1093/mnras/stad1123},
  \href {https://ui.adsabs.harvard.edu/abs/2023MNRAS.522.5013B} {522, 5013}

\bibitem[\protect\citeauthoryear{{Bose} et~al.,}{{Bose}
  et~al.}{2017}]{Bose:2017aa}
{Bose} S.,  et~al., 2017, \mn@doi [\mnras] {10.1093/mnras/stw2686}, \href
  {http://adsabs.harvard.edu/abs/2017MNRAS.464.4520B} {464, 4520}

\bibitem[\protect\citeauthoryear{{Bose}, {Deason}  \& {Frenk}}{{Bose}
  et~al.}{2018}]{Bose:2018aa}
{Bose} S.,  {Deason} A.~J.,   {Frenk} C.~S.,  2018, \mn@doi [\apj]
  {10.3847/1538-4357/aacbc4}, \href
  {https://ui.adsabs.harvard.edu/abs/2018ApJ...863..123B} {863, 123}

\bibitem[\protect\citeauthoryear{{Bower}, {Benson}, {Malbon}, {Helly}, {Frenk},
  {Baugh}, {Cole}  \& {Lacey}}{{Bower} et~al.}{2006}]{Bower:2006aa}
{Bower} R.~G.,  {Benson} A.~J.,  {Malbon} R.,  {Helly} J.~C.,  {Frenk} C.~S.,
  {Baugh} C.~M.,  {Cole} S.,   {Lacey} C.~G.,  2006, \mn@doi [\mnras]
  {10.1111/j.1365-2966.2006.10519.x}, \href
  {http://adsabs.harvard.edu/abs/2006MNRAS.370..645B} {370, 645}

\bibitem[\protect\citeauthoryear{{Bower}, {Schaye}, {Frenk}, {Theuns},
  {Schaller}, {Crain}  \& {McAlpine}}{{Bower} et~al.}{2017}]{Bower:2017vf}
{Bower} R.~G.,  {Schaye} J.,  {Frenk} C.~S.,  {Theuns} T.,  {Schaller} M.,
  {Crain} R.~A.,   {McAlpine} S.,  2017, \mn@doi [\mnras]
  {10.1093/mnras/stw2735}, \href
  {https://ui.adsabs.harvard.edu/abs/2017MNRAS.465...32B} {465, 32}

\bibitem[\protect\citeauthoryear{{Boylan-Kolchin}, {Springel}, {White}  \&
  {Jenkins}}{{Boylan-Kolchin} et~al.}{2010}]{Boylan-Kolchin:2010aa}
{Boylan-Kolchin} M.,  {Springel} V.,  {White} S.~D.~M.,   {Jenkins} A.,  2010,
  \mn@doi [\mnras] {10.1111/j.1365-2966.2010.16774.x}, \href
  {http://adsabs.harvard.edu/abs/2010MNRAS.406..896B} {406, 896}

\bibitem[\protect\citeauthoryear{{Brook} et~al.,}{{Brook}
  et~al.}{2011}]{Brook:2011aa}
{Brook} C.~B.,  et~al., 2011, \mn@doi [\mnras]
  {10.1111/j.1365-2966.2011.18545.x}, \href
  {https://ui.adsabs.harvard.edu/abs/2011MNRAS.415.1051B} {415, 1051}

\bibitem[\protect\citeauthoryear{{Brook}, {Stinson}, {Gibson}, {Ro{\v{s}}kar},
  {Wadsley}  \& {Quinn}}{{Brook} et~al.}{2012}]{Brook:2012aa}
{Brook} C.~B.,  {Stinson} G.,  {Gibson} B.~K.,  {Ro{\v{s}}kar} R.,  {Wadsley}
  J.,   {Quinn} T.,  2012, \mn@doi [\mnras] {10.1111/j.1365-2966.2011.19740.x},
  \href {https://ui.adsabs.harvard.edu/abs/2012MNRAS.419..771B} {419, 771}

\bibitem[\protect\citeauthoryear{{Bullock} \& {Johnston}}{{Bullock} \&
  {Johnston}}{2005}]{Bullock:2005aa}
{Bullock} J.~S.,  {Johnston} K.~V.,  2005, \mn@doi [\apj] {10.1086/497422},
  \href {http://adsabs.harvard.edu/abs/2005ApJ...635..931B} {635, 931}

\bibitem[\protect\citeauthoryear{{Bullock}, {Kravtsov}  \&
  {Weinberg}}{{Bullock} et~al.}{2001}]{Bullock:2001aa}
{Bullock} J.~S.,  {Kravtsov} A.~V.,   {Weinberg} D.~H.,  2001, \mn@doi [\apj]
  {10.1086/318681}, \href {http://adsabs.harvard.edu/abs/2001ApJ...548...33B}
  {548, 33}

\bibitem[\protect\citeauthoryear{{Callingham} et~al.,}{{Callingham}
  et~al.}{2019}]{Callingham:2019aa}
{Callingham} T.~M.,  et~al., 2019, \mn@doi [\mnras] {10.1093/mnras/stz365},
  \href {https://ui.adsabs.harvard.edu/abs/2019MNRAS.484.5453C} {484, 5453}

\bibitem[\protect\citeauthoryear{{Carlin} et~al.,}{{Carlin}
  et~al.}{2016}]{Carlin:2016aa}
{Carlin} J.~L.,  et~al., 2016, \mn@doi [\apjl] {10.3847/2041-8205/828/1/L5},
  \href {https://ui.adsabs.harvard.edu/abs/2016ApJ...828L...5C} {828, L5}

\bibitem[\protect\citeauthoryear{{Chandro-G{\'o}mez}
  et~al.,}{{Chandro-G{\'o}mez} et~al.}{2025}]{Chandro-Gomez:2025aa}
{Chandro-G{\'o}mez} {\'A}.,  et~al., 2025, \mn@doi [\mnras]
  {10.1093/mnras/staf519}, \href
  {https://ui.adsabs.harvard.edu/abs/2025MNRAS.539..776C} {539, 776}

\bibitem[\protect\citeauthoryear{{Chiti} et~al.,}{{Chiti}
  et~al.}{2021}]{Chiti:2021aa}
{Chiti} A.,  et~al., 2021, \mn@doi [Nature Astronomy]
  {10.1038/s41550-020-01285-w}, \href
  {https://ui.adsabs.harvard.edu/abs/2021NatAs...5..392C} {5, 392}

\bibitem[\protect\citeauthoryear{{Cockcroft} et~al.,}{{Cockcroft}
  et~al.}{2013}]{Cockcroft:2013aa}
{Cockcroft} R.,  et~al., 2013, \mn@doi [\mnras] {10.1093/mnras/sts112}, \href
  {http://adsabs.harvard.edu/abs/2013MNRAS.428.1248C} {428, 1248}

\bibitem[\protect\citeauthoryear{{Cole}, {Lacey}, {Baugh}  \& {Frenk}}{{Cole}
  et~al.}{2000}]{Cole:2000aa}
{Cole} S.,  {Lacey} C.~G.,  {Baugh} C.~M.,   {Frenk} C.~S.,  2000, \mn@doi
  [\mnras] {10.1046/j.1365-8711.2000.03879.x}, \href
  {http://adsabs.harvard.edu/abs/2000MNRAS.319..168C} {319, 168}

\bibitem[\protect\citeauthoryear{{Conroy} et~al.,}{{Conroy}
  et~al.}{2024}]{Conroy:2024aa}
{Conroy} C.,  et~al., 2024, \mn@doi [\apj] {10.3847/1538-4357/ad414a}, \href
  {https://ui.adsabs.harvard.edu/abs/2024ApJ...968..129C} {968, 129}

\bibitem[\protect\citeauthoryear{{Cooper} et~al.,}{{Cooper}
  et~al.}{2010}]{Cooper:2010aa}
{Cooper} A.~P.,  et~al., 2010, \mn@doi [\mnras]
  {10.1111/j.1365-2966.2010.16740.x}, \href
  {http://adsabs.harvard.edu/abs/2010MNRAS.406..744C} {406, 744}

\bibitem[\protect\citeauthoryear{{Cooper}, {D'Souza}, {Kauffmann}, {Wang},
  {Boylan-Kolchin}, {Guo}, {Frenk}  \& {White}}{{Cooper}
  et~al.}{2013}]{Cooper:2013aa}
{Cooper} A.~P.,  {D'Souza} R.,  {Kauffmann} G.,  {Wang} J.,  {Boylan-Kolchin}
  M.,  {Guo} Q.,  {Frenk} C.~S.,   {White} S.~D.~M.,  2013, \mn@doi [\mnras]
  {10.1093/mnras/stt1245}, \href
  {http://adsabs.harvard.edu/abs/2013MNRAS.434.3348C} {434, 3348}

\bibitem[\protect\citeauthoryear{{Cooper}, {Gao}, {Guo}, {Frenk}, {Jenkins},
  {Springel}  \& {White}}{{Cooper} et~al.}{2015a}]{Cooper:2015ab}
{Cooper} A.~P.,  {Gao} L.,  {Guo} Q.,  {Frenk} C.~S.,  {Jenkins} A.,
  {Springel} V.,   {White} S.~D.~M.,  2015a, \mn@doi [\mnras]
  {10.1093/mnras/stv1042}, \href
  {http://ui.adsabs.harvard.edu/abs/2015MNRAS.451.2703C} {451, 2703}

\bibitem[\protect\citeauthoryear{{Cooper}, {Parry}, {Lowing}, {Cole}  \&
  {Frenk}}{{Cooper} et~al.}{2015b}]{Cooper:2015aa}
{Cooper} A.~P.,  {Parry} O.~H.,  {Lowing} B.,  {Cole} S.,   {Frenk} C.,  2015b,
  \mn@doi [\mnras] {10.1093/mnras/stv2057}, \href
  {http://adsabs.harvard.edu/abs/2015MNRAS.454.3185C} {454, 3185}

\bibitem[\protect\citeauthoryear{{Cooper}, {Cole}, {Frenk}, {Le Bret}  \&
  {Pontzen}}{{Cooper} et~al.}{2017}]{Cooper:2017wv}
{Cooper} A.~P.,  {Cole} S.,  {Frenk} C.~S.,  {Le Bret} T.,   {Pontzen} A.,
  2017, \mn@doi [\mnras] {10.1093/mnras/stx955}, \href
  {https://ui.adsabs.harvard.edu/abs/2017MNRAS.469.1691C} {469, 1691}

\bibitem[\protect\citeauthoryear{{Crain} et~al.,}{{Crain}
  et~al.}{2015}]{Crain:2015aa}
{Crain} R.~A.,  et~al., 2015, \mn@doi [\mnras] {10.1093/mnras/stv725}, \href
  {https://ui.adsabs.harvard.edu/abs/2015MNRAS.450.1937C} {450, 1937}

\bibitem[\protect\citeauthoryear{{D'Souza} \& {Bell}}{{D'Souza} \&
  {Bell}}{2018}]{DSouza:2018aa}
{D'Souza} R.,  {Bell} E.~F.,  2018, \mn@doi [\mnras] {10.1093/mnras/stx3081},
  \href {https://ui.adsabs.harvard.edu/abs/2018MNRAS.474.5300D} {474, 5300}

\bibitem[\protect\citeauthoryear{{D'Souza}, {Kauffman}, {Wang}  \&
  {Vegetti}}{{D'Souza} et~al.}{2014}]{DSouza:2014aa}
{D'Souza} R.,  {Kauffman} G.,  {Wang} J.,   {Vegetti} S.,  2014, \mn@doi
  [\mnras] {10.1093/mnras/stu1194}, \href
  {http://adsabs.harvard.edu/abs/2014MNRAS.443.1433D} {443, 1433}

\bibitem[\protect\citeauthoryear{{D'Souza}, {Vegetti}  \&
  {Kauffmann}}{{D'Souza} et~al.}{2015}]{DSouza:2015aa}
{D'Souza} R.,  {Vegetti} S.,   {Kauffmann} G.,  2015, \mn@doi [\mnras]
  {10.1093/mnras/stv2234}, \href
  {http://adsabs.harvard.edu/abs/2015MNRAS.454.4027D} {454, 4027}

\bibitem[\protect\citeauthoryear{{Davis}, {Efstathiou}, {Frenk}  \&
  {White}}{{Davis} et~al.}{1985}]{Davis:1985tb}
{Davis} M.,  {Efstathiou} G.,  {Frenk} C.~S.,   {White} S.~D.~M.,  1985,
  \mn@doi [\apj] {10.1086/163168}, \href
  {https://ui.adsabs.harvard.edu/abs/1985ApJ...292..371D} {292, 371}

\bibitem[\protect\citeauthoryear{{Deason}, {Belokurov}, {Evans}  \&
  {Johnston}}{{Deason} et~al.}{2013}]{Deason:2013aa}
{Deason} A.~J.,  {Belokurov} V.,  {Evans} N.~W.,   {Johnston} K.~V.,  2013,
  \mn@doi [\apj] {10.1088/0004-637X/763/2/113}, \href
  {http://adsabs.harvard.edu/abs/2013ApJ...763..113D} {763, 113}

\bibitem[\protect\citeauthoryear{{Deason} et~al.,}{{Deason}
  et~al.}{2021}]{Deason:2021aa}
{Deason} A.~J.,  et~al., 2021, \mn@doi [\mnras] {10.1093/mnras/staa3984}, \href
  {https://ui.adsabs.harvard.edu/abs/2021MNRAS.501.5964D} {501, 5964}

\bibitem[\protect\citeauthoryear{{Deason}, {Bose}, {Fattahi}, {Amorisco},
  {Hellwing}  \& {Frenk}}{{Deason} et~al.}{2022}]{Deason:2022aa}
{Deason} A.~J.,  {Bose} S.,  {Fattahi} A.,  {Amorisco} N.~C.,  {Hellwing} W.,
  {Frenk} C.~S.,  2022, \mn@doi [\mnras] {10.1093/mnras/stab3524}, \href
  {https://ui.adsabs.harvard.edu/abs/2022MNRAS.511.4044D} {511, 4044}

\bibitem[\protect\citeauthoryear{{Duffy}, {Schaye}, {Kay}, {Dalla Vecchia},
  {Battye}  \& {Booth}}{{Duffy} et~al.}{2010}]{Duffy:2010aa}
{Duffy} A.~R.,  {Schaye} J.,  {Kay} S.~T.,  {Dalla Vecchia} C.,  {Battye}
  R.~A.,   {Booth} C.~M.,  2010, \mn@doi [\mnras]
  {10.1111/j.1365-2966.2010.16613.x}, \href
  {https://ui.adsabs.harvard.edu/abs/2010MNRAS.405.2161D} {405, 2161}

\bibitem[\protect\citeauthoryear{{Dutton}, {van den Bosch}, {Dekel}  \&
  {Courteau}}{{Dutton} et~al.}{2007}]{Dutton07}
{Dutton} A.~A.,  {van den Bosch} F.~C.,  {Dekel} A.,   {Courteau} S.,  2007,
  \mn@doi [\apj] {10.1086/509314}, \href
  {http://adsabs.harvard.edu/abs/2007ApJ...654...27D} {654, 27}

\bibitem[\protect\citeauthoryear{{Elias}, {Sales}, {Creasey}, {Cooper},
  {Bullock}, {Rich}  \& {Hernquist}}{{Elias} et~al.}{2018}]{Elias:2018aa}
{Elias} L.~M.,  {Sales} L.~V.,  {Creasey} P.,  {Cooper} M.~C.,  {Bullock}
  J.~S.,  {Rich} R.~M.,   {Hernquist} L.,  2018, \mn@doi [\mnras]
  {10.1093/mnras/sty1718}, \href
  {https://ui.adsabs.harvard.edu/abs/2018MNRAS.479.4004E} {479, 4004}

\bibitem[\protect\citeauthoryear{{Errani}, {Pe{\~n}arrubia}, {Laporte}  \&
  {G{\'o}mez}}{{Errani} et~al.}{2017}]{Errani:2017aa}
{Errani} R.,  {Pe{\~n}arrubia} J.,  {Laporte} C. F.~P.,   {G{\'o}mez} F.~A.,
  2017, \mn@doi [\mnras] {10.1093/mnrasl/slw211}, \href
  {https://ui.adsabs.harvard.edu/abs/2017MNRAS.465L..59E} {465, L59}

\bibitem[\protect\citeauthoryear{{Errani}, {Navarro}, {Pe{\~n}arrubia},
  {Famaey}  \& {Ibata}}{{Errani} et~al.}{2023}]{Errani:2023aa}
{Errani} R.,  {Navarro} J.~F.,  {Pe{\~n}arrubia} J.,  {Famaey} B.,   {Ibata}
  R.,  2023, \mn@doi [\mnras] {10.1093/mnras/stac3499}, \href
  {https://ui.adsabs.harvard.edu/abs/2023MNRAS.519..384E} {519, 384}

\bibitem[\protect\citeauthoryear{{Errani}, {Ibata}, {Navarro}, {Pe{\~n}arrubia}
   \& {Walker}}{{Errani} et~al.}{2024}]{Errani:2024aa}
{Errani} R.,  {Ibata} R.,  {Navarro} J.~F.,  {Pe{\~n}arrubia} J.,   {Walker}
  M.~G.,  2024, \mn@doi [\apj] {10.3847/1538-4357/ad402d}, \href
  {https://ui.adsabs.harvard.edu/abs/2024ApJ...968...89E} {968, 89}

\bibitem[\protect\citeauthoryear{{Font}, {Johnston}, {Bullock}  \&
  {Robertson}}{{Font} et~al.}{2006}]{Font:2006aa}
{Font} A.~S.,  {Johnston} K.~V.,  {Bullock} J.~S.,   {Robertson} B.~E.,  2006,
  \mn@doi [\apj] {10.1086/498970}, \href
  {http://adsabs.harvard.edu/abs/2006ApJ...638..585F} {638, 585}

\bibitem[\protect\citeauthoryear{{Font}, {McCarthy}, {Crain}, {Theuns},
  {Schaye}, {Wiersma}  \& {Dalla Vecchia}}{{Font} et~al.}{2011a}]{Font:2011uz}
{Font} A.~S.,  {McCarthy} I.~G.,  {Crain} R.~A.,  {Theuns} T.,  {Schaye} J.,
  {Wiersma} R.~P.~C.,   {Dalla Vecchia} C.,  2011a, \mn@doi [\mnras]
  {10.1111/j.1365-2966.2011.19227.x}, \href
  {https://ui.adsabs.harvard.edu/abs/2011MNRAS.416.2802F} {416, 2802}

\bibitem[\protect\citeauthoryear{{Font} et~al.,}{{Font}
  et~al.}{2011b}]{Font:2011aa}
{Font} A.~S.,  et~al., 2011b, \mn@doi [\mnras]
  {10.1111/j.1365-2966.2011.19339.x}, \href
  {http://adsabs.harvard.edu/abs/2011MNRAS.417.1260F} {417, 1260}

\bibitem[\protect\citeauthoryear{{Font} et~al.,}{{Font}
  et~al.}{2020}]{Font:2020aa}
{Font} A.~S.,  et~al., 2020, \mn@doi [\mnras] {10.1093/mnras/staa2463}, \href
  {https://ui.adsabs.harvard.edu/abs/2020MNRAS.498.1765F} {498, 1765}

\bibitem[\protect\citeauthoryear{{Frenk}, {White}, {Davis}  \&
  {Efstathiou}}{{Frenk} et~al.}{1988}]{Frenk88}
{Frenk} C.~S.,  {White} S.~D.~M.,  {Davis} M.,   {Efstathiou} G.,  1988,
  \mn@doi [\apj] {10.1086/166213}, \href
  {http://adsabs.harvard.edu/abs/1988ApJ...327..507F} {327, 507}

\bibitem[\protect\citeauthoryear{{Gilbert} et~al.,}{{Gilbert}
  et~al.}{2022}]{Gilbert:2022aa}
{Gilbert} K.~M.,  et~al., 2022, \mn@doi [\apj] {10.3847/1538-4357/ac3480},
  \href {https://ui.adsabs.harvard.edu/abs/2022ApJ...924..116G} {924, 116}

\bibitem[\protect\citeauthoryear{{Gilhuly} et~al.,}{{Gilhuly}
  et~al.}{2022}]{Gilhuly:2022aa}
{Gilhuly} C.,  et~al., 2022, \mn@doi [\apj] {10.3847/1538-4357/ac6750}, \href
  {https://ui.adsabs.harvard.edu/abs/2022ApJ...932...44G} {932, 44}

\bibitem[\protect\citeauthoryear{{Goater} et~al.,}{{Goater}
  et~al.}{2024}]{Goater:2024aa}
{Goater} A.,  et~al., 2024, \mn@doi [\mnras] {10.1093/mnras/stad3354}, \href
  {https://ui.adsabs.harvard.edu/abs/2024MNRAS.527.2403G} {527, 2403}

\bibitem[\protect\citeauthoryear{{G{\'o}mez} et~al.,}{{G{\'o}mez}
  et~al.}{2017}]{Gomez:2017aa}
{G{\'o}mez} F.~A.,  et~al., 2017, \mn@doi [\mnras] {10.1093/mnras/stx2149},
  \href {https://ui.adsabs.harvard.edu/abs/2017MNRAS.472.3722G} {472, 3722}

\bibitem[\protect\citeauthoryear{Gommers et~al.,}{Gommers
  et~al.}{2024}]{scipy_13352243}
Gommers R.,  et~al., 2024, scipy/scipy: SciPy 1.14.1,
  \mn@doi{10.5281/zenodo.13352243}, \url
  {https://doi.org/10.5281/zenodo.13352243}

\bibitem[\protect\citeauthoryear{{Gonzalez-Perez}, {Lacey}, {Baugh}, {Lagos},
  {Helly}, {Campbell}  \& {Mitchell}}{{Gonzalez-Perez}
  et~al.}{2014}]{Gonzalez-Perez:2014aa}
{Gonzalez-Perez} V.,  {Lacey} C.~G.,  {Baugh} C.~M.,  {Lagos} C.~D.~P.,
  {Helly} J.,  {Campbell} D.~J.~R.,   {Mitchell} P.~D.,  2014, \mn@doi [\mnras]
  {10.1093/mnras/stt2410}, \href
  {http://adsabs.harvard.edu/abs/2014MNRAS.439..264G} {439, 264}

\bibitem[\protect\citeauthoryear{{Gozman} et~al.,}{{Gozman}
  et~al.}{2023}]{Gozman:2023aa}
{Gozman} K.,  et~al., 2023, \mn@doi [\apj] {10.3847/1538-4357/acbe3a}, \href
  {https://ui.adsabs.harvard.edu/abs/2023ApJ...947...21G} {947, 21}

\bibitem[\protect\citeauthoryear{{Grand} et~al.,}{{Grand}
  et~al.}{2017}]{Grand:2017aa}
{Grand} R.~J.~J.,  et~al., 2017, \mn@doi [\mnras] {10.1093/mnras/stx071}, \href
  {http://adsabs.harvard.edu/abs/2017MNRAS.467..179G} {467, 179}

\bibitem[\protect\citeauthoryear{{Grand}, {Fragkoudi}, {G{\'o}mez}, {Jenkins},
  {Marinacci}, {Pakmor}  \& {Springel}}{{Grand} et~al.}{2024}]{Grand:2024aa}
{Grand} R. J.~J.,  {Fragkoudi} F.,  {G{\'o}mez} F.~A.,  {Jenkins} A.,
  {Marinacci} F.,  {Pakmor} R.,   {Springel} V.,  2024, \mn@doi [\mnras]
  {10.1093/mnras/stae1598}, \href
  {https://ui.adsabs.harvard.edu/abs/2024MNRAS.532.1814G} {532, 1814}

\bibitem[\protect\citeauthoryear{{Green}, {van den Bosch}  \& {Jiang}}{{Green}
  et~al.}{2021}]{Green:2021aa}
{Green} S.~B.,  {van den Bosch} F.~C.,   {Jiang} F.,  2021, \mn@doi [\mnras]
  {10.1093/mnras/stab696}, \href
  {https://ui.adsabs.harvard.edu/abs/2021MNRAS.503.4075G} {503, 4075}

\bibitem[\protect\citeauthoryear{{Green}, {van den Bosch}  \& {Jiang}}{{Green}
  et~al.}{2022}]{Green:2022aa}
{Green} S.~B.,  {van den Bosch} F.~C.,   {Jiang} F.,  2022, \mn@doi [\mnras]
  {10.1093/mnras/stab3130}, \href
  {https://ui.adsabs.harvard.edu/abs/2022MNRAS.509.2624G} {509, 2624}

\bibitem[\protect\citeauthoryear{{Guo} \& {White}}{{Guo} \&
  {White}}{2008}]{Guo:2008aa}
{Guo} Q.,  {White} S.~D.~M.,  2008, \mn@doi [\mnras]
  {10.1111/j.1365-2966.2007.12619.x}, \href
  {http://adsabs.harvard.edu/abs/2008MNRAS.384....2G} {384, 2}

\bibitem[\protect\citeauthoryear{{Guo} \& {White}}{{Guo} \&
  {White}}{2014}]{Guo14}
{Guo} Q.,  {White} S.,  2014, \mn@doi [\mnras] {10.1093/mnras/stt2116}, \href
  {http://adsabs.harvard.edu/abs/2014MNRAS.437.3228G} {437, 3228}

\bibitem[\protect\citeauthoryear{{Guo}, {White}, {Li}  \&
  {Boylan-Kolchin}}{{Guo} et~al.}{2010}]{Guo:2010aa}
{Guo} Q.,  {White} S.,  {Li} C.,   {Boylan-Kolchin} M.,  2010, \mn@doi [\mnras]
  {10.1111/j.1365-2966.2010.16341.x}, \href
  {http://adsabs.harvard.edu/abs/2010MNRAS.404.1111G} {404, 1111}

\bibitem[\protect\citeauthoryear{{Guo} et~al.,}{{Guo}
  et~al.}{2011a}]{Guo:2011ab}
{Guo} Q.,  et~al., 2011a, \mn@doi [\mnras] {10.1111/j.1365-2966.2010.18114.x},
  \href {http://adsabs.harvard.edu/abs/2011MNRAS.413..101G} {413, 101}

\bibitem[\protect\citeauthoryear{{Guo}, {Cole}, {Eke}  \& {Frenk}}{{Guo}
  et~al.}{2011b}]{Guo:2011aa}
{Guo} Q.,  {Cole} S.,  {Eke} V.,   {Frenk} C.,  2011b, \mn@doi [\mnras]
  {10.1111/j.1365-2966.2011.19270.x}, \href
  {http://adsabs.harvard.edu/abs/2011MNRAS.417..370G} {417, 370}

\bibitem[\protect\citeauthoryear{{Guzman} et~al.,}{{Guzman}
  et~al.}{2022}]{guzman:2022aa}
{Guzman} R.,  et~al., 2022, Analysis of Resolved Remnants of Accreted galaxies
  as a Key Instrument for Halo Surveys,
  {\href{https://www.cosmos.esa.int/documents/7423467/7423486/ESA-F2-ARRAKIHS-Phase-2-PUBLIC-v0.9.2.pdf/61b363d7-2a06-1196-5c40-c85aa90c2113?t=1667557422996}{ARRAKIHS
  Proposal}}

\bibitem[\protect\citeauthoryear{{Harmsen}, {Monachesi}, {Bell}, {de Jong},
  {Bailin}, {Radburn-Smith}  \& {Holwerda}}{{Harmsen}
  et~al.}{2017}]{Harmsen:2017aa}
{Harmsen} B.,  {Monachesi} A.,  {Bell} E.~F.,  {de Jong} R.~S.,  {Bailin} J.,
  {Radburn-Smith} D.~J.,   {Holwerda} B.~W.,  2017, \mn@doi [\mnras]
  {10.1093/mnras/stw2992}, \href
  {http://adsabs.harvard.edu/abs/2017MNRAS.466.1491H} {466, 1491}

\bibitem[\protect\citeauthoryear{Harris et~al.,}{Harris et~al.}{2020}]{numpy}
Harris C.~R.,  et~al., 2020, \mn@doi [Nature] {10.1038/s41586-020-2649-2}, 585,
  357

\bibitem[\protect\citeauthoryear{{Hellwing}, {Frenk}, {Cautun}, {Bose},
  {Helly}, {Jenkins}, {Sawala}  \& {Cytowski}}{{Hellwing}
  et~al.}{2016}]{Hellwing:2016aa}
{Hellwing} W.~A.,  {Frenk} C.~S.,  {Cautun} M.,  {Bose} S.,  {Helly} J.,
  {Jenkins} A.,  {Sawala} T.,   {Cytowski} M.,  2016, \mn@doi [\mnras]
  {10.1093/mnras/stw214}, \href
  {http://adsabs.harvard.edu/abs/2016MNRAS.457.3492H} {457, 3492}

\bibitem[\protect\citeauthoryear{{Higgs}, {McConnachie}, {Annau}, {Irwin},
  {Battaglia}, {C{\^o}t{\'e}}, {Lewis}  \& {Venn}}{{Higgs}
  et~al.}{2021}]{Higgs:2021aa}
{Higgs} C.~R.,  {McConnachie} A.~W.,  {Annau} N.,  {Irwin} M.,  {Battaglia} G.,
   {C{\^o}t{\'e}} P.,  {Lewis} G.~F.,   {Venn} K.,  2021, \mn@doi [\mnras]
  {10.1093/mnras/stab002}, \href
  {https://ui.adsabs.harvard.edu/abs/2021MNRAS.503..176H} {503, 176}

\bibitem[\protect\citeauthoryear{{Hunt} et~al.,}{{Hunt}
  et~al.}{2024}]{Hunt:2024aa}
{Hunt} L.~K.,  et~al., 2024, \mn@doi [arXiv e-prints]
  {10.48550/arXiv.2405.13499}, \href
  {https://ui.adsabs.harvard.edu/abs/2024arXiv240513499H} {p. arXiv:2405.13499}

\bibitem[\protect\citeauthoryear{Hunter}{Hunter}{2007}]{Hunter:2007}
Hunter J.~D.,  2007, \mn@doi [Computing in Science \& Engineering]
  {10.1109/MCSE.2007.55}, 9, 90

\bibitem[\protect\citeauthoryear{{Irwin} \& {Hatzidimitriou}}{{Irwin} \&
  {Hatzidimitriou}}{1995}]{Irwin:1995aa}
{Irwin} M.,  {Hatzidimitriou} D.,  1995, \mn@doi [\mnras]
  {10.1093/mnras/277.4.1354}, \href
  {https://ui.adsabs.harvard.edu/abs/1995MNRAS.277.1354I} {277, 1354}

\bibitem[\protect\citeauthoryear{{Jang}, {de Jong}, {Holwerda}, {Monachesi},
  {Bell}  \& {Bailin}}{{Jang} et~al.}{2020a}]{Jang:2020ab}
{Jang} I.~S.,  {de Jong} R.~S.,  {Holwerda} B.~W.,  {Monachesi} A.,  {Bell}
  E.~F.,   {Bailin} J.,  2020a, \mn@doi [\aap] {10.1051/0004-6361/201936994},
  \href {https://ui.adsabs.harvard.edu/abs/2020A&A...637A...8J} {637, A8}

\bibitem[\protect\citeauthoryear{{Jang} et~al.,}{{Jang}
  et~al.}{2020b}]{Jang:2020aa}
{Jang} I.~S.,  et~al., 2020b, \mn@doi [\aap] {10.1051/0004-6361/202038651},
  \href {https://ui.adsabs.harvard.edu/abs/2020A&A...640L..19J} {640, L19}

\bibitem[\protect\citeauthoryear{{Jensen}, {Hayes}, {Sestito}, {McConnachie},
  {Waller}, {Smith}, {Navarro}  \& {Venn}}{{Jensen}
  et~al.}{2024}]{Jensen:2024aa}
{Jensen} J.,  {Hayes} C.~R.,  {Sestito} F.,  {McConnachie} A.~W.,  {Waller} F.,
   {Smith} S. E.~T.,  {Navarro} J.,   {Venn} K.~A.,  2024, \mn@doi [\mnras]
  {10.1093/mnras/stad3322}, \href
  {https://ui.adsabs.harvard.edu/abs/2024MNRAS.527.4209J} {527, 4209}

\bibitem[\protect\citeauthoryear{{Jiang}, {Helly}, {Cole}  \& {Frenk}}{{Jiang}
  et~al.}{2014}]{Jiang:2014aa}
{Jiang} L.,  {Helly} J.~C.,  {Cole} S.,   {Frenk} C.~S.,  2014, \mn@doi
  [\mnras] {10.1093/mnras/stu390}, \href
  {https://ui.adsabs.harvard.edu/abs/2014MNRAS.440.2115J} {440, 2115}

\bibitem[\protect\citeauthoryear{{Kado-Fong} et~al.,}{{Kado-Fong}
  et~al.}{2022}]{Kado-Fong:2022aa}
{Kado-Fong} E.,  et~al., 2022, \mn@doi [\apj] {10.3847/1538-4357/ac6c88}, \href
  {https://ui.adsabs.harvard.edu/abs/2022ApJ...931..152K} {931, 152}

\bibitem[\protect\citeauthoryear{{Kang} \& {Ricotti}}{{Kang} \&
  {Ricotti}}{2019}]{Kang:2019aa}
{Kang} H.~D.,  {Ricotti} M.,  2019, \mn@doi [\mnras] {10.1093/mnras/stz1886},
  \href {https://ui.adsabs.harvard.edu/abs/2019MNRAS.488.2673K} {488, 2673}

\bibitem[\protect\citeauthoryear{{Kim} \& {Peter}}{{Kim} \&
  {Peter}}{2021}]{Kim:2021aa}
{Kim} S.~Y.,  {Peter} A. H.~G.,  2021, \mn@doi [arXiv e-prints]
  {10.48550/arXiv.2106.09050}, \href
  {https://ui.adsabs.harvard.edu/abs/2021arXiv210609050K} {p. arXiv:2106.09050}

\bibitem[\protect\citeauthoryear{{Kim} et~al.,}{{Kim}
  et~al.}{2024}]{Kim:2024aa}
{Kim} S.~Y.,  et~al., 2024, \mn@doi [arXiv e-prints]
  {10.48550/arXiv.2408.15214}, \href
  {https://ui.adsabs.harvard.edu/abs/2024arXiv240815214K} {p. arXiv:2408.15214}

\bibitem[\protect\citeauthoryear{{Kluge}, {Bender}, {Riffeser}, {Goessl},
  {Hopp}, {Schmidt}  \& {Ries}}{{Kluge} et~al.}{2021}]{Kluge:2021aa}
{Kluge} M.,  {Bender} R.,  {Riffeser} A.,  {Goessl} C.,  {Hopp} U.,  {Schmidt}
  M.,   {Ries} C.,  2021, \mn@doi [\apjs] {10.3847/1538-4365/abcda6}, \href
  {https://ui.adsabs.harvard.edu/abs/2021ApJS..252...27K} {252, 27}

\bibitem[\protect\citeauthoryear{Kluyver et~al.,}{Kluyver
  et~al.}{2016}]{kluyver2016jupyter}
Kluyver T.,  et~al., 2016, in Loizides F.,  Schmidt B.,  eds, Positioning and
  Power in Academic Publishing: Players, Agents and Agendas. IOS Press, pp 87
  -- 90

\bibitem[\protect\citeauthoryear{{Komatsu} et~al.,}{{Komatsu}
  et~al.}{2011}]{Komatsu:2011aa}
{Komatsu} E.,  et~al., 2011, \mn@doi [\apjs] {10.1088/0067-0049/192/2/18},
  \href {http://adsabs.harvard.edu/abs/2011ApJS..192...18K} {192, 18}

\bibitem[\protect\citeauthoryear{{Lacey} et~al.,}{{Lacey}
  et~al.}{2016}]{Lacey:2016aa}
{Lacey} C.~G.,  et~al., 2016, \mn@doi [\mnras] {10.1093/mnras/stw1888}, \href
  {http://adsabs.harvard.edu/abs/2016MNRAS.462.3854L} {462, 3854}

\bibitem[\protect\citeauthoryear{{Lagos}, {Lacey}, {Baugh}, {Bower}  \&
  {Benson}}{{Lagos} et~al.}{2011}]{Lagos:2011aa}
{Lagos} C.~D.~P.,  {Lacey} C.~G.,  {Baugh} C.~M.,  {Bower} R.~G.,   {Benson}
  A.~J.,  2011, \mn@doi [\mnras] {10.1111/j.1365-2966.2011.19160.x}, \href
  {http://adsabs.harvard.edu/abs/2011MNRAS.416.1566L} {416, 1566}

\bibitem[\protect\citeauthoryear{{Lauer} et~al.,}{{Lauer}
  et~al.}{2007}]{Lauer07}
{Lauer} T.~R.,  et~al., 2007, \mn@doi [\apj] {10.1086/518223}, \href
  {http://adsabs.harvard.edu/abs/2007ApJ...662..808L} {662, 808}

\bibitem[\protect\citeauthoryear{{Le Bret}, {Pontzen}, {Cooper}, {Frenk},
  {Zolotov}, {Brooks}, {Governato}  \& {Parry}}{{Le Bret}
  et~al.}{2017}]{Le-Bret:2017aa}
{Le Bret} T.,  {Pontzen} A.,  {Cooper} A.~P.,  {Frenk} C.,  {Zolotov} A.,
  {Brooks} A.~M.,  {Governato} F.,   {Parry} O.~H.,  2017, \mn@doi [\mnras]
  {10.1093/mnras/stx552}, \href
  {https://ui.adsabs.harvard.edu/abs/2017MNRAS.468.3212L} {468, 3212}

\bibitem[\protect\citeauthoryear{{Li} \& {White}}{{Li} \&
  {White}}{2008}]{Li:2008ab}
{Li} Y.-S.,  {White} S.~D.~M.,  2008, \mn@doi [\mnras]
  {10.1111/j.1365-2966.2007.12748.x}, \href
  {http://adsabs.harvard.edu/abs/2008MNRAS.384.1459L} {384, 1459}

\bibitem[\protect\citeauthoryear{{Li}, {Helmi}, {De Lucia}  \& {Stoehr}}{{Li}
  et~al.}{2009}]{Li:2009aa}
{Li} Y.-S.,  {Helmi} A.,  {De Lucia} G.,   {Stoehr} F.,  2009, \mn@doi [\mnras]
  {10.1111/j.1745-3933.2009.00690.x}, \href
  {http://adsabs.harvard.edu/abs/2009MNRAS.397L..87L} {397, L87}

\bibitem[\protect\citeauthoryear{{Liao} \& {Cooper}}{{Liao} \&
  {Cooper}}{2023}]{Liao:2023aa}
{Liao} L.-W.,  {Cooper} A.~P.,  2023, \mn@doi [\mnras]
  {10.1093/mnras/stac3327}, \href
  {https://ui.adsabs.harvard.edu/abs/2023MNRAS.518.3999L} {518, 3999}

\bibitem[\protect\citeauthoryear{{Licquia}, {Newman}  \& {Bershady}}{{Licquia}
  et~al.}{2016}]{Licquia:2016aa}
{Licquia} T.~C.,  {Newman} J.~A.,   {Bershady} M.~A.,  2016, \mn@doi [\apj]
  {10.3847/1538-4357/833/2/220}, \href
  {https://ui.adsabs.harvard.edu/abs/2016ApJ...833..220L} {833, 220}

\bibitem[\protect\citeauthoryear{{Lovell}, {Cautun}, {Frenk}, {Hellwing}  \&
  {Newton}}{{Lovell} et~al.}{2021}]{Lovell:2021aa}
{Lovell} M.~R.,  {Cautun} M.,  {Frenk} C.~S.,  {Hellwing} W.~A.,   {Newton} O.,
   2021, \mn@doi [\mnras] {10.1093/mnras/stab2452}, \href
  {https://ui.adsabs.harvard.edu/abs/2021MNRAS.507.4826L} {507, 4826}

\bibitem[\protect\citeauthoryear{{Ludlow}, {Fall}, {Wilkinson}, {Schaye}  \&
  {Obreschkow}}{{Ludlow} et~al.}{2023}]{Ludlow:2023aa}
{Ludlow} A.~D.,  {Fall} S.~M.,  {Wilkinson} M.~J.,  {Schaye} J.,   {Obreschkow}
  D.,  2023, \mn@doi [\mnras] {10.1093/mnras/stad2615}, \href
  {https://ui.adsabs.harvard.edu/abs/2023MNRAS.525.5614L} {525, 5614}

\bibitem[\protect\citeauthoryear{{Mackey} \& {Gilmore}}{{Mackey} \&
  {Gilmore}}{2003}]{Mackey:2003aa}
{Mackey} A.~D.,  {Gilmore} G.~F.,  2003, \mn@doi [\mnras]
  {10.1046/j.1365-8711.2003.07001.x}, \href
  {https://ui.adsabs.harvard.edu/abs/2003MNRAS.345..747M} {345, 747}

\bibitem[\protect\citeauthoryear{{Mart{\'\i}nez-Delgado}
  et~al.,}{{Mart{\'\i}nez-Delgado} et~al.}{2023}]{Martinez-Delgado:2023aa}
{Mart{\'\i}nez-Delgado} D.,  et~al., 2023, \mn@doi [\aap]
  {10.1051/0004-6361/202245011}, \href
  {https://ui.adsabs.harvard.edu/abs/2023A&A...671A.141M} {671, A141}

\bibitem[\protect\citeauthoryear{{McConnachie}}{{McConnachie}}{2012}]{McConnachie:2012aa}
{McConnachie} A.~W.,  2012, \mn@doi [\aj] {10.1088/0004-6256/144/1/4}, \href
  {http://adsabs.harvard.edu/abs/2012AJ....144....4M} {144, 4}

\bibitem[\protect\citeauthoryear{{McMonigal} et~al.,}{{McMonigal}
  et~al.}{2016a}]{McMonigal:2016ab}
{McMonigal} B.,  et~al., 2016a, \mn@doi [\mnras] {10.1093/mnras/stv2690}, \href
  {http://adsabs.harvard.edu/abs/2016MNRAS.456..405M} {456, 405}

\bibitem[\protect\citeauthoryear{{McMonigal} et~al.,}{{McMonigal}
  et~al.}{2016b}]{McMonigal:2016aa}
{McMonigal} B.,  et~al., 2016b, \mn@doi [\mnras] {10.1093/mnras/stw1657}, \href
  {http://adsabs.harvard.edu/abs/2016MNRAS.461.4374M} {461, 4374}

\bibitem[\protect\citeauthoryear{{Merritt}, {van Dokkum}, {Abraham}  \&
  {Zhang}}{{Merritt} et~al.}{2016}]{Merritt:2016aa}
{Merritt} A.,  {van Dokkum} P.,  {Abraham} R.,   {Zhang} J.,  2016, \mn@doi
  [\apj] {10.3847/0004-637X/830/2/62}, \href
  {https://ui.adsabs.harvard.edu/abs/2016ApJ...830...62M} {830, 62}

\bibitem[\protect\citeauthoryear{{Merritt}, {Pillepich}, {van Dokkum},
  {Nelson}, {Hernquist}, {Marinacci}  \& {Vogelsberger}}{{Merritt}
  et~al.}{2020}]{Merritt:2020aa}
{Merritt} A.,  {Pillepich} A.,  {van Dokkum} P.,  {Nelson} D.,  {Hernquist} L.,
   {Marinacci} F.,   {Vogelsberger} M.,  2020, \mn@doi [\mnras]
  {10.1093/mnras/staa1164}, \href
  {https://ui.adsabs.harvard.edu/abs/2020MNRAS.495.4570M} {495, 4570}

\bibitem[\protect\citeauthoryear{{Miro-Carretero} et~al.,}{{Miro-Carretero}
  et~al.}{2024}]{Miro-Carretero:2024aa}
{Miro-Carretero} J.,  et~al., 2024, \mn@doi [arXiv e-prints]
  {10.48550/arXiv.2409.03585}, \href
  {https://ui.adsabs.harvard.edu/abs/2024arXiv240903585M} {p. arXiv:2409.03585}

\bibitem[\protect\citeauthoryear{{Mitchell}, {Lacey}, {Baugh}  \&
  {Cole}}{{Mitchell} et~al.}{2013}]{Mitchell:2013aa}
{Mitchell} P.~D.,  {Lacey} C.~G.,  {Baugh} C.~M.,   {Cole} S.,  2013, \mn@doi
  [\mnras] {10.1093/mnras/stt1280}, \href
  {https://ui.adsabs.harvard.edu/abs/2013MNRAS.435...87M} {435, 87}

\bibitem[\protect\citeauthoryear{{Mo}, {Mao}  \& {White}}{{Mo}
  et~al.}{1998}]{Mo:1998tf}
{Mo} H.~J.,  {Mao} S.,   {White} S. D.~M.,  1998, \mn@doi [\mnras]
  {10.1046/j.1365-8711.1998.01227.x}, \href
  {https://ui.adsabs.harvard.edu/abs/1998MNRAS.295..319M} {295, 319}

\bibitem[\protect\citeauthoryear{{Monachesi}, {Bell}, {Radburn-Smith},
  {Bailin}, {de Jong}, {Holwerda}, {Streich}  \& {Silverstein}}{{Monachesi}
  et~al.}{2016a}]{Monachesi:2016aa}
{Monachesi} A.,  {Bell} E.~F.,  {Radburn-Smith} D.~J.,  {Bailin} J.,  {de Jong}
  R.~S.,  {Holwerda} B.,  {Streich} D.,   {Silverstein} G.,  2016a, \mn@doi
  [\mnras] {10.1093/mnras/stv2987}, \href
  {http://adsabs.harvard.edu/abs/2016MNRAS.457.1419M} {457, 1419}

\bibitem[\protect\citeauthoryear{{Monachesi}, {G{\'o}mez}, {Grand},
  {Kauffmann}, {Marinacci}, {Pakmor}, {Springel}  \& {Frenk}}{{Monachesi}
  et~al.}{2016b}]{Monachesi:2016ab}
{Monachesi} A.,  {G{\'o}mez} F.~A.,  {Grand} R.~J.~J.,  {Kauffmann} G.,
  {Marinacci} F.,  {Pakmor} R.,  {Springel} V.,   {Frenk} C.~S.,  2016b,
  \mn@doi [\mnras] {10.1093/mnrasl/slw052}, \href
  {http://adsabs.harvard.edu/abs/2016MNRAS.459L..46M} {459, L46}

\bibitem[\protect\citeauthoryear{{Monachesi} et~al.,}{{Monachesi}
  et~al.}{2019}]{Monachesi:2019aa}
{Monachesi} A.,  et~al., 2019, \mn@doi [\mnras] {10.1093/mnras/stz538}, \href
  {https://ui.adsabs.harvard.edu/abs/2019MNRAS.485.2589M} {485, 2589}

\bibitem[\protect\citeauthoryear{{Mosleh}, {Williams}  \& {Franx}}{{Mosleh}
  et~al.}{2013}]{Mosleh:2013aa}
{Mosleh} M.,  {Williams} R.~J.,   {Franx} M.,  2013, \mn@doi [\apj]
  {10.1088/0004-637X/777/2/117}, \href
  {http://adsabs.harvard.edu/abs/2013ApJ...777..117M} {777, 117}

\bibitem[\protect\citeauthoryear{{Moster}, {Naab}  \& {White}}{{Moster}
  et~al.}{2018}]{Moster:2018aa}
{Moster} B.~P.,  {Naab} T.,   {White} S. D.~M.,  2018, \mn@doi [\mnras]
  {10.1093/mnras/sty655}, \href
  {https://ui.adsabs.harvard.edu/abs/2018MNRAS.477.1822M} {477, 1822}

\bibitem[\protect\citeauthoryear{{Navarro}, {Eke}  \& {Frenk}}{{Navarro}
  et~al.}{1996}]{Navarro:1996aa}
{Navarro} J.~F.,  {Eke} V.~R.,   {Frenk} C.~S.,  1996, \mn@doi [\mnras]
  {10.1093/mnras/283.3.72L}, \href
  {http://adsabs.harvard.edu/abs/1996MNRAS.283L..72N} {283, L72}

\bibitem[\protect\citeauthoryear{{Ogami} et~al.,}{{Ogami}
  et~al.}{2024}]{Ogami:2024aa}
{Ogami} I.,  et~al., 2024, \mn@doi [\apj] {10.3847/1538-4357/ad5445}, \href
  {https://ui.adsabs.harvard.edu/abs/2024ApJ...971..107O} {971, 107}

\bibitem[\protect\citeauthoryear{{Orkney} et~al.,}{{Orkney}
  et~al.}{2021}]{Orkney:2021aa}
{Orkney} M. D.~A.,  et~al., 2021, \mn@doi [\mnras] {10.1093/mnras/stab1066},
  \href {https://ui.adsabs.harvard.edu/abs/2021MNRAS.504.3509O} {504, 3509}

\bibitem[\protect\citeauthoryear{{Patel}, {Besla}, {Mandel}  \& {Sohn}}{{Patel}
  et~al.}{2018}]{Patel:2018aa}
{Patel} E.,  {Besla} G.,  {Mandel} K.,   {Sohn} S.~T.,  2018, \mn@doi [\apj]
  {10.3847/1538-4357/aab78f}, \href
  {http://adsabs.harvard.edu/abs/2018ApJ...857...78P} {857, 78}

\bibitem[\protect\citeauthoryear{{Peebles}}{{Peebles}}{2020}]{Peebles:2020aa}
{Peebles} P.~J.~E.,  2020, \mn@doi [\mnras] {10.1093/mnras/staa2649}, \href
  {https://ui.adsabs.harvard.edu/abs/2020MNRAS.498.4386P} {498, 4386}

\bibitem[\protect\citeauthoryear{{Perez} \& {Granger}}{{Perez} \&
  {Granger}}{2007}]{2007CSE.....9c..21P}
{Perez} F.,  {Granger} B.~E.,  2007, \mn@doi [Computing in Science and
  Engineering] {10.1109/MCSE.2007.53}, \href
  {https://ui.adsabs.harvard.edu/abs/2007CSE.....9c..21P} {9, 21}

\bibitem[\protect\citeauthoryear{{Pillepich} et~al.,}{{Pillepich}
  et~al.}{2014}]{Pillepich:2014aa}
{Pillepich} A.,  et~al., 2014, \mn@doi [\mnras] {10.1093/mnras/stu1408}, \href
  {https://ui.adsabs.harvard.edu/abs/2014MNRAS.444..237P} {444, 237}

\bibitem[\protect\citeauthoryear{{Pontzen} \& {Governato}}{{Pontzen} \&
  {Governato}}{2012}]{Pontzen:2012aa}
{Pontzen} A.,  {Governato} F.,  2012, \mn@doi [\mnras]
  {10.1111/j.1365-2966.2012.20571.x}, \href
  {http://adsabs.harvard.edu/abs/2012MNRAS.421.3464P} {421, 3464}

\bibitem[\protect\citeauthoryear{{Posti}, {Marasco}, {Fraternali}  \&
  {Famaey}}{{Posti} et~al.}{2019}]{Posti:2019aa}
{Posti} L.,  {Marasco} A.,  {Fraternali} F.,   {Famaey} B.,  2019, \mn@doi
  [\aap] {10.1051/0004-6361/201935982}, \href
  {https://ui.adsabs.harvard.edu/abs/2019A&A...629A..59P} {629, A59}

\bibitem[\protect\citeauthoryear{{Proctor}, {Ludlow}, {Lagos}  \&
  {Robotham}}{{Proctor} et~al.}{2024a}]{Proctor:2024ab}
{Proctor} K.~L.,  {Ludlow} A.~D.,  {Lagos} C. d.~P.,   {Robotham} A. S.~G.,
  2024a, \mn@doi [arXiv e-prints] {10.48550/arXiv.2407.11444}, \href
  {https://ui.adsabs.harvard.edu/abs/2024arXiv240711444P} {p. arXiv:2407.11444}

\bibitem[\protect\citeauthoryear{{Proctor}, {Lagos}, {Ludlow}  \&
  {Robotham}}{{Proctor} et~al.}{2024b}]{Proctor:2024aa}
{Proctor} K.~L.,  {Lagos} C. d.~P.,  {Ludlow} A.~D.,   {Robotham} A. S.~G.,
  2024b, \mn@doi [\mnras] {10.1093/mnras/stad3341}, \href
  {https://ui.adsabs.harvard.edu/abs/2024MNRAS.527.2624P} {527, 2624}

\bibitem[\protect\citeauthoryear{{Pu}, {Cooper}, {Grand}, {G{\'o}mez}  \&
  {Monachesi}}{{Pu} et~al.}{2025}]{Pu:2025aa}
{Pu} S.-Y.,  {Cooper} A.~P.,  {Grand} R. J.~J.,  {G{\'o}mez} F.~A.,
  {Monachesi} A.,  2025, \mn@doi [\apj] {10.3847/1538-4357/ada382}, \href
  {https://ui.adsabs.harvard.edu/abs/2025ApJ...980...63P} {980, 63}

\bibitem[\protect\citeauthoryear{{Pucha} et~al.,}{{Pucha}
  et~al.}{2019}]{Pucha:2019ab}
{Pucha} R.,  et~al., 2019, \mn@doi [\apj] {10.3847/1538-4357/ab29fb}, \href
  {https://ui.adsabs.harvard.edu/abs/2019ApJ...880..104P} {880, 104}

\bibitem[\protect\citeauthoryear{{Purcell}, {Bullock}  \& {Zentner}}{{Purcell}
  et~al.}{2007}]{Purcell:2007aa}
{Purcell} C.~W.,  {Bullock} J.~S.,   {Zentner} A.~R.,  2007, \mn@doi [\apj]
  {10.1086/519787}, \href {http://adsabs.harvard.edu/abs/2007ApJ...666...20P}
  {666, 20}

\bibitem[\protect\citeauthoryear{{Radburn-Smith} et~al.,}{{Radburn-Smith}
  et~al.}{2011}]{Radburn-Smith:2011aa}
{Radburn-Smith} D.~J.,  et~al., 2011, \mn@doi [\apjs]
  {10.1088/0067-0049/195/2/18}, \href
  {http://adsabs.harvard.edu/abs/2011ApJS..195...18R} {195, 18}

\bibitem[\protect\citeauthoryear{{Rey}, {Pontzen}  \& {Saintonge}}{{Rey}
  et~al.}{2019a}]{Rey:2019aa}
{Rey} M.~P.,  {Pontzen} A.,   {Saintonge} A.,  2019a, \mn@doi [\mnras]
  {10.1093/mnras/stz552}, \href
  {http://adsabs.harvard.edu/abs/2019MNRAS.485.1906R} {485, 1906}

\bibitem[\protect\citeauthoryear{{Rey}, {Pontzen}, {Agertz}, {Orkney}, {Read},
  {Saintonge}  \& {Pedersen}}{{Rey} et~al.}{2019b}]{Rey:2019ab}
{Rey} M.~P.,  {Pontzen} A.,  {Agertz} O.,  {Orkney} M. D.~A.,  {Read} J.~I.,
  {Saintonge} A.,   {Pedersen} C.,  2019b, \mn@doi [\apjl]
  {10.3847/2041-8213/ab53dd}, \href
  {https://ui.adsabs.harvard.edu/abs/2019ApJ...886L...3R} {886, L3}

\bibitem[\protect\citeauthoryear{{Richings} et~al.,}{{Richings}
  et~al.}{2020}]{Richings:2020aa}
{Richings} J.,  et~al., 2020, \mn@doi [\mnras] {10.1093/mnras/stz3448}, \href
  {https://ui.adsabs.harvard.edu/abs/2020MNRAS.492.5780R} {492, 5780}

\bibitem[\protect\citeauthoryear{{Ricotti}, {Polisensky}  \&
  {Cleland}}{{Ricotti} et~al.}{2022}]{Ricotti:2022aa}
{Ricotti} M.,  {Polisensky} E.,   {Cleland} E.,  2022, \mn@doi [\mnras]
  {10.1093/mnras/stac1485}, \href
  {https://ui.adsabs.harvard.edu/abs/2022MNRAS.515..302R} {515, 302}

\bibitem[\protect\citeauthoryear{{Ristea}, {Cortese}, {Groves},
  {Fraser-McKelvie}, {Obreschkow}  \& {Glazebrook}}{{Ristea}
  et~al.}{2024}]{Ristea:2024aa}
{Ristea} A.,  {Cortese} L.,  {Groves} B.,  {Fraser-McKelvie} A.,  {Obreschkow}
  D.,   {Glazebrook} K.,  2024, \mn@doi [\mnras] {10.1093/mnras/stae2085},
  \href {https://ui.adsabs.harvard.edu/abs/2024MNRAS.534..995R} {534, 995}

\bibitem[\protect\citeauthoryear{{Santos-Santos}, {Frenk}, {Navarro}, {Cole}
  \& {Helly}}{{Santos-Santos} et~al.}{2025}]{Santos-Santos:2025aa}
{Santos-Santos} I. M.~E.,  {Frenk} C.~S.,  {Navarro} J.~F.,  {Cole} S.,
  {Helly} J.,  2025, \mn@doi [\mnras] {10.1093/mnras/staf749}, \href
  {https://ui.adsabs.harvard.edu/abs/2025MNRAS.tmp..721S} {}

\bibitem[\protect\citeauthoryear{{Scannapieco} et~al.,}{{Scannapieco}
  et~al.}{2012}]{Scannapieco:2012aa}
{Scannapieco} C.,  et~al., 2012, \mn@doi [\mnras]
  {10.1111/j.1365-2966.2012.20993.x}, \href
  {http://adsabs.harvard.edu/abs/2012MNRAS.423.1726S} {423, 1726}

\bibitem[\protect\citeauthoryear{{Shen}, {Mo}, {White}, {Blanton}, {Kauffmann},
  {Voges}, {Brinkmann}  \& {Csabai}}{{Shen} et~al.}{2003}]{Shen:2003aa}
{Shen} S.,  {Mo} H.~J.,  {White} S.~D.~M.,  {Blanton} M.~R.,  {Kauffmann} G.,
  {Voges} W.,  {Brinkmann} J.,   {Csabai} I.,  2003, \mn@doi [\mnras]
  {10.1046/j.1365-8711.2003.06740.x}, \href
  {http://adsabs.harvard.edu/abs/2003MNRAS.343..978S} {343, 978}

\bibitem[\protect\citeauthoryear{{Shipp} et~al.,}{{Shipp}
  et~al.}{2024}]{Shipp:2024aa}
{Shipp} N.,  et~al., 2024, arXiv e-prints, \href
  {https://ui.adsabs.harvard.edu/abs/2024arXiv241009143S} {p. arXiv:2410.09143}

\bibitem[\protect\citeauthoryear{{Simha} \& {Cole}}{{Simha} \&
  {Cole}}{2017}]{Simha:2017aa}
{Simha} V.,  {Cole} S.,  2017, \mn@doi [\mnras] {10.1093/mnras/stx1942}, \href
  {https://ui.adsabs.harvard.edu/abs/2017MNRAS.472.1392S} {472, 1392}

\bibitem[\protect\citeauthoryear{{Smercina}, {Bell}, {Price}, {D'Souza},
  {Slater}, {Bailin}, {Monachesi}  \& {Nidever}}{{Smercina}
  et~al.}{2018}]{Smercina:2018aa}
{Smercina} A.,  {Bell} E.~F.,  {Price} P.~A.,  {D'Souza} R.,  {Slater} C.~T.,
  {Bailin} J.,  {Monachesi} A.,   {Nidever} D.,  2018, \mn@doi [\apj]
  {10.3847/1538-4357/aad2d6}, \href
  {http://adsabs.harvard.edu/abs/2018ApJ...863..152S} {863, 152}

\bibitem[\protect\citeauthoryear{{Smercina} et~al.,}{{Smercina}
  et~al.}{2020}]{Smercina:2020aa}
{Smercina} A.,  et~al., 2020, \mn@doi [\apj] {10.3847/1538-4357/abc485}, \href
  {https://ui.adsabs.harvard.edu/abs/2020ApJ...905...60S} {905, 60}

\bibitem[\protect\citeauthoryear{{Smercina} et~al.,}{{Smercina}
  et~al.}{2023}]{Smercina:2023aa}
{Smercina} A.,  et~al., 2023, \mn@doi [\apjl] {10.3847/2041-8213/acd5d1}, \href
  {https://ui.adsabs.harvard.edu/abs/2023ApJ...949L..37S} {949, L37}

\bibitem[\protect\citeauthoryear{{Springel}}{{Springel}}{2005}]{Springel05}
{Springel} V.,  2005, \mn@doi [\mnras] {10.1111/j.1365-2966.2005.09655.x},
  \href {http://adsabs.harvard.edu/abs/2005MNRAS.364.1105S} {364, 1105}

\bibitem[\protect\citeauthoryear{{Springel}, {White}, {Tormen}  \&
  {Kauffmann}}{{Springel} et~al.}{2001}]{Springel:2001aa}
{Springel} V.,  {White} S.~D.~M.,  {Tormen} G.,   {Kauffmann} G.,  2001,
  \mn@doi [\mnras] {10.1046/j.1365-8711.2001.04912.x}, \href
  {http://adsabs.harvard.edu/abs/2001MNRAS.328..726S} {328, 726}

\bibitem[\protect\citeauthoryear{{Springel} et~al.,}{{Springel}
  et~al.}{2005}]{Springel:2005aa}
{Springel} V.,  et~al., 2005, \mn@doi [\nat] {10.1038/nature03597}, \href
  {https://ui.adsabs.harvard.edu/abs/2005Natur.435..629S} {435, 629}

\bibitem[\protect\citeauthoryear{{Tanaka}, {Chiba}, {Hayashi}, {Komiyama},
  {Okamoto}, {Cooper}, {Okamoto}  \& {Spitler}}{{Tanaka}
  et~al.}{2018}]{Tanaka:2018aa}
{Tanaka} M.,  {Chiba} M.,  {Hayashi} K.,  {Komiyama} Y.,  {Okamoto} T.,
  {Cooper} A.~P.,  {Okamoto} S.,   {Spitler} L.,  2018, \mn@doi [\apj]
  {10.3847/1538-4357/aad9fe}, \href
  {http://adsabs.harvard.edu/abs/2018ApJ...865..125T} {865, 125}

\bibitem[\protect\citeauthoryear{{Tanakul}, {Yang}  \& {Sarajedini}}{{Tanakul}
  et~al.}{2017}]{Tanakul:2017aa}
{Tanakul} N.,  {Yang} S.-C.,   {Sarajedini} A.,  2017, \mn@doi [\mnras]
  {10.1093/mnras/stx515}, \href
  {https://ui.adsabs.harvard.edu/abs/2017MNRAS.468..870T} {468, 870}

\bibitem[\protect\citeauthoryear{{Tau} et~al.,}{{Tau}
  et~al.}{2024a}]{Tau:2024aa}
{Tau} E.~A.,  et~al., 2024a, \mn@doi [arXiv e-prints]
  {10.48550/arXiv.2412.13807}, \href
  {https://ui.adsabs.harvard.edu/abs/2024arXiv241213807T} {p. arXiv:2412.13807}

\bibitem[\protect\citeauthoryear{{Tau}, {Vivas}  \&
  {Mart{\'\i}nez-V{\'a}zquez}}{{Tau} et~al.}{2024b}]{Tau:2024ab}
{Tau} E.~A.,  {Vivas} A.~K.,   {Mart{\'\i}nez-V{\'a}zquez} C.~E.,  2024b,
  \mn@doi [\aj] {10.3847/1538-3881/ad1509}, \href
  {https://ui.adsabs.harvard.edu/abs/2024AJ....167...57T} {167, 57}

\bibitem[\protect\citeauthoryear{Team}{Team}{02  }]{pytables}
Team P.~D.,  2002--, {PyTables}: Hierarchical Datasets in {Python}, \url
  {http://www.pytables.org/}

\bibitem[\protect\citeauthoryear{{Trujillo} et~al.,}{{Trujillo}
  et~al.}{2021}]{Trujillo:2021aa}
{Trujillo} I.,  et~al., 2021, \mn@doi [\aap] {10.1051/0004-6361/202141603},
  \href {https://ui.adsabs.harvard.edu/abs/2021A&A...654A..40T} {654, A40}

\bibitem[\protect\citeauthoryear{Van~Rossum \& Drake}{Van~Rossum \&
  Drake}{2009}]{python}
Van~Rossum G.,  Drake F.~L.,  2009, Python 3 Reference Manual.
CreateSpace, Scotts Valley, CA

\bibitem[\protect\citeauthoryear{Virtanen et~al.,}{Virtanen
  et~al.}{2020}]{2020SciPy-NMeth}
Virtanen P.,  et~al., 2020, \mn@doi [Nature Methods]
  {10.1038/s41592-019-0686-2}, \href {https://rdcu.be/b08Wh} {17, 261}

\bibitem[\protect\citeauthoryear{{Wagg} \& {Broekgaarden}}{{Wagg} \&
  {Broekgaarden}}{2024}]{software-citation-station-paper}
{Wagg} T.,  {Broekgaarden} F.~S.,  2024, arXiv e-prints, \href
  {https://ui.adsabs.harvard.edu/abs/2024arXiv240604405W} {p. arXiv:2406.04405}

\bibitem[\protect\citeauthoryear{Wagg, Broekgaarden  \& G{\"u}ltekin}{Wagg
  et~al.}{2024}]{software-citation-station-zenodo}
Wagg T.,  Broekgaarden F.,   G{\"u}ltekin K.,  2024,
  TomWagg/software-citation-station: v1.2, \mn@doi{10.5281/zenodo.13225824},
  \url {https://doi.org/10.5281/zenodo.13225824}

\bibitem[\protect\citeauthoryear{{Wang} et~al.,}{{Wang}
  et~al.}{2019}]{Wang:2019aa}
{Wang} W.,  et~al., 2019, \mn@doi [\mnras] {10.1093/mnras/stz1339}, \href
  {https://ui.adsabs.harvard.edu/abs/2019MNRAS.487.1580W} {487, 1580}

\bibitem[\protect\citeauthoryear{{Wang}, {Han}, {Cautun}, {Li}  \&
  {Ishigaki}}{{Wang} et~al.}{2020}]{Wang:2020aa}
{Wang} W.,  {Han} J.,  {Cautun} M.,  {Li} Z.,   {Ishigaki} M.~N.,  2020,
  \mn@doi [Science China Physics, Mechanics, and Astronomy]
  {10.1007/s11433-019-1541-6}, \href
  {https://ui.adsabs.harvard.edu/abs/2020SCPMA..63j9801W} {63, 109801}

\bibitem[\protect\citeauthoryear{{Wang}, {Cooper}, {Bose}, {Frenk}  \&
  {Hellwing}}{{Wang} et~al.}{2023}]{Wang:2023aa}
{Wang} C.-W.,  {Cooper} A.~P.,  {Bose} S.,  {Frenk} C.~S.,   {Hellwing} W.~A.,
  2023, \mn@doi [\apj] {10.3847/1538-4357/ad011d}, \href
  {https://ui.adsabs.harvard.edu/abs/2023ApJ...958..166W} {958, 166}

\bibitem[\protect\citeauthoryear{{White} \& {Frenk}}{{White} \&
  {Frenk}}{1991}]{White:1991aa}
{White} S.~D.~M.,  {Frenk} C.~S.,  1991, \mn@doi [\apj] {10.1086/170483}, \href
  {http://adsabs.harvard.edu/abs/1991ApJ...379...52W} {379, 52}

\bibitem[\protect\citeauthoryear{{Williams} et~al.,}{{Williams}
  et~al.}{2021}]{Williams:2021aa}
{Williams} B.~F.,  et~al., 2021, \mn@doi [\apjs] {10.3847/1538-4365/abdf4e},
  \href {https://ui.adsabs.harvard.edu/abs/2021ApJS..253...53W} {253, 53}

\bibitem[\protect\citeauthoryear{{Zaritsky} \& {Behroozi}}{{Zaritsky} \&
  {Behroozi}}{2023}]{Zaritsky:2023aa}
{Zaritsky} D.,  {Behroozi} P.,  2023, \mn@doi [\mnras]
  {10.1093/mnras/stac3610}, \href
  {https://ui.adsabs.harvard.edu/abs/2023MNRAS.519..871Z} {519, 871}

\bibitem[\protect\citeauthoryear{{Zaritsky} et~al.,}{{Zaritsky}
  et~al.}{2024}]{Zaritsky:2024aa}
{Zaritsky} D.,  et~al., 2024, \mn@doi [\aj] {10.3847/1538-3881/ad543f}, \href
  {https://ui.adsabs.harvard.edu/abs/2024AJ....168...69Z} {168, 69}

\bibitem[\protect\citeauthoryear{{Zhang}, {Mackey}  \& {Da Costa}}{{Zhang}
  et~al.}{2021}]{Zhang:2021ab}
{Zhang} S.,  {Mackey} D.,   {Da Costa} G.~S.,  2021, \mn@doi [\mnras]
  {10.1093/mnras/stab2642}, \href
  {https://ui.adsabs.harvard.edu/abs/2021MNRAS.508.2098Z} {508, 2098}

\bibitem[\protect\citeauthoryear{{Zhu}, {Marinacci}, {Maji}, {Li}, {Springel}
  \& {Hernquist}}{{Zhu} et~al.}{2016}]{Zhu:2016aa}
{Zhu} Q.,  {Marinacci} F.,  {Maji} M.,  {Li} Y.,  {Springel} V.,   {Hernquist}
  L.,  2016, \mn@doi [\mnras] {10.1093/mnras/stw374}, \href
  {http://adsabs.harvard.edu/abs/2016MNRAS.458.1559Z} {458, 1559}

\bibitem[\protect\citeauthoryear{{Zolotov}, {Willman}, {Brooks}, {Governato},
  {Brook}, {Hogg}, {Quinn}  \& {Stinson}}{{Zolotov}
  et~al.}{2009}]{Zolotov:2009aa}
{Zolotov} A.,  {Willman} B.,  {Brooks} A.~M.,  {Governato} F.,  {Brook} C.~B.,
  {Hogg} D.~W.,  {Quinn} T.,   {Stinson} G.,  2009, \mn@doi [\apj]
  {10.1088/0004-637X/702/2/1058}, \href
  {http://adsabs.harvard.edu/abs/2009ApJ...702.1058Z} {702, 1058}

\bibitem[\protect\citeauthoryear{{van Dokkum}, {Abraham}  \& {Merritt}}{{van
  Dokkum} et~al.}{2014}]{van-Dokkum:2014aa}
{van Dokkum} P.~G.,  {Abraham} R.,   {Merritt} A.,  2014, \mn@doi [\apjl]
  {10.1088/2041-8205/782/2/L24}, \href
  {http://adsabs.harvard.edu/abs/2014ApJ...782L..24V} {782, L24}

\bibitem[\protect\citeauthoryear{{van den Bosch} \& {Ogiya}}{{van den Bosch} \&
  {Ogiya}}{2018}]{van-den-Bosch:2018ab}
{van den Bosch} F.~C.,  {Ogiya} G.,  2018, \mn@doi [\mnras]
  {10.1093/mnras/sty084}, \href
  {https://ui.adsabs.harvard.edu/abs/2018MNRAS.475.4066V} {475, 4066}

\bibitem[\protect\citeauthoryear{{van den Bosch}, {Ogiya}, {Hahn}  \&
  {Burkert}}{{van den Bosch} et~al.}{2018}]{van-den-Bosch:2018aa}
{van den Bosch} F.~C.,  {Ogiya} G.,  {Hahn} O.,   {Burkert} A.,  2018, \mn@doi
  [\mnras] {10.1093/mnras/stx2956}, \href
  {https://ui.adsabs.harvard.edu/abs/2018MNRAS.474.3043V} {474, 3043}

\bibitem[\protect\citeauthoryear{{van der Marel}, {Fardal}, {Sohn}, {Patel},
  {Besla}, {del Pino}, {Sahlmann}  \& {Watkins}}{{van der Marel}
  et~al.}{2019}]{van-der-marel:2019aa}
{van der Marel} R.~P.,  {Fardal} M.~A.,  {Sohn} S.~T.,  {Patel} E.,  {Besla}
  G.,  {del Pino} A.,  {Sahlmann} J.,   {Watkins} L.~L.,  2019, \mn@doi [\apj]
  {10.3847/1538-4357/ab001b}, \href
  {https://ui.adsabs.harvard.edu/abs/2019ApJ...872...24V} {872, 24}

\makeatother
\end{thebibliography}
\bsp

\appendix

\section{Particle Data Release}
\label{appendix:datarelease}

\begin{table*}
\caption{Stellar population particle ("tag") data model for \coco{}-\stings{}. For details, please see documentation provided alongside the data files.}
\label{tab:datamodel}
\begin{tabular}{lll}
\hline
Data set & Unit & Description\\
\hline
\texttt{Coordinates}    & Mpc        & Particle coordinates relative to the centre of the host halo potential. \\
\texttt{DMIDs}          & Integer    & \coco{} $N$-body  particle ID associated with tag. \\
\texttt{HSML}           & Mpc	     & Particle smoothing length. \\ 
\texttt{LookbackFormed} & Gyr        & Lookback time at which stellar population tag formed (i.e. stellar population age). \\
\texttt{Mass}          & $\mathrm{M_{\sun}}$ & 	Present-day mass of stellar population tag. \\ 
\texttt{Mmetal}        & $\mathrm{M_{\sun}}$ & 	Present-day mass of metals in stellar population tag. \\ 
\texttt{ParticleIDs}   & Integer    &	Unique integer ID of stellar population tag. \\
\texttt{SubhaloNr}	   & Integer    & ID of subhalo to which tagged DM particle is bound (-1: unbound particles). \\
\texttt{TPSnapshot}	   & Integer    & Snapshot of simulation at which tag was assigned. \\
\texttt{Velocities}	   & km/s	    & Particle velocities relative to the centre of the host halo potential. \\ 

\hline
\end{tabular}
\end{table*}

All the \stings{} particle tagging models described in this work will be made available incrementally. With this paper, we release the stellar particle data for the approximately 6400 central galaxies with virial mass $M_\mathrm{vir} > 10^{10}\,\Msol$ in our fiducial model \citepalias[based on the \galform{} model of][and using $\fmb = 3\%$]{Lacey:2016aa}. Instructions for downloading the data, further documentation of its contents and updates can be found at \url{http://github.com/nthu-ga/stings-data}. Additional particle properties and alternative semi-analytic models will be added in future.

We provide a FITS table of properties for each central galaxy in the release. For each galaxy in that table, we provide particle data for a cubic region cut out from the original simulation, extending to $\pm1.2 R_\mathrm{vir}$. As described in the main text, $R_\mathrm{vir}$ is calculated for Dhaloes, not $N$-body friends-of-friends groups, and therefore is not identical to $R_{200}$ reported by FoF/SUBFIND. We calculate $R_\mathrm{200}$ assuming $M_\mathrm{vir} \approx M_{200}$, although this is strictly only necessary for central haloes `promoted' from subhaloes to `primary' haloes by the DHalo algorithm. 

Table~\ref{tab:datamodel} lists the properties available for the particles in a cutout. The cutouts are provided in HDF5 format, and follow the specification of the Gadget 2 snapshot format \citep{Springel:2005aa}. These files should be readable by any Gadget-compatible analysis package. They contain only "type 4" (star) particles, which correspond to individual stellar mass tags. Coordinates (positions and velocities) are specified relative to the central galaxy (the centre of the potential, identified by SUBFIND). Particles associated with satellite galaxies are also included and can be identified by distinct values in the \texttt{SubhaloNr} dataset. Further details and examples are provided in the accompanying documentation. 

\section{Details of the model}
\label{appendix:technical}

\subsection{Failed Milky Ways: galaxies with exceptionally low \insitu{} star formation efficiency}
\label{appendix:mstar_mhalo}

In Section~\ref{sec:mstar_mhalo} we discuss the SMHM relation predicted by our fiducial  \galform{} model. This appendix examines differences between our SMHM relation and the predictions of other simulations, and the implications for the results we present in the main text. We focus on the seemingly greater number of "failed Milky Way" galaxies in the \citetalias{Lacey:2016aa} \galform{} model. 

\apc{The SMHM relation can be used directly to calibrate numerical and physical parameters of subgrid physical models in cosmological simulations, or as a means of comparison between models calibrated in other ways (for example, using stellar mass or luminosity functions).}
The SMHM relation is approximately a \apc{steep power law at low mass and a shallower power law at high mass}, changing slope around $M_\mathrm{vir}\sim10^{12}\,\Msol{}$. \apc{This corresponds approximately to $M^{\star}$, the characteristic scale above which the galaxy stellar mass function declines exponentially}. In most simulations, the shape and amplitude of the SMHM relation around the SMHM break scale are determined by evolution in the relative efficiency of supernova feedback and AGN feedback \citep{Bower:2017vf}. %
The calibration of \apc{simulations to observations} in this regime is particularly sensitive to `design choices' \citep{Mitchell:2013aa}. For example, model-builders may choose to calibrate either the `raw' stellar mass predicted by their simulation, a `stellar mass observable' derived from simulated photometry, or the simulated luminosity itself. If using mock observables, they must decide (for example)
how to account for the effects dust extinction in the simulated and real photometry.
Other data sets (such as the size--mass or mass--metallicity relations) may be included in a joint calibration, in which case their relative weight must be decided. Choices also have to be made regarding the statistical techniques used to determined  agreement between the model and the calibration data. 

Different groups address these issues in different ways. Consequently, most modern cosmological simulations can be said to be `calibrated', often to essentially the same observational data, while still predicting significantly different underlying SMHM relations.
Further discussion of these issues, in the context of the model we use, can be found in \citet[][see e.g. their section 6.2]{Lacey:2016aa}. The \citetalias{Lacey:2016aa} model itself is calibrated to the $z=0$ luminosity function in several bands, with weight given to other data sets, including infrared photometry of dust-rich galaxies at high redshift.

In the context of this paper, the issue of calibration is important because many predictions for the properties accreted stellar haloes have a strong and straightforward dependence on the form of the SMHM relation, as discussed in Sections~\ref{sec:intro} and \ref{sec:accreted_fraction_trends} (see also \citealt{Purcell:2007aa}; \citetalias{Cooper:2013aa}). 
Model-to-model differences in many of the quantities we discuss, including accreted mass fractions, density profiles and stellar populations, can be understood first and foremost as the result of different SMHM relations.

In this respect (as we say in Section~\ref{sec:mstar_mhalo}), it is significant that the \citetalias{Lacey:2016aa} model predicts a pronounced under-abundance of galaxies with stellar mass $10\lesssim \log_{10} M_{\star}/\Msol \lesssim11$ (see e.g. their fig.~31) relative to the stellar mass function obtained from SDSS by \citet{Li:2009aa}. The significance of this result, and its origin in terms of the ingredients of the \citetalias{Lacey:2016aa} model, are beyond the scope of our work here. It is possible (although by no means certain) that it can be attributed to the way in which the \citetalias{Lacey:2016aa} model treats AGN feedback. Since the \citetalias{Lacey:2016aa} model is in reasonable agreement with other observables, and this effect applies only to a subset of haloes, the existence of such a population does not seem to be ruled out by observations. On the other hand, the effect is not seen in other cosmological models of galaxy formation, such as TNG and Eagle.

\begin{figure}
  \includegraphics[width=84mm, trim=0.0cm 0cm 0.0cm 0cm,clip=True]{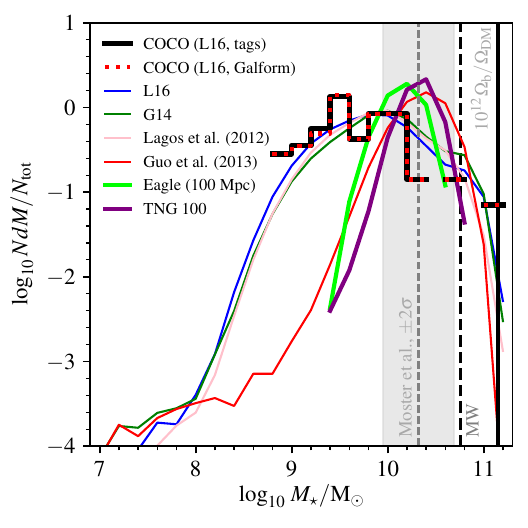}

  \caption{Distribution of central galaxy stellar mass in 'Milky Way-like' haloes of mass $11.8<\log_{10} M_{\mathrm{vir}}/\Msol{} < 12.2$ for a range of models, given in the legend. The theoretical maximum stellar mass for $\log_{10} M_{\mathrm{vir}}/\Msol{} = 12$ in our fiducial cosmology is shown by a solid vertical line. We compare these to our simulations (the distribution of which does not differ significantly between the tagged-particle realization and the underlying  \galform{} model). The thick vertical dashed line indicates the approximate stellar mass of the Milky Way \citep{Licquia:2016aa}. The grey shaded region shows the $2\sigma$ range around the mean (think vertical dashed line) of an empirical fit to observed stellar masses by \citet{Moster:2018aa}. As described in the text, thin coloured lines correspond to other semi-analytic models and thick coloured lines to hydrodynamical simulations.}.
\label{fig:appendix:mhalo_mw_compare}
\end{figure}

Fig.~\ref{fig:appendix:mhalo_mw_compare} shows the distribution of central galaxy stellar mass over a narrow range of Milky Way-like virial mass, $11.8<\log_{10} M_{\mathrm{vir}}/\Msol{} < 12.2$. Our fiducial \citetalias{Lacey:2016aa}-\coco{} model is shown by the black histogram. This distribution from the particle tagging results is almost identical to that from the underlying semi-analytic realization with the \coco{} merger trees\footnote{Some discrepancy is possible due to the different treatment of ongoing mass loss from satellites, but this seems to be small.} (red dotted line). The figure shows that our stellar mass distribution is consistent with the results of the \citetalias{Lacey:2016aa} model applied to a much larger cosmological volume (blue line). It also shows that the prediction from the \citetalias{Lacey:2016aa} model is very similar to that of other \galform{} models \citep{Lagos:2011aa, Gonzalez-Perez:2014aa}. We note, however, that \citetalias{Lacey:2016aa}-\coco{} has a small but noticeable deficit relative to all the \galform{} models, including the large-volume \citetalias{Lacey:2016aa} model, at masses above the peak of the empirical relation  \citep[inferred from a simultaneous fit to a variety of data sets by][]{Moster:2018aa}. There are fractionally fewer high stellar mass galaxies in \coco{} Milky Way analogue haloes than expected from the \citetalias{Lacey:2016aa} model. We attribute this to sample variance in the \coco{} volume (which happens to exaggerate the effect we now discuss).

Much more significantly, all \galform{} models with similar implementations of AGN feeback predict an average stellar mass in Milky Way-mass haloes that is $\gtrsim2\sigma$ below the mean empirical value, as determined by \citet[][their section 4.2]{Moster:2018aa}. A large fraction of such haloes in \citetalias{Lacey:2016aa} host galaxies of mass $9 \lesssim \log_{10} M_{\star}/\Msol \lesssim 10$. These are the `failed Milky Ways' we highlight in the main text - a population of MW analogues by virial mass having exceptionally low \insitu{} star formation efficiency. In contrast, TNG-100 and Eagle show a narrower distribution, peaking approximately around the empirical mean stellar mass. The semi-analytic model of \citet[][used in \citetalias{Cooper:2013aa}]{Guo:2011aa} resembles these hydrodynamic models, and, like them, shows a tail of relatively inefficient galaxy formation (at the $<10$~per~cent level). All the models predict an inefficient population, but make different predictions for the fraction of galaxies in that population.

The three  \galform{} models shown in Fig.~\ref{fig:appendix:mhalo_mw_compare} are therefore in  
greater tension with results from abundance matching and the empirical methods of \citet{Moster:2018aa} and \citep{Behroozi:2019aa}. At least part of this discrepancy is likely to be due to a `design choice': the \galform{} models were calibrated to luminosity functions rather than the stellar mass function or empirical SMHM relation directly \citep[see][]{Mitchell:2013aa}.
Further observational constraints on the SMHM relation in this regime are clearly important. Semi-analytic models like \galform{} are constructed in terms of explicit physical models, rather than empirical scaling relations. Observations that can distinguish between these models, for example by measuring the abundance of highly inefficient systems around the knee of the mass function, may therefore constrain the parameter space of the physical processes responsible for the differences in Fig.~\ref{fig:appendix:mhalo_mw_compare} - for example, with regard to the physics of gas accretion, star formation or AGN feedback. Our \citetalias{Lacey:2016aa}-\coco{} simulations demonstrate that the nature of the failed Milky Way's, at a given stellar or virial mass, would likely be apparent in their ratio of \insitu{} to accreted stellar material, and perhaps in other observables sensitive to that ratio, such as morphology.\footnote{It may also be apparent in other observables thought to correlate with virial mass, such as the satellite luminosity function, stellar velocity dispersion, or globular cluster count.} 

Although we have examined the assembly histories of the failed Milky Ways, we find no straightforward explanation for why they appear more common in \citetalias{Lacey:2016aa}-\coco{} than in other models. Their dark matter haloes grow fractionally later on average, with a median half-mass assembly redshift $z_{50}\simeq 2.6$, compared to $z_{50}\simeq 3.1$ for haloes in the same virial mass range and within $2\sigma$ of the abundance-matching SMHM expectation. Their stellar mass growth histories are otherwise qualitatively similar to those of their `regular' counterparts, in that the failed Milky Ways form roughly the same fraction of their final stellar mass by a given time. However, their final stellar mass is lower in absolute terms. The superficial reason for this seems to be a reduction in their radiative cooling rate at redshifts $z\lesssim5$, such that they build up substantially less cold ISM around redshifts $\sim2-3$. The underlying cause is not easy to identify from the \galform{} model outputs we have in hand. It may involve a combination of the transition from cold-flow accretion to a hydrostatic gas halo; the dependence of supernova feedback on galaxy size; gas accretion in mergers; disc instabilities; and the early onset of thermostatic AGN feedback. All of these are potentially linked to the difference in the halo overdensity suggested by the later collapse time, an effect which may be slightly enhanced in the \coco{} volume. We  leave further investigation of the origin of this population to future work.

\subsection{Definitions of galaxies}
\label{sec:appendix:definition_dhalo}

Operationally, we define a `galaxy' as the set of stellar mass tags associated with the particles bound to a particular (sub)halo, according the {\sc Subfind} algorithm. The \galform{} model uses a more complex definition, in two respects:

\begin{itemize}
    \item The \dhalo{} algorithm \citep{Jiang:2014aa} is used to distinguish central and satellite haloes. This algorithm effectively modifies the results of {\sc Subfind}, with some "satellite" {\sc Subfind} subhaloes being promoted to central galaxies in \dhalo{}s, and vice versa. In some cases the \dhalo{} algorithm also splits FoF groups into multiple central haloes, which may change the central galaxy and host halo to which a given satellite galaxy is associated.
    
    \item The algorithm used in  \galform{} to evolve the properties of galaxies through junctions between merger tree branches (i.e. to compute the evolution and outcome of mergers between haloes) has an edge-case behaviour, which \refrep{(particularly in high-resolution `zoom` simulations)} leads to a small fraction of galaxies `jumping' artificially from the merger tree branch in which they form to the branch of another galaxy in the same \dhalo{}. \refrep{This behaviour is related to more general issues associated with uncertainty in substructure-finding algorithms, which propagate to the construction of the merger trees themselves, as  described in detail by \citet{Chandro-Gomez:2025aa}}. The \refrep{impact of these effects are difficult to quantify and depend} on the specific \galform{} \refrep{model} and on the resolution of the underlying $N$-body merger trees. As a consequence, at $z=0$, a minority of galaxies identified by \galform{} as satellites will, in fact, occupy the most massive haloes of their FoF groups. Moreover, a small number of massive haloes identified by {\sc Subfind} will not be associated with any \galform{} galaxy. Cases in which a galaxy assigned to a satellite subhalo in an FoF group but nevertheless identified by \galform{} as central, are in most cases the intended result of the \dhalo{} algorithm.
\end{itemize}

Our definition of central galaxies follows that of the \dhalo{} algorithm, and hence corresponds to the definition used by \citetalias{Lacey:2016aa}. The uncertainty in the assignment of a handful of galaxies as either central or satellite, perhaps over a few crucial time-steps, may contribute in part to the small fraction of galaxies with extremely low efficiency discussed in the previous section, although this is by no means a straightforward or complete explanation for that population. 

Note that, as in C10, we disable the option in \galform{} to generate `orphan' galaxies (in other work also called `type 2' galaxies). These are a longstanding feature of semi-analytic models. The rate of convergence of such models with increasing $N$-body resolution can be improved by assigning an individual $N$-body particle to represent the orbit of a subhalo after tidal stripping reduces its mass below the detection limit of {\sc Subfind} (galaxies tracked in this way are said to be the `orphan' of their `lost' parent subhalo). This approach is usually combined with a semi-analytic calculation of the physical tidal disruption or merger time of the galaxy \citep[e.g.][]{Guo:2011ab,Guo14,Lacey:2016aa,Simha:2017aa}. The need for orphan galaxies to ensure convergence is reduced but not eliminated at higher $N$-body resolution. \citet{Santos-Santos:2025aa} provide a thorough description of the significance of orphan galaxies in Milky Way-scale zoom simulations.

In a particle tagging realization, the natural resolution limit is set by the particle distribution in exactly the same way as a hydrodynamical simulation. For consistency, and because our underlying simulation can already resolve the stripping of all potentially star forming haloes down to a few per cent of their peak mass, we require the semi-analytic model to `merge' galaxies as soon as their subhaloes are lost, rather than to track their subsequent evolution as orphans. Note that the particle tagging approach provides its own (in principle much more accurate) assessment of the distribution of stellar mass during and after tidal stripping and merging. It does not rely on the artificial semi-analytic distinction between these processes.

\section{Size distribution of M33 analogues}
\label{appendix:m33_sizes}

\begin{figure}
  \includegraphics[width=84mm, trim=0.0cm 0cm 0.0cm 0cm,clip=True]{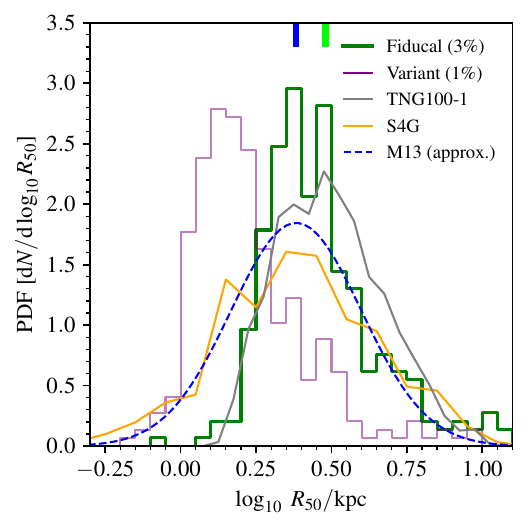}

  \caption{Distributions of projected half-light radii from our fiducial model (green) compared to data from the S4G survey and \citet{Mosleh:2013aa} and the Illustris TNG100-1 simulation.The green and blue markers indicate the mean of our fiducial \coco{} and \citep{Mosleh:2013aa} distributions, respectively. All these distributions are similar, and clearly inconsistent with the distribution for a model variant with $\fmb=1$~per~cent (purple line). }.

\label{fig:appendix:m33_sizemass}
\end{figure}

As a further test of our particle tagging model, Fig.~\ref{fig:appendix:m33_sizemass} shows the distribution of projected half-light radii for galaxies with similar stellar mass to M33. Our fiducial model produces a size distribution with a similar average to that of observed galaxies in this mass range from the S4G survey and the results of \citet{Mosleh:2013aa}. The shape of these distributions are also very similar at larger size, but the distribution from our simulation is less symmetric, falling off more rapidly than the observations towards smaller sizes. The same behaviour is seen for the distribution from TNG100-1. Conversely, a choice of $\fmb=1$~per~cent produces much more compact galaxies with a significantly different distribution of sizes. 

\section{Density profiles of Milky Way analogues}
\label{appendix:mhalo_mw_compare}

\begin{figure*}
  \includegraphics[width=160mm, trim=0.0cm 0cm 0.0cm 0cm,clip=True]{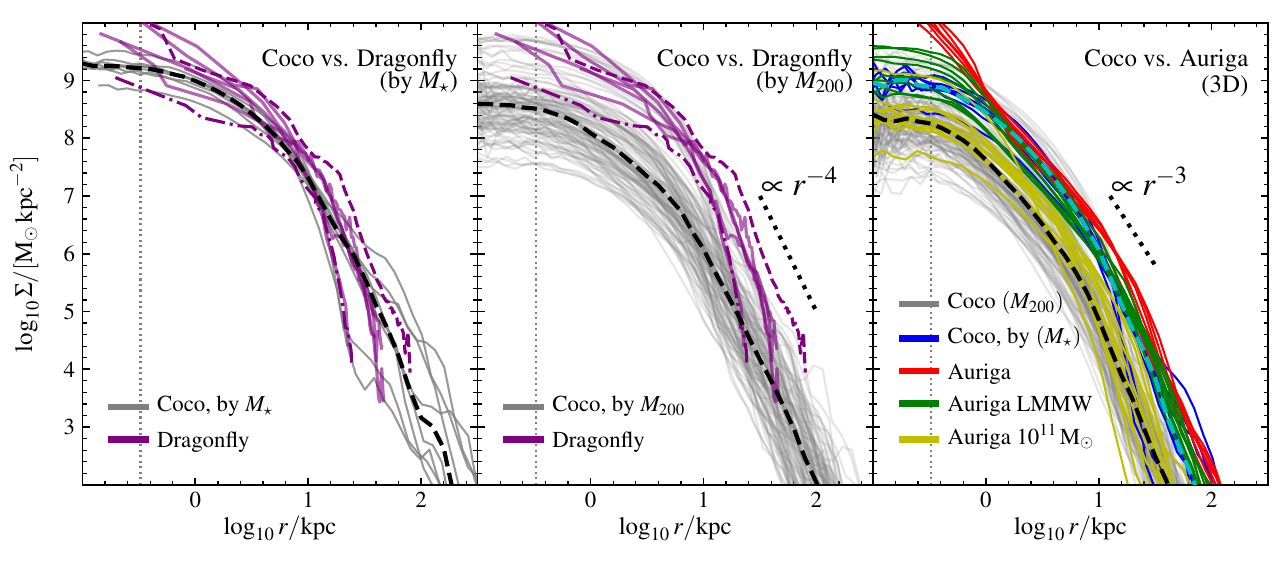}
  \caption{Projected stellar mass surface density profiles 
  around \coco{} Milky Way analogue galaxies generated with our fiducial model (grey lines for individual galaxies, black dashed lines for their medians) compared to data from observations and the Auriga simulations \citep{Grand:2024aa}. Indicative power law slopes are shown with black dotted lines. Left panel: \coco{} galaxies selected in the stellar mass range $10.66 < \log_{10} M_{\star}/\Msol < 10.85$ compared to observed profiles reported by \citealt[][]{Merritt:2020aa}; purple lines. The galaxies NGC 4258 and NGC 1024 are shown with dashed and dot--dashed lines respectively. Middle panel: \coco{} galaxies selected in the virial mass range $11.8 < \log_{10} M_{200}/\Msol < 12.5$ compared to the same Dragonfly Survey data as the left panel. Right panel: Three-dimensional profiles for \coco{} selected by virial mass as in the middle panel, compared to the original level 3 Auriga simulations (red) and two additional suites of Auriga simulations provided as part of the data release described by \citet{Grand:2024aa}: low-mass Milky Ways (LMMW; green) and galaxies with virial mass $M_{200}\sim10^{11}\,\mathrm{\Msol}$ (yellow). The set of \coco{} MW analogues by stellar mass, shown in the left panel, are shown in this panel with blue lines.}
\label{fig:mw_profiles_compare_dragonfly_sims}
\end{figure*}

The resolution of the \coco{} simulations is comparable to that of other high-resolution cosmological box models, such as TNG50, and the high or intermediate resolution levels of some suites of `zoom' simulations such as Auriga \citep{Grand:2017aa} and Artemis \citep{Font:2020aa}. As a straightforward means of comparing our particle-tagging results with those from state-of-the-art hydrodynamical simulations, we examine the stellar mass density profiles of Milky Way analogues in the Auriga simulations (\citealt{Grand:2024aa}; see also \citealt{Tau:2024aa}).

We first compare the predictions of \coco{} to observations in the Milky Way regime. The left-hand and central panels of Fig.~\ref{fig:mw_profiles_compare_dragonfly_sims} compare the stellar mass surface density profiles of Milky Way analogue galaxies in \coco{} to those measured by the Dragonfly Nearby Galaxy Survey \citep[DNGS;][]{Merritt:2020aa}. Similarity to the Milky Way is defined either by a range in stellar mass (left panel) or virial mass (central panel). 

Only nine \coco{} galaxies are selected as stellar mass analogues. Their profiles are similar to those of DNGS galaxies (left panel). In contrast, the profiles of most \coco{} MW analogues defined by virial mass have lower amplitude at all radii (central panel). This is simply because the average stellar mass of the sample selected by virial mass in \coco{} is lower than that of the galaxies in the DNGS sample. The \coco{} galaxies in the virial mass-selected sample that most closely resemble those in DNGS are those with the highest \insitu{} star formation efficiency. The stellar mass surface density  profiles of such galaxies are comparable to those found in DNGS at radii $\lesssim 10$ kpc, because \insitu{} stars dominate that region. At larger radii (where the \coco{} profiles are dominated by accreted stars), most \coco{} profiles have density $\sim1$~dex below the DNGS average, although the envelope of the stellar mass selected subset corresponds well to the DNGS results.

Similarly, in the DNGS profiles, the transition to a steeper outer slope seems to occur at larger radii than the typical \insitu{}/accreted transition in \coco{} virial mass selected analogues. The most notable example is NGC 4258 (dashed purple line), which has a shallow ($\Sigma \propto r^{-2}$) slope out to $\sim25$~kpc, followed by a sharp break to a steeper slope ($\Sigma \propto r^{-4}$) around $\sim15$ kpc. Breaks in surface density profiles arise from the transition between an accretion dominated profile and at the apocentres of the debris from individual accreted progenitors (e.g.\ \citetalias{Cooper:2010aa}; \citealt{Deason:2013aa}). For galaxies with high total stellar mass, the \insitu{}-to-accreted transition is smoothest (i.e. the break is weakest, occurring at high surface density and having a less pronounced change in slope) in galaxies that have a higher accreted stellar mass fraction, corresponding to a more massive dark matter halo. In such cases, breaks are more likely to arise from the apocentres of one or more halo debris components. Of course, profiles like that of NGC 4258 could also arise from the combination of an atypically extended \insitu{} component and a diffuse stellar halo.

We note that \citet{Merritt:2020aa} compared the DNGS sample to the Illustris TNG100 simulation. They demonstrated that TNG100 galaxies matched to the stellar masses of each DGNS galaxy have higher surface mass density than observed at radii larger than $\sim20$~kpc. Consequently, DNGS surface density profiles are best matched by TNG100 galaxies that have atypically low accreted mass fractions for their total stellar mass. 

At fixed stellar mass, low accreted mass fractions can arise from assembly histories with relatively fewer (and less massive) satellite accretion events, which become more common towards the lower end of the corresponding distribution of virial mass. As discussed in the main text, they may also arise from atypically efficient \insitu{} star formation; depending on the galaxy formation model, these effects may be correlated. 
Consequently, it is perhaps not surprising that the best matching TNG galaxies are those that form \insitu{} stars most efficiently, and hence are in the low-mass tail of the conditional virial mass distribution for their stellar mass. 

The left-hand panel of Fig.~\ref{fig:mw_profiles_compare_dragonfly_sims} shows that we do not find a systematic excess density at large radi in our model, relative to DNGS galaxies, when selecting galaxies by stellar mass. \citeauthor{Merritt:2020aa} termed this the `missing outskirts problem'; we do not find a problem of this kind in our simulations. However, we stress that our comparison is more simplistic than that of \citeauthor{Merritt:2020aa}, where each DNGS galaxy was matched individually to a bespoke set of similar TNG100 galaxies, selected to have no massive neighbours and to be disc-dominated (i.e. to have a large fraction of stars with high orbital angular momentum). Specifically, the stellar mass distribution of each comparison set was matched (by rejection sampling) to the Gaussian PDF of its corresponding DNGS stellar mass, in order to avoid bias towards less massive simulated galaxies. In contrast, we simply compare with the entire sample of MW analogues defined by stellar mass, selected in a $1\sigma$ range around the average DNGS stellar mass (which is approximately equivalent to the stellar mass of the MW, if NGC 1042 is excluded). Although this narrow mass range does not sample the full range of simulated galaxies that could reasonably be compared to each DNGS system, it does result in a roughly symmetric distribution of stellar mass. 

Several galaxies in our comparison sample have higher amplitude than any DNGS galaxy beyond $\sim30$~kpc. These correspond to the most massive dark matter haloes in the sample, which may be much more massive than the MW halo and the haloes of most of the galaxies in DNGS (see Figs.~\ref{fig:mw_images} and \ref{fig:mstar_mhalo}). Massive haloes would likely be removed from the sample in a more thorough comparison that included requirements on  isolation and angular momentum, as in \citet{Merritt:2020aa}.

We now turn to comparisons with hydrodynamical simulations. The right-hand panel of Fig.~\ref{fig:mw_profiles_compare_dragonfly_sims} compares our \coco{} results to the Auriga simulations, for a sample selected by virial mass as in the middle panel. However, following common practice for the presentation of simulated density profiles, we present these comparisons in terms of the three-dimensional rather than projected density.

Compared to our models, the original set of Auriga simulations (at the higher `level 3' resolution; red lines) have much less galaxy-to-galaxy scatter between their density profiles. They follow the upper envelope of the distribution predicted in \coco{}. In that respect are consistent with the subset of \coco{} systems in this virial mass range that are also Milky Way analogues by stellar mass (blue lines, with the median shown in cyan). The  Auriga galaxies mostly fall above the mean $M_\star-M_{200}$ relation \citep[see e.g.\ fig.~6 of][]{Grand:2024aa}. Many of the simulations therefore fall in the same high-efficiency region of that diagram as the Milky Way. Overall, the principle difference of Auriga with respect to \coco{} is the lack of scatter towards lower efficiency, rather than the distribution of the accreted material. However, the greater efficiency of star formation in halo progenitors does have an effect on the amplitude of the density of accreted stars $\gtrsim20\,\mathrm{kpc}$. We discuss this in more detail below, in the context of systems with lower virial mass.

The first Auriga data release \citep{Grand:2024aa} provides additional sets of simulations which apply the same Auriga galaxy formation model to systems with lower virial mass than the original sample. The ranges of virial masss for these additional sets are $5\times10^{11}<M_{200}/\Msol<10^{12}$ (\texttt{LowMassMWs}), $5\times10^{10}<M_{200}/\Msol<5\times10^{11}$ (\texttt{haloes\_1e11msol}), and $5\times10^{9}<M_{200}/\Msol<5\times10^{10}$ (\texttt{haloes\_1e10msol}). In the right-hand panel of Fig.~\ref{fig:mw_profiles_compare_dragonfly_sims} we show the stellar mass surface density profiles of galaxies in the \texttt{LowMassMWs} and \texttt{haloes\_1e11msol} sets. The \texttt{LowMassMWs} profiles match well to the stellar mass selected Milky Way sample in \coco{}, although they correspond to haloes with $\sim0.5$~dex lower virial mass. The \texttt{haloes\_1e11msol} profiles correspond approximately to the median of the virial mass selected Milky Way sample in \coco{}; in this case the difference in virial mass is $\sim1$~dex. This reinforces the point above that the principle difference appears to be the total amount of accreted stellar mass rather than any fundamental difference in shapes of these projected density profiles.

\begin{figure}
  \includegraphics[width=84mm, trim=0.0cm 0cm 0.0cm 0cm,clip=True]{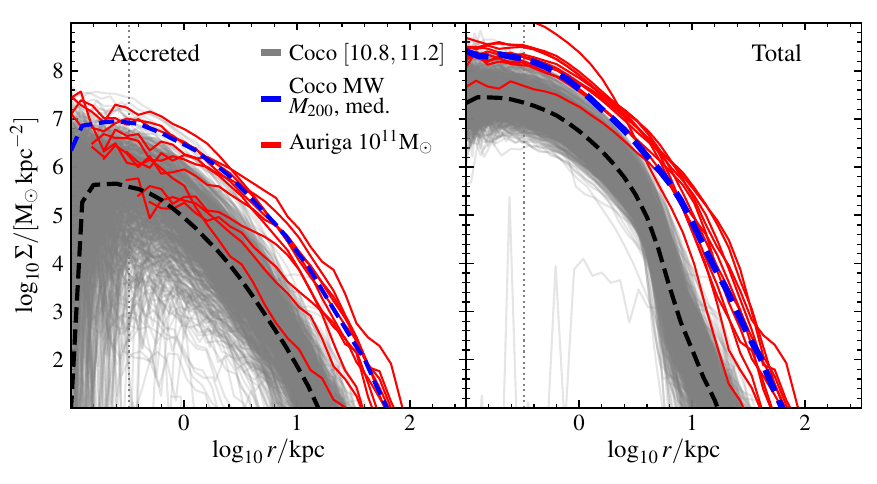}
  \includegraphics[width=84mm, trim=0.0cm 0cm 0.0cm 0cm,clip=True]{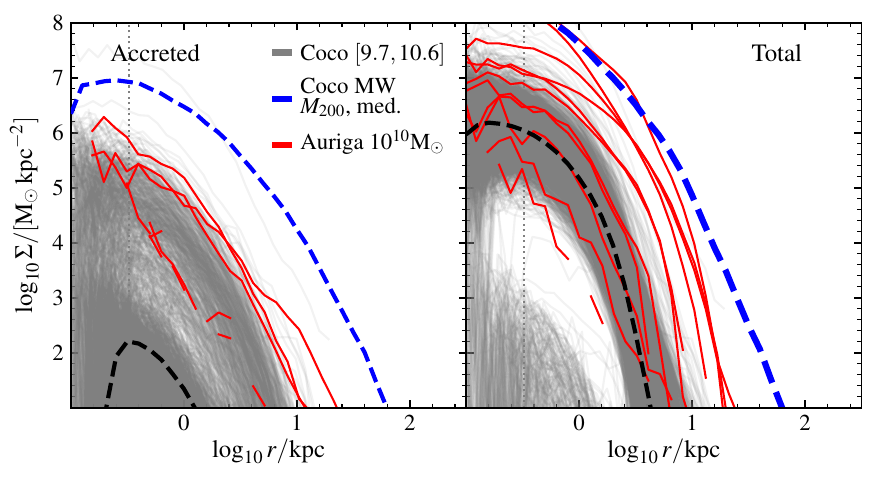}
  
  \caption{Comparison of three-dimensional stellar mass density profiles for \coco{} galaxies (grey) in two different ranges of viral mass $\log_{10}\mathrm{M_{200}}$: $[10.8,11.2]$ (top) and $[9.7,10.6]$ (bottom). Black dashed lines indicate the medians of these distributions and blue dashed lines show the median for \coco{} Milky Way analogues selected by $M_{200}$ (as in the central panel of Fig.~\ref{fig:mw_profiles_compare_dragonfly_sims}). These ranges correspond approximately to those of the Auriga  $M_{200}\sim10^{11}\,\mathrm{\Msol}$ and $M_{200}\sim10^{10}\,\mathrm{\Msol}$ simulation suites, respectively \citep{Grand:2024aa, Tau:2024aa}. Density profiles from Auriga are shown in red. Left-hand panels show the density of the accreted component only, right-hand panels show the density of all stellar mass.}.
\label{fig:appendix:coco_auriga_lowmass}\vspace{-0.2in}
\end{figure}

Fig.~\ref{fig:appendix:coco_auriga_lowmass} examines this further. We show comparisons between the \texttt{haloes\_1e11msol} (top) and \texttt{haloes\_1e10msol} (bottom) Auriga sets and approximately virial mass-matched samples of galaxies from \coco{}. The left-hand panels show the density of accreted stars only, while the right-hand panels show the total density (in spherical shells). The Auriga total mass profiles in the Fig.~\ref{fig:mw_profiles_compare_dragonfly_sims}simulations (top right) are therefore the same as those shown in the right-hand panel of Fig.~\ref{fig:mw_profiles_compare_dragonfly_sims}, where they were shown to match well to the profiles predicted in \coco{} for more massive systems. Here they are compared with systems of the same virial mass, which, as expected, have a lower amplitude. The same is true if we examine only the accreted component (upper left panel); in most cases, these profiles match well to those of \coco{} Milky Way analogues, but are much higher than the corresponding profiles for galaxies of similar mass to the \texttt{haloes\_1e10msol} set. The shapes of the spherically averaged profiles also agree well between \coco{} and Auriga. Finally, the least massive Auriga galaxies, \texttt{haloes\_1e10msol}, in many cases do not have significant accreted mass. The same is true for \coco{}, which much better samples the range of variation in this regime. The lower left panel of Fig.~\ref{fig:appendix:coco_auriga_lowmass} clearly shows that an insignificant and centrally concentrated accreted component is typical for these galaxies, although a fraction of systems have a more massive and extensive stellar halo. The predictions of Auriga and \coco{} agree much more closely at this mass scale. This agreement is also seen in the total stellar mass surface density profiles (lower right panel), although the majority of the Auriga galaxies have higher than the average \coco{} \insitu{} mass. The profiles of the \coco{} galaxies with higher \insitu{} mass fractions nevertheless match well to those from Auriga. 

From this comparison, we conclude that our particle tagging method applied to \coco{} predicts stellar mass density profiles with very similar spherically averaged density profiles to those predicted by the more self-consistent Auriga hydrodynamical simulations. This suggests that the differences in the typical amplitudes of these profiles at fixed virial mass are due mostly to differences between the SMHM relation predicted by Auriga and that predicted by the \citetalias{Lacey:2016aa}  \galform{} model, rather than to the limitations of the particle tagging method.

\label{lastpage}
\end{document}